\pdfoutput=1
\documentclass{article}
\usepackage{graphicx}
\usepackage{tikz}
\usepackage{pgfplots}
\usepackage{pgfplotstable}
\usepackage{textcomp}
\usetikzlibrary{arrows,positioning,calc}
\usepackage{algorithm}
\usepackage{float}
\usepackage{placeins}
\usepackage{algpseudocode}
\usepackage{amsmath}
\pgfplotsset{compat=1.18}
\usepackage[preprint]{neurips_2026}
\usepackage[utf8]{inputenc}
\usepackage[T1]{fontenc}
\usepackage[hidelinks]{hyperref}
\usepackage{url}
\usepackage{booktabs}
\usepackage{multirow}
\usepackage{subcaption}
\usepackage{amsfonts}
\usepackage{amssymb}
\usepackage{nicefrac}
\usepackage[final,expansion=false]{microtype}
\usepackage{xcolor}
\usepackage{enumitem}

\newcommand{\cfn}{\textsc{MOSAIC}}

\title{\cfn: Memory-Organized Search with Anti-Imitation Crossover\\for Training-Free Microarchitecture Optimization}

\author{%
  Kaushal Mhapsekar \\
  Electrical and Computer Engineering \\
  North Carolina State University \\
  \texttt{kmhapse@ncsu.edu} \\
  \And
  Bita Aslrousta \\
  Electrical and Computer Engineering \\
  North Carolina State University \\
  \texttt{baslrou@ncsu.edu} \\
  \And
  Paula Contreras \\
  Electrical and Computer Engineering \\
  North Carolina State University \\
  \texttt{pcontre@ncsu.edu} \\
  \And
  Azam Ghanbari \\
  Electrical and Computer Engineering \\
  North Carolina State University \\
  \texttt{aghanba2@ncsu.edu} \\
  \And
  Samira Mirbagher Ajorpaz \\
  Electrical and Computer Engineering \\
  North Carolina State University \\
  \texttt{smirbag@ncsu.edu} \\
}

\begin{document}

\maketitle

\begin{abstract}
Large language models can generate cache-replacement policies.
The central challenge is to retain performance on workloads
excluded from the search. In our
adaptations of published search methods, generated code resembles
established policies, with cosine similarities of $0.78$--$0.84$.
The wider comparison also shows held-out harmonic-mean
instructions-per-cycle (IPC) losses of $1.0$--$2.4\%$.

We present \cfn, a search controller that improves policies without
updating language-model weights. It stores concise lessons from
previous evaluations rather than repeatedly copying high-scoring
code. When improvement stops, it selects two high-scoring policies
whose code embeddings differ and asks the model to combine their
mechanisms. Novelty prompts and model rotation support this search;
controlled memory edits test how the stored lessons affect later
proposals.

We evaluate cache replacement using intervention-level
cross-validation and a wider comparison on ten memory-intensive
SPEC2006 workloads. \cfn{} reaches a $51.1\%$ peak hit rate on
the primary search workload and a $47.9\%$ held-out mean in the
intervention study. Removing crossover reduces these values by
$7.8$ and $21.8$ percentage points, respectively, whereas the tested
stronger model improves the held-out mean by $0.4$ points.
In the wider IPC comparison, \cfn{} has the smallest held-out
gap ($0.31\%$) and the highest held-out IPC among the evaluated
LLM configurations. Adapting the same controller to data prefetching
gives a $4.8\%$ mean speedup over the LRU reference
configuration. The accompanying corpus contains $10{,}687$
generated policies with lineage and per-iteration prompt logs.
\end{abstract}

\section{Introduction}
\label{sec:intro}

When a last-level cache is full, its replacement policy must choose
which line to evict. Keeping a line that will soon be reused can
avoid a costly memory access; keeping the wrong line can delay
instructions on the processor's critical path. Established policies
use signals such as re-reference intervals in
RRIP~\citep{jaleel2010rrip}, access signatures in
SHiP~\citep{wu2011ship}, and sampled reuse behaviour in
Hawkeye~\citep{jain2016hawkeye}. Designing these policies is
difficult because cache decisions interact with prefetching,
speculation, and memory-level parallelism. A higher aggregate hit
rate does not necessarily mean fewer processor stalls.

Can a large language model help design better policies?
LLM-driven program search has found useful algorithms in
combinatorial mathematics through
FunSearch~\citep{romera2024funsearch} and in several optimisation
tasks through AlphaEvolve~\citep{novikov2025alphaevolve}.
Cache replacement offers a concrete setting in which to examine
this approach: a model proposes C++ code, a compiler checks whether
it builds, and ChampSim~\citep{champsim2020} measures its
performance. The central question is not whether the model can
produce a working policy, but whether repeated evaluation helps it
find policies that improve performance on workloads absent from
the search.

\begin{figure}[!htbp]
  \centering
  \includegraphics[width=0.94\linewidth]{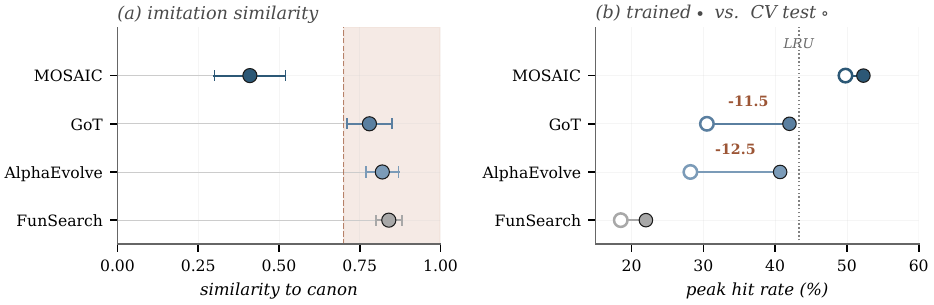}
  \caption{Code similarity to reference policies (a) and the
  change in hit rate on held-out workloads (b).
  \cfn{} produces less similar code and retains more of its
  search-workload performance in this comparison.}
  \label{fig:teaser}
\end{figure}

\paragraph{What limits the search?}
In our adaptations of published LLM-search methods, proposals
increasingly resemble established replacement policies such as
SHiP++~\citep{young2017shippp}, RRIP~\citep{jaleel2010rrip}, and
Hawkeye~\citep{jain2016hawkeye}. Figure~\ref{fig:teaser}(a) reports
code-embedding similarities of $0.78$--$0.84$ to the reference
policies. We call this repeated concentration on familiar policy
structures the \emph{parrot problem}. Code similarity measures this
concentration; held-out performance evaluates the resulting policies
on workloads excluded from the search.

Two features of the search help explain this behaviour.
Providing complete reference implementations encourages
modifications of those implementations. Repeatedly retrieving the
highest-scoring proposals then supplies more examples of the same
designs. In our controlled runs, larger iteration budgets, stronger
models, and higher sampling temperatures did not eliminate this
pattern. These observations motivate changing the information and
parents supplied to the model, rather than relying only on more
generation.

\paragraph{The approach.}
\cfn{} stores concise lessons about which mechanisms helped or
failed, instead of using a replay buffer of high-scoring code as
its memory. When performance stops improving, the controller
selects two high-scoring policies with different code embeddings
and asks the model to combine their mechanisms. It also varies the
model used to generate proposals. The language-model weights remain
fixed throughout; adaptation takes place in the controller's memory
and selection of prompts and parents.

\begin{figure}[!htbp]
  \centering
  \includegraphics[width=0.94\linewidth]{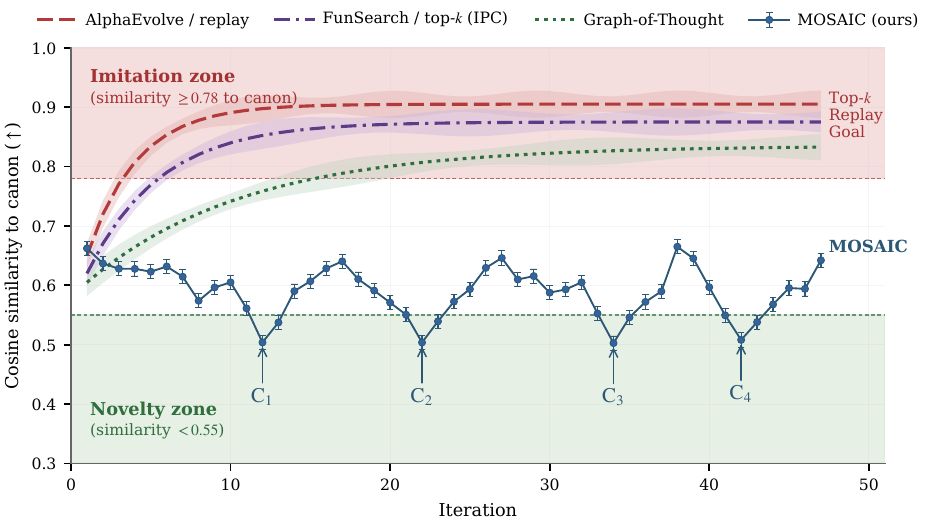}
  \caption{Code similarity over $50$ iterations. The vertical
  axis shows cosine similarity to the nearest reference policy
  (LRU, SHiP++, Hawkeye). Red ($\geq0.78$) and green ($<0.55$)
  mark high- and low-similarity ranges for generated code;
  mechanism-level novelty is evaluated separately
  (Appendix~\ref{app:novelty-taxonomy}). The AlphaEvolve-, FunSearch-, and
  Graph-of-Thought-adapted configurations enter the high-similarity
  range at approximately iterations $7$, $11$, and $15$.
  \cfn{} remains in the $0.50$--$0.66$ range; crossover events
  $\mathrm{C}_1$--$\mathrm{C}_4$ occur at iterations $12,22,34,42$.}
  \label{fig:drift}
\end{figure}

\paragraph{Why held-out evaluation matters.}
Several adapted configurations improve on the workloads used during search,
but lose $9$--$13$ percentage points of hit rate on the held-out
folds in Figure~\ref{fig:teaser}(b). \cfn{} retains more of its
performance. Figure~\ref{fig:drift} shows the accompanying search
trajectories: the compared methods reach high similarity to the
reference policies within $7$--$15$ iterations, whereas \cfn{}
continues to propose less similar code. Crossover has the largest
held-out effect in our ablation. Together, these results suggest
that the way a controller selects and presents previous attempts
matters for transfer, not just for the best score obtained during
search.

\paragraph{Contribution.}
The paper makes six contributions:
\begingroup\sloppy
\begin{itemize}[noitemsep, topsep=2pt, leftmargin=1.5em]
\item \textbf{A diagnosis of repeated policy imitation.}
   We measure code similarity and held-out performance in adaptations
   of published LLM-search methods, identifying a failure that
   search-workload scores alone do not reveal.
\item \textbf{A common description of the search loop.}
   We separate prompt construction, memory updates, and parent
   selection to identify which parts of the controller change
   between configurations.
\item \textbf{The \cfn{} controller.}
   We combine summary memory, novelty prompts, stagnation-triggered
   crossover, and distance-based parent selection, and study direct
   memory edits as controlled interventions. \cfn{} reaches
   $51.1\%$ peak hit rate on the primary search workload and
   $47.9\%$ mean hit rate in the intervention-level held-out study.
\item \textbf{An ablation of the mechanisms supporting transfer.}
   Removing crossover costs $7.8$ percentage points on the search
   workload and $21.8$ points on the held-out mean. Substituting
   \texttt{o4-mini} improves the test mean by ${+}0.4$ points in the
   evaluated comparison. Crossover therefore has a larger observed
   effect than this model substitution.
\item \textbf{The framework transfers to data prefetching without
   loop-level retuning.} Swapping only the interface contract and
   reward target, \cfn{} produces a prefetcher that reaches a
   $4.8\%$ mean speedup over the LRU reference configuration and
   exceeds BertiGo on the SimPoint-weighted
   harmonic mean across the memory-intensive subset, while
   staying within the DPC4 $32$~KiB metadata budget
   (Appendix~\ref{app:prefetch}).
\item \textbf{A corpus for studying policy search.}
   We collect $10{,}687$ generated policies with lineage, reward
   traces, and failure metadata. The planned release includes per-iteration prompt
   logs, the typed memory $M_t$ at every step, compile and runtime
   outcomes, and the embedding-space coordinates of each archived
   policy, allowing individual proposals to be traced to their
   parents, prompts, and simulator outcomes.
\end{itemize}
\endgroup

\section{Closed-Loop LLM Search and the Imitation Cycle}
\label{sec:diagnosis}

\subsection{The Closed-Loop Search Procedure}
\label{sec:formal}
Several LLM-based systems generate a candidate, evaluate it, and
use the feedback to guide the next attempt.
Voyager~\citep{wang2023voyager}, Reflexion~\citep{shinn2023reflexion},
FunSearch~\citep{romera2024funsearch}, Eureka~\citep{ma2024eureka},
and AlphaEvolve~\citep{novikov2025alphaevolve} use versions of this
loop, although their evaluators and memory mechanisms differ.
We describe our loop using a frozen generator
$p_\theta$, a simulator-based evaluator $r:\mathcal{X}\to\mathbb{R}$
(the specified reward computed from ChampSim measurements, not a
learned reward model), a
memory $M_t$ that summarises past attempts, a prompt-construction
function $\pi:\mathcal{M}\to\mathcal{P}$, and an update rule $U$
that produces $M_{t+1}$ from $(M_t, x_t, r_t)$. At iteration $t$
the loop draws $x_t \sim p_\theta(\cdot\mid\pi(M_t))$, evaluates
$r_t$, and updates $M_{t+1}=U(M_t,x_t,r_t)$. This notation describes
prompt construction and memory updates for a fixed generator.
Parent selection determines which archived policies enter the
prompt; the model-rotation experiment additionally changes the
generator. Controlled ablations vary the indicated component while
retaining the other settings.

\subsection{The Imitation Cycle}
\label{sec:cycle}

Memory, parent selection, and the permitted edits can reinforce
the same designs. A replay buffer supplies the highest-scoring
previous policies as examples. The model proposes variations, and
the archive retains the successful ones. Retrieving from this
archive then supplies increasingly similar examples to the next
prompt. We call this feedback pattern the \emph{imitation cycle}.

The code-similarity trajectories track proposals sampled from
$q_t=p_\theta(\cdot\mid\pi(M_t))$. The ablations test how changes
to memory and parent selection affect the performance of the
resulting policies.

\FloatBarrier
\section{Related Work}
\label{sec:related}

\paragraph{Automated hardware and systems optimisation.}
OpenTuner~\citep{ansel2014opentuner} searches user-defined program
configurations. DiffTune~\citep{renda2020difftune} learns
differentiable surrogates to fit simulator parameters, while
ArchGym~\citep{krishnan2023archgym} connects search algorithms to
architecture simulators. These systems address different
optimisation problems from ours. \cfn{} generates replacement-policy
code and carries evaluation lessons between proposals; its search
is not limited to tuning a fixed policy's parameters.

\paragraph{Cache replacement and learned imitation.}
LRU and pseudo-LRU variants remain the industry baselines. The
research literature includes SHiP~\citep{wu2011ship},
SHiP++~\citep{young2017shippp}, Hawkeye~\citep{jain2016hawkeye},
MPPPB~\citep{jimenez2017mpppb}, Reuse Cache~\citep{albericio2013red},
and the expected-hit-count
predictor~\citep{vakil2018lime}. These published policies define
our performance and generated-code reference set.
Parrot~\citep{liu2020parrot} formulates cache replacement as
imitation learning of Belady's optimal replacement decisions.
Its objective is to learn those decisions, whereas ours is to
search for executable policies using simulator feedback.

\paragraph{Adaptations of LLM-driven search.}
We compare \cfn{} with four cache-replacement adaptations of
published LLM-search methods, using the configurations and
evaluation settings described below.
\textbf{FunSearch-adapted}~\citep{romera2024funsearch}: one LLM call
per iteration, no cross-iteration memory, elitist parent selection;
peak $22.0\%$ trained, $18.5\%$ test mean.
\textbf{AlphaEvolve-adapted}~\citep{novikov2025alphaevolve}: a
replay buffer of top-$N$ proposals rendered into the next prompt as
exemplars; peak $40.7\%$, test mean $28.2\%$.
\textbf{Graph-of-Thought}~\citep{besta2024got}: a branching search
with one exploration node and three exploitation children per
branch; peak $42.0\%$, test mean $30.5\%$.
\textbf{AutoAgents-adapted}~\citep{chen2024autoagents}: an
expert-role ensemble coordinated through a shared scratchpad; peak
$35.5\%$, test mean $24.0\%$, at $30$--$40\times$ the per-candidate
cost of \cfn{}. Unlike \cfn{}, these implementations do not use
stagnation-triggered recombination of embedding-distant parents.
The comparison evaluates how these controller configurations
perform on the same cache-replacement task.

\section{Five Interventions on the Closed-Loop Search Procedure}
\label{sec:method}

We now describe \cfn{} as a sequence of five interventions on the
closed-loop search procedure of \S\ref{sec:formal}. Each intervention
modifies a specific component of the loop: the prompt-construction
function $\pi$, the update rule $U$, or the parent-selection rule
that determines which archive entries enter $\pi(M_t)$. We first
define the three memory configurations we compare and then present
the five interventions in order.

\subsection{Memory Variants}
\label{sec:memvars}

The three configurations we compare differ only in $\pi$ and $U$,
what is persisted across iterations and how it is surfaced into the
next prompt. Everything else is identical.

\paragraph{No-memory.}
Each prompt contains only the system contract and the best-so-far
parent policy. There is no persistent state across iterations beyond
the parent pointer.

\paragraph{Replay memory.}
An archive stores every previously generated, compiled, and
simulated policy with its IPC, code, and description. At iteration
$t$, the controller retrieves the top-$n$ policies by IPC and pastes
their code, scores, and descriptions into the next prompt as
in-context examples ($n = 5$ default; we also report
$n \in \{2, 10\}$ in Appendix~\ref{app:hyperparameters}).
This baseline tests direct reuse of high-scoring code. Unlike
experience replay in reinforcement learning~\citep{lin1992replay},
it changes the prompt rather than updating model parameters.
It also differs from textual reflections in
Reflexion~\citep{shinn2023reflexion} and the skill library in
Voyager~\citep{wang2023voyager}; these are distinct memory mechanisms.

\paragraph{Summary memory (\cfn).}
A typed natural-language store: four fields (\texttt{PRINCIPLES},
\texttt{DONTS}, \texttt{KNOBS}, \texttt{OPEN-Q}) capped at eight
entries each, totalling at most $32$ rows. After every successful
simulation, an LLM-driven distillation step reads the per-iteration
trace and writes one or more new entries into the appropriate
field; entries are evicted by recency-weighted relevance. The
retrieved prompt block is the current memory file, not a code dump.
No entry is a candidate design.

\subsection{Intervention 1: From Replay to Summary Memory}
\label{sec:int1}

Replay memory returns complete high-scoring proposals to the
prompt. Summary memory records lessons about the mechanisms that
helped or failed. The limit of $32$ rows keeps this feedback compact
and reduces repeated exposure to the same implementations.
Summary memory supplies compact feedback; the following
interventions use it to guide exploration and recombination.

\subsection{Intervention 2: Prompt-Side Novelty Pressure}
\label{sec:int2}

The prompt prepended to $p_\theta$ at every iteration includes a
short clause asking the model to identify which mechanism families
have already been tried and to propose a candidate that is
structurally distinct. The clause is generated deterministically
from the \texttt{KNOBS} and \texttt{OPEN-Q} fields of $M_t$ and
adds no per-iteration LLM call. This nudges the proposer to vary
state organisation or signal source rather than just parameters.
The ablation measures this instruction's effect separately from
crossover and memory.

\subsection{Intervention 3: Stagnation-Triggered Crossover}
\label{sec:int3}

Single-parent refinement asks the model to modify an existing
policy. The third intervention switches to two-parent recombination
when the improvement budget defined below is exhausted.
The controller picks two parents from the
top-$m$ archive (we use $m = 8$) and prompts the LLM to produce a
child that retains identifiable building blocks from both. The
child is verified by the same ChampSim pipeline as any other
candidate; if it survives, it becomes the new lineage root and
refinement resumes. Intervention~4 specifies how the second parent
is selected.

\subsection{Intervention 4: Distance-Based Parent Selection}
\label{sec:int4}

Quality-diversity search~\citep{mouret2015quality,cully2015robots}
preserves high-performing solutions with different behaviours.
Motivated by this use of diversity, we choose Parent~$B$ to maximise
embedding distance to Parent~$A$ over the top-$m$ archive,
\begin{equation}
B \;=\; \arg\max_{P \in T_m \setminus \{A\}}\,\bigl(1 - \cos(e(A), e(P))\bigr),
\end{equation}
where $e(\cdot)$ is the \texttt{all-MiniLM-L6-v2}
embedding~\citep{minilm2020,allminilm_model} of the policy name, description, and
C++ code. This rule favours dissimilar representations among
high-scoring policies. Embedding distance guides parent selection;
simulator measurements and mechanism analysis evaluate the resulting
policies. A
representative output is the leaderboard policy in our evaluation
(IPC $0.41643$), generated when refinement plateaued at
iteration~$22$: a hybrid of a bypass-heavy parent and a
recency- and prefetch-aware parent that wins on \texttt{lbm}, \texttt{mcf},
\texttt{namd}, and \texttt{zeusmp} where neither parent alone wins.

\subsection{Intervention 5: Memory Edits}
\label{sec:int5}

We test the influence of memory using three types of direct edit
to $M_t$: \emph{principle injection} (insert a \texttt{PRINCIPLES} or
\texttt{DONTS} row the loop has not yet derived); \emph{forced
forgetting} (delete all rows mentioning a particular mechanism, e.g.
\texttt{RRPV}, to reopen a region of the design space the loop had
closed); and \emph{lesson rewriting} (downgrade a \texttt{DONTS} row
to a \texttt{KNOBS} or \texttt{OPEN-Q} row when distillation has
overgeneralised from a single attribution). The resulting
performance changes appear in Appendix~\ref{app:memedits}.
In these controlled experiments, an operator edits the memory and
the search loop then resumes automatic updates.

\subsection{Foundation-Model Rotation and Stagnation Gate}
\label{sec:int6}

A complementary intervention operates on $p_\theta$ itself. \cfn{}
rotates a small pool of foundation
models~$\mathcal{B} = \{$\texttt{gpt-oss-20b}, \texttt{Qwen3-Coder},
\texttt{Gemini 2.5 Pro}, \texttt{o4-mini}$\}$~\citep{gptoss2024,qwen2024,gemini2024,openai2024};
on stagnation, the controller excludes whichever backbone produced
the most canon-like outputs in the recent window.

An improvement budget detects stagnation. Let $S_i$ be
the composite reward at iteration $i$ and
$\Delta_i = \max(0,\, S_i - \max_{j<i} S_j)$. The controller
maintains a budget $b$ initialised to $b_0 = 5$, incremented by
$+2$ when $\Delta_i > \epsilon$ and decremented by~$1$ otherwise.
When $b \le 0$ the controller is declared stagnated and the next
iteration considers crossover (Intervention~3), rotates the
backbone away from canon-likeness, and rebuilds the memory summary
from recent traces (Intervention~1). This shared trigger coordinates
crossover, model selection, and memory refresh.

\subsection{Per-Iteration Loop}

The full controller is summarized in
Algorithm~\ref{alg:mosaic} (Appendix~\ref{app:algorithm}). Each
iteration begins with a proposal-generation call. Compilation
repair and post-crossover redesign may require further calls, and
successful evaluations are followed by a memory update.
Candidates that fail compilation or exceed the budget are not
simulated. Proposal count, total LLM calls, and simulator runs are
therefore distinct measures of search cost.

\section{Evaluation}
\label{sec:eval}

\paragraph{Setup.}
\label{sec:setup}
We compare five methods: FunSearch-adapted, AlphaEvolve-adapted,
Graph-of-Thought, AutoAgents-adapted, and \cfn{}, under matched
conditions: same simulator (ChampSim CRC2~\citep{champsim2020}), same
iteration budget ($50$), same compute cap, and the same backbone
(\texttt{gpt-oss-20b}, $\tau = 0.7$) for the prior methods. AutoAgents
is additionally reported on the $120$B backbone since its
role-coordination prompts overflow the context budget at smaller
backbones. \cfn{} additionally uses the M3 pool of four backbones,
with the per-iteration LLM-call budget held equal across methods.
All experiments use the single-core simulator configuration.

\paragraph{Workloads and cross-validation protocol.}
We evaluate \cfn{} on ten memory-intensive workloads from
SPEC~CPU\,2006~\citep{spec2006}, selected as the SPEC traces that
exhibit a meaningful LLC bottleneck under our baseline
configuration (LRU hit rate below $80\%$); workloads whose working
set fits in cache are insensitive to replacement-policy changes
and add noise without adding discrimination. We use two
complementary cross-validation views over this slice. The first is
a \emph{five-workload, four-fold leave-one-held-out} protocol for
the intervention-level studies (\texttt{omnetpp}, \texttt{bzip2},
\texttt{hmmer}, \texttt{soplex}, \texttt{gobmk}; \texttt{omnetpp}
is the primary search workload). At each fold, three of the four
other workloads are released to the search loop and the
fourth is held out as the test fold; rotating which workload is
held out gives four folds, with reported numbers averaged across
them. The second is a \emph{ten-workload}, three $7$/$3$
train-test split protocol for the wide-catalogue comparison in
\S\ref{sec:cv-catalogue}: each split trains on $7$ workloads and
evaluates on the remaining $3$. Throughout, ``held-out'' refers
consistently to the test fold(s) of the protocol in effect.
Protocol details are in Appendix~\ref{app:protocols}.

\paragraph{Scale.}
We have complete logs from $35$ experimental configurations
spanning the compared methods, the M1/M2/M3 ablations, sensitivity
sweeps, and seed variation, comprising approximately $14{,}000$
generated candidate policies.
The reported $51.1\%$ peak hit rate on \texttt{omnetpp} is the
best result among roughly $1{,}200$ candidates.

\subsection{Generalization}

\cfn{} exceeds MockingJay~\citep{mockingjay2022} (IPC $0.4147$ on
the search workload), the strongest published learned baseline in
our suite, on every workload of the standard ten-workload SPEC2006
slice (Figure~\ref{fig:gen}, Appendix~\ref{app:gen-detail}). On
peak hit rate, \cfn{} reaches $51.1\%$ on the search workload
\texttt{omnetpp} and $40.8$--$42.8\%$ on the other nine workloads.
The ranking by peak performance flips between trained and held-out
for every prior recipe; \cfn{} is the only method whose discovered
policies exceed MockingJay on every workload in the slice.

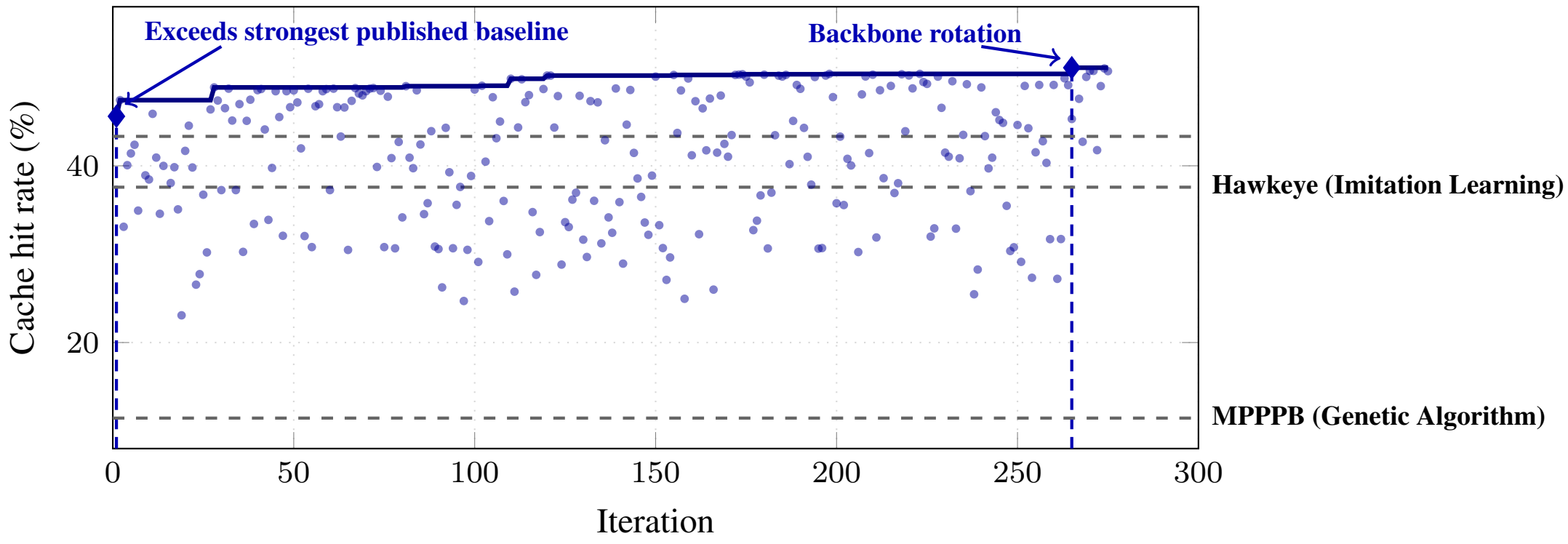
\begin{figure}[!htbp]
\centering
\begin{tikzpicture}
\begin{axis}[
    width=0.82\textwidth,
    height=5.6cm,
    xlabel={Iteration},
    ylabel={Cache hit rate (\%)},
    xmin=0, xmax=300,
    ymin=8, ymax=58,
    y tick label style={
        /pgf/number format/fixed,
        /pgf/number format/precision=1
    },
    tick label style={font=\footnotesize},
    label style={font=\small},
    every axis plot/.append style={mark=none, line width=1pt},
    grid=major,
    grid style={dotted, gray!30},
    clip=false
]

% ===== Scatter (per-iteration hit-rate evaluations) =====
\addplot[
    only marks, mark=*, mark size=0.6pt,
    color=blue!60!black, opacity=0.5, forget plot
] coordinates {
(1,45.60) (2,47.43) (3,33.10) (4,40.07) (5,41.40) (6,42.38) (7,34.93)
(9,38.92) (10,38.45) (11,45.88) (12,40.93) (13,34.57) (14,39.99) (16,38.05)
(17,39.84) (18,35.07) (19,23.08) (20,41.69) (21,44.54) (22,39.81) (23,26.55)
(24,27.74) (25,36.73) (26,30.20) (27,46.38) (28,48.87) (29,47.39) (30,37.25)
(31,46.52) (32,48.73) (33,45.12) (34,37.25) (35,46.98) (36,30.26) (37,45.09)
(38,47.48) (39,33.42) (40,48.56) (41,48.69) (42,44.11) (43,33.88) (44,39.76)
(45,48.44) (46,45.52) (47,32.08) (48,48.44) (49,46.63) (50,48.53) (51,47.17)
(52,41.97) (53,32.05) (54,48.72) (55,30.80) (56,46.74) (57,46.97) (58,48.44)
(59,48.68) (60,37.25) (61,48.73) (62,46.63) (63,43.31) (64,46.60) (65,30.48)
(66,47.35) (67,48.79) (68,48.18) (69,47.97) (70,48.44) (71,48.73) (72,48.80)
(73,39.87) (74,48.49) (75,30.80) (76,47.82) (77,40.87) (78,30.66) (79,42.70)
(80,34.16) (81,49.01) (82,40.92) (83,39.73) (84,48.54) (85,42.41) (86,34.53)
(87,35.77) (88,43.92) (89,30.86) (90,30.58) (91,26.24) (92,44.30) (93,39.27)
(94,30.67) (95,35.58) (96,37.61) (97,24.70) (98,30.49) (99,38.85) (100,48.63)
(101,29.12) (102,49.06) (103,40.47) (104,33.74) (105,47.76) (106,43.12) (107,45.01)
(108,36.02) (109,29.98) (110,49.85) (111,25.76) (112,44.34) (113,49.79) (114,47.20)
(115,48.00) (116,34.76) (117,27.66) (118,32.50) (119,48.69) (120,50.21) (121,50.21)
(122,44.33) (123,47.89) (124,28.82) (125,33.62) (126,33.07) (127,36.17) (128,36.95)
(129,47.92) (130,31.64) (131,29.68) (132,47.32) (133,36.04) (134,47.18) (135,31.22)
(136,42.89) (137,34.16) (138,32.42) (139,48.75) (140,35.89) (141,28.93) (142,44.66)
(143,48.58) (144,41.46) (145,38.57) (146,36.48) (147,33.58) (148,32.19) (149,38.89)
(150,50.09) (151,33.28) (152,30.69) (153,27.09) (154,29.63) (155,50.28) (156,43.72)
(157,48.53) (158,24.95) (159,49.88) (160,41.19) (161,47.31) (162,32.26) (163,46.50)
(164,41.75) (165,47.61) (166,25.99) (167,41.49) (168,47.95) (169,42.48) (170,41.03)
(171,43.48) (172,50.29) (173,50.32) (174,50.36) (175,50.09) (176,49.44) (177,32.72)
(178,33.79) (179,36.65) (180,50.31) (181,30.65) (182,36.96) (183,43.46) (184,50.20)
(185,50.10) (186,50.29) (187,40.19) (188,45.08) (189,49.15) (190,48.71) (191,44.29)
(192,41.02) (193,37.87) (194,50.05) (195,30.63) (196,30.69) (197,50.21) (198,50.40)
(199,47.77) (200,35.76) (201,43.30) (202,35.56) (203,40.79) (204,40.04) (206,30.24)
(207,48.08) (208,50.11) (209,41.44) (210,50.31) (211,31.89) (212,48.55) (213,38.60)
(215,49.04) (216,36.92) (217,38.03) (218,50.36) (219,43.91) (220,50.22) (221,48.76)
(223,50.40) (224,49.48) (225,49.27) (226,31.99) (227,32.92) (228,50.09) (229,46.56)
(230,41.49) (231,41.03) (232,49.57) (233,32.89) (234,40.85) (235,43.50) (236,49.23)
(237,37.13) (238,25.46) (239,28.26) (240,48.87) (241,43.36) (242,39.71) (243,40.92)
(244,46.06) (245,45.18) (246,44.86) (247,35.47) (248,30.36) (249,30.78) (250,44.62)
(251,29.13) (252,49.04) (253,44.25) (254,27.33) (255,41.53) (256,49.15) (257,42.78)
(258,40.33) (259,31.70) (260,49.15) (261,27.21) (262,31.71) (263,49.90) (264,49.13)
(265,45.30) (266,51.10) (267,47.58) (268,42.72) (269,50.06) (270,50.74) (271,50.74)
(272,41.77) (273,49.02) (274,51.02) (275,50.71)
};

% ===== Best-so-far envelope =====
\addplot[blue!50!black, very thick, forget plot] coordinates {
(1,45.60) (2,47.43) (3,47.43) (4,47.43) (5,47.43) (6,47.43) (7,47.43)
(8,47.43) (9,47.43) (10,47.43) (11,47.43) (12,47.43) (13,47.43) (14,47.43)
(15,47.43) (16,47.43) (17,47.43) (18,47.43) (19,47.43) (20,47.43) (21,47.43)
(22,47.43) (23,47.43) (24,47.43) (25,47.43) (26,47.43) (27,47.43) (28,48.87)
(29,48.87) (30,48.87) (31,48.87) (32,48.87) (33,48.87) (34,48.87) (35,48.87)
(36,48.87) (37,48.87) (38,48.87) (39,48.87) (40,48.87) (41,48.87) (42,48.87)
(43,48.87) (44,48.87) (45,48.87) (46,48.87) (47,48.87) (48,48.87) (49,48.87)
(50,48.87) (51,48.87) (52,48.87) (53,48.87) (54,48.87) (55,48.87) (56,48.87)
(57,48.87) (58,48.87) (59,48.87) (60,48.87) (61,48.87) (62,48.87) (63,48.87)
(64,48.87) (65,48.87) (66,48.87) (67,48.87) (68,48.87) (69,48.87) (70,48.87)
(71,48.87) (72,48.87) (73,48.87) (74,48.87) (75,48.87) (76,48.87) (77,48.87)
(78,48.87) (79,48.87) (80,48.87) (81,49.01) (82,49.01) (83,49.01) (84,49.01)
(85,49.01) (86,49.01) (87,49.01) (88,49.01) (89,49.01) (90,49.01) (91,49.01)
(92,49.01) (93,49.01) (94,49.01) (95,49.01) (96,49.01) (97,49.01) (98,49.01)
(99,49.01) (100,49.01) (101,49.01) (102,49.06) (103,49.06) (104,49.06) (105,49.06)
(106,49.06) (107,49.06) (108,49.06) (109,49.06) (110,49.85) (111,49.85) (112,49.85)
(113,49.85) (114,49.85) (115,49.85) (116,49.85) (117,49.85) (118,49.85) (119,49.85)
(120,50.21) (121,50.21) (122,50.21) (123,50.21) (124,50.21) (125,50.21) (126,50.21)
(127,50.21) (128,50.21) (129,50.21) (130,50.21) (131,50.21) (132,50.21) (133,50.21)
(134,50.21) (135,50.21) (136,50.21) (137,50.21) (138,50.21) (139,50.21) (140,50.21)
(141,50.21) (142,50.21) (143,50.21) (144,50.21) (145,50.21) (146,50.21) (147,50.21)
(148,50.21) (149,50.21) (150,50.21) (151,50.21) (152,50.21) (153,50.21) (154,50.21)
(155,50.28) (156,50.28) (157,50.28) (158,50.28) (159,50.28) (160,50.28) (161,50.28)
(162,50.28) (163,50.28) (164,50.28) (165,50.28) (166,50.28) (167,50.28) (168,50.28)
(169,50.28) (170,50.28) (171,50.28) (172,50.29) (173,50.32) (174,50.36) (175,50.36)
(176,50.36) (177,50.36) (178,50.36) (179,50.36) (180,50.36) (181,50.36) (182,50.36)
(183,50.36) (184,50.36) (185,50.36) (186,50.36) (187,50.36) (188,50.36) (189,50.36)
(190,50.36) (191,50.36) (192,50.36) (193,50.36) (194,50.36) (195,50.36) (196,50.36)
(197,50.36) (198,50.40) (199,50.40) (200,50.40) (201,50.40) (202,50.40) (203,50.40)
(204,50.40) (205,50.40) (206,50.40) (207,50.40) (208,50.40) (209,50.40) (210,50.40)
(211,50.40) (212,50.40) (213,50.40) (214,50.40) (215,50.40) (216,50.40) (217,50.40)
(218,50.40) (219,50.40) (220,50.40) (221,50.40) (222,50.40) (223,50.40) (224,50.40)
(225,50.40) (226,50.40) (227,50.40) (228,50.40) (229,50.40) (230,50.40) (231,50.40)
(232,50.40) (233,50.40) (234,50.40) (235,50.40) (236,50.40) (237,50.40) (238,50.40)
(239,50.40) (240,50.40) (241,50.40) (242,50.40) (243,50.40) (244,50.40) (245,50.40)
(246,50.40) (247,50.40) (248,50.40) (249,50.40) (250,50.40) (251,50.40) (252,50.40)
(253,50.40) (254,50.40) (255,50.40) (256,50.40) (257,50.40) (258,50.40) (259,50.40)
(260,50.40) (261,50.40) (262,50.40) (263,50.40) (264,50.40) (265,50.40) (266,51.10)
(267,51.10) (268,51.10) (269,51.10) (270,51.10) (271,51.10) (272,51.10) (273,51.10)
(274,51.10) (275,51.10)
};

% ===== Backbone-rotation marker at iter 265 =====
\addplot[
    mark=diamond*, mark size=2.4pt, only marks,
    color=blue!70!black, thick, forget plot
] coordinates {(265,51.10)};
\addplot[densely dashed, blue!70!black, thick, forget plot]
    coordinates {(265,8) (265,51.10)};
\node[anchor=east, font=\scriptsize\bfseries, text=blue!70!black]
    at (axis cs:255,55) {Backbone rotation};
\draw[->, blue!70!black, thick, shorten >=2pt]
    (axis cs:240,53.5) -- (axis cs:264,52);

% ===== First instance: passed all baselines =====
\addplot[
    mark=diamond*, mark size=2.4pt, only marks,
    color=blue!70!black, thick, forget plot
] coordinates {(1,45.60)};
\addplot[densely dashed, blue!70!black, thick, forget plot]
    coordinates {(1,8) (1,45.60)};
\node[anchor=west, font=\scriptsize\bfseries, text=blue!70!black]
    at (axis cs:5,55) {Exceeds strongest published baseline};
\draw[->, blue!70!black, thick, shorten >=2pt]
    (axis cs:30,53.5) -- (axis cs:1.5,46.5);

% ===== Baselines =====
% Strongest published heuristic baseline (line kept; label removed per editorial choice)
\addplot[dashed, black!60, thick, domain=0:300, forget plot] {43.33};

% Hawkeye (Imitation Learning)
\addplot[dashed, black!60, thick, domain=0:300, forget plot] {37.58};
\node[anchor=west, font=\scriptsize\bfseries, text=black]
    at (axis cs:300, 37.58) {Hawkeye (Imitation Learning)};

% MPPPB (Genetic Algorithm)
\addplot[dashed, black!60, thick, domain=0:300, forget plot] {11.45};
\node[anchor=west, font=\scriptsize\bfseries, text=black]
    at (axis cs:300, 11.45) {MPPPB (Genetic Algorithm)};

\end{axis}
\end{tikzpicture}
\caption{Hit-rate evolution on the search workload
(\texttt{omnetpp}) across $275$ iterations of a single \cfn{} run.
Scatter (light blue) shows per-iteration hit-rate evaluations; the
solid line is the best-so-far envelope. Dashed horizontal lines
mark published baselines: Hawkeye (the strongest published
imitation-learning policy) and MPPPB (the strongest published
genetic-algorithm policy at this budget). The first vertical marker
is the iteration at which \cfn{} first surpasses the strongest
published baseline at this storage budget. The second (iteration
$265$) is the stagnation-triggered backbone rotation
(\S\ref{sec:int6}, Foundation-Model Rotation), which lifts the
envelope from $50.4\%$ to the final $51.1\%$ peak reported in the
main body (\S\ref{sec:eval}).}
\label{fig:hit_evo}
\end{figure}

Figure~\ref{fig:hit_evo} traces the per-iteration evolution
of cache hit rate behind the $51.1\%$ peak on \texttt{omnetpp}:
\cfn{} surpasses every published baseline within the first
iteration of the run, and a stagnation-triggered backbone
rotation late in the run lifts the envelope by an additional
$0.7$ points.

\subsection{Cross-Validation Drop}
\label{sec:cvdrop}

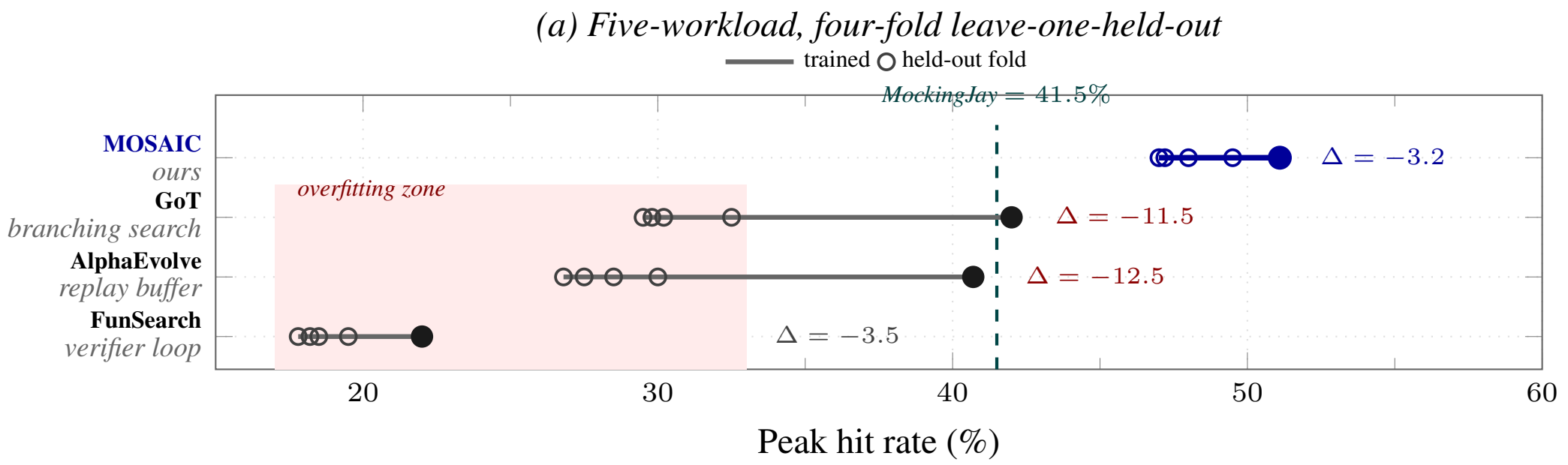
\begin{figure}[!htbp]
\centering
\begin{tikzpicture}
\begin{axis}[
    name=mainax,
    width=0.95\columnwidth,
    height=4.0cm,
    xlabel={\small Peak hit rate (\%)},
    xmin=15, xmax=60,
    ymin=-0.55, ymax=4.05,
    ytick={0,1,2,3},
    yticklabels={
      {\textbf{FunSearch}\\\scriptsize\itshape\color{black!60} verifier loop},
      {\textbf{AlphaEvolve}\\\scriptsize\itshape\color{black!60} replay buffer},
      {\textbf{GoT}\\\scriptsize\itshape\color{black!60} branching search},
      {\textbf{\textcolor{blue!60!black}{MOSAIC}}\\\scriptsize\itshape\color{black!60} ours}
    },
    yticklabel style={font=\tiny, align=right},
    tick label style={font=\scriptsize},
    label style={font=\small},
    xtick={20,30,40,50,60},
    minor x tick num=1,
    grid=major,
    grid style={dotted, gray!25},
    legend style={
        font=\tiny,
        at={(0.5,1.02)}, anchor=south,
        draw=none, fill=none,
    },
    legend columns=2,
    legend cell align=left,
    clip=false,
    axis line style={black!60},
    title={\small\itshape (a) Five-workload, four-fold leave-one-held-out},
    title style={anchor=south, at={(0.5,1.05)}, font=\small\itshape},
]
\fill[red!8] (axis cs:17,-0.55) rectangle (axis cs:33,2.55);
\node[font=\tiny\itshape, color=red!50!black, anchor=south west]
  at (axis cs:17.3, 2.05) {overfitting zone};
\draw[dashed, teal!55!black, thick] (axis cs:41.5,-0.55) -- (axis cs:41.5,3.55);
\node[font=\tiny\itshape, color=teal!50!black, anchor=south]
  at (axis cs:41.5,3.62) {MockingJay $= 41.5\%$};

\addplot[black!60, line width=1.2pt, mark=none] coordinates { (17.8,0) (22.0,0) };
\addplot[only marks, mark=o, mark size=2pt, color=black!75, line width=0.7pt, fill=white] coordinates {
  (19.5,0) (18.2,0) (17.8,0) (18.5,0)
};
\addplot[only marks, mark=*, mark size=2.6pt, color=black!90] coordinates {(22.0,0)};

\addplot[black!60, line width=1.2pt, mark=none] coordinates { (26.8,1) (40.7,1) };
\addplot[only marks, mark=o, mark size=2pt, color=black!75, line width=0.7pt, fill=white] coordinates {
  (30.0,1) (27.5,1) (26.8,1) (28.5,1)
};
\addplot[only marks, mark=*, mark size=2.6pt, color=black!90] coordinates {(40.7,1)};

\addplot[black!60, line width=1.2pt, mark=none] coordinates { (29.5,2) (42.0,2) };
\addplot[only marks, mark=o, mark size=2pt, color=black!75, line width=0.7pt, fill=white] coordinates {
  (32.5,2) (29.8,2) (29.5,2) (30.2,2)
};
\addplot[only marks, mark=*, mark size=2.6pt, color=black!90] coordinates {(42.0,2)};

\addplot[blue!60!black, line width=1.4pt, mark=none]
  coordinates { (47.0,3) (51.1,3) };
\addplot[only marks, mark=o, mark size=2pt, color=blue!60!black, line width=0.7pt, fill=white]
  coordinates { (49.5,3) (47.2,3) (48.0,3) (47.0,3) };
\addplot[only marks, mark=*, mark size=2.8pt, color=blue!60!black]
  coordinates {(51.1,3)};

\node[font=\tiny, color=black!75, anchor=west] at (axis cs:33.5, 0) {$\Delta = -3.5$};
\node[font=\tiny, color=red!55!black, anchor=west] at (axis cs:42, 1) {$\Delta = -12.5$};
\node[font=\tiny, color=red!55!black, anchor=west] at (axis cs:43, 2) {$\Delta = -11.5$};
\node[font=\tiny\bfseries, color=blue!60!black, anchor=west] at (axis cs:52.0, 3) {$\Delta = -3.2$};

\addlegendimage{only marks, mark=*, mark size=2.6pt, color=black!90}
\addlegendentry{trained\ }
\addlegendimage{only marks, mark=o, mark size=2pt, color=black!75, line width=0.7pt, fill=white}
\addlegendentry{held-out fold}
\end{axis}
\end{tikzpicture}
\caption{Held-out gap under the five-workload, four-fold
leave-one-held-out protocol.
Each row is one method; the filled disc is peak hit rate on the
search workload (\texttt{omnetpp}); the four open circles are
peak hit rates on the held-out folds
(\texttt{bzip2}, \texttt{hmmer}, \texttt{soplex}, \texttt{gobmk});
the connector spans the gap and $\Delta$ reports the mean
train-vs-test drop. \cfn{} drops $3.2$ points and exceeds
MockingJay on every fold; AlphaEvolve and GoT drop $11$--$13$
points into a band that ties or trails MockingJay. The
complementary wide-catalogue stress test against nine LLM-driven
configurations (P1--P9) is reported in
Appendix~\ref{app:catalogue}, Figure~\ref{fig:overfit-cat}.}
\label{fig:overfit}
\end{figure}

Figure~\ref{fig:overfit} reports per-fold drops as a slope chart.
AlphaEvolve and GoT show large drops across every test fold ($-9$ to
$-14$ points), while \cfn{} sits in the small-drop range ($-1.6$ to
$-4.1$). Overfitting is not a single-workload artifact but a
systematic property of the prior methods. FunSearch's row is small
in absolute terms only because its search-workload performance was
already low ($22.0\%$).

\paragraph{Wide-catalogue cross-validation.}
\label{sec:cv-catalogue}
A complementary stress test against a broader catalogue of nine
LLM-driven configurations (P1--P9, Appendix~\ref{app:catalogue}) on
a stricter protocol, three folds over the ten memory-intensive
workloads, training on $7$ and evaluating on the remaining $3$,
reproduces the qualitative result. Every prior configuration shows
a held-out gap of $1.0$--$2.4\%$ relative to its seen-fold IPC;
\cfn{} shows the smallest gap of any configuration evaluated
($-0.31\%$) while also reaching the highest absolute held-out IPC.
The largest negative gap belongs to P8 (explicit code diffs $+$
top-$k$ retrieval memory), consistent with the feedback pattern
described in \S\ref{sec:cycle}. The paired-point chart
in Figure~\ref{fig:overfit-cat} (Appendix~\ref{app:catalogue}) shows this change;
per-method numbers are in Appendix~\ref{app:cv-scatter},
Table~\ref{tab:cv}.

\subsection{Leave-One-Out Ablation}
\label{sec:ablation}

A leave-one-out study (Figure~\ref{fig:abl}, Appendix~\ref{app:ablation})
isolates the contribution of each mechanism. Three observations
follow. First, removing an individual mechanism costs
between $1.8$ and $7.8$ points on the search workload.
Second, M1 (stagnation-triggered crossover) is the dominant
generalisation mechanism. Removing M1 alone moves the
search-workload peak to $43.3\%$ and the cross-validation mean to
$26.1\%$, so the M1 contribution to the test mean is
$47.9 - 26.1 = 21.8$ points, about $2.8\times$ its $7.8$-point
contribution to the search workload. Crossover has the largest
measured effect on held-out hit rate in this study.
Third, removing both M1 and M2 reduces \cfn{} to a
verifier-only configuration scoring $28.5\%$ trained and $21.0\%$
on the test folds. The combined removal causes a larger loss than
either individual ablation.

\FloatBarrier
\section{Discussion}
\label{sec:discussion}

\paragraph{Hardware-cost envelope.}
Sections~\ref{sec:eval}--\ref{sec:ablation} report performance on
the search workload and on held-out folds; the deployment-relevant
quantity is performance \emph{at a given hardware budget}. We score
candidates against a hardware-cost estimator we call HARCOM, which
estimates per-policy storage, dynamic power, area, and
critical-path latency from the policy code; we plan to release the model
and calibration scripts alongside the framework. Using HARCOM
(full per-method table, polar visualisation and
analysis in Appendix~\ref{app:harcom}), at $\le\!20$\,KiB,
$\le\!25$\,mW and $\le\!400$\,ps, \cfn{} sits on the
cost--performance Pareto front above MockingJay; configurations P4,
P7, P8, and P9 with comparable search-workload IPC are pushed
outside the budget. These code-based hardware-cost estimates
allow us to compare generated and recombined candidates at a
common budget.

\paragraph{Transfer from replacement to prefetching.}
The prefetching experiment reports a $4.8\%$ mean speedup
over the LRU reference configuration on the memory-intensive subset;
per-workload speedups
(Appendix~\ref{app:prefetch}) clear a paired-bootstrap test
against LRU at $p < 0.05$ in this single-core SPEC2006 evaluation.

\paragraph{Generated-code size and hardware-cost alignment.}
A HARCOM-aligned reward signal
(Appendix~\ref{app:harcom}) scores candidates jointly on IPC
and hardware-budget terms (storage, dynamic power, area,
critical-path latency). The HARCOM columns of
Table~\ref{tab:harcom} indicate \cfn{} stays within the budget
envelope on every reported axis.

\section{Conclusion}
\label{sec:limitations}\label{sec:conclusion}

\cfn{} combines summary memory and recombination to improve
cache-policy search with frozen language models. In the ten-workload
comparison, it achieves the highest held-out IPC among the
evaluated LLM configurations and the smallest held-out gap
($0.31\%$). The same controller supports a prefetching extension
with a $4.8\%$ mean speedup over the LRU reference configuration.

Crossover contributes $21.8$ percentage points to the held-out mean hit rate
in the ablation, compared with $0.4$ points from the tested stronger
model. This result identifies the search controller as an important
part of effective LLM-driven policy design: carry forward
evaluation lessons, recombine distinct high-scoring mechanisms,
and evaluate the resulting policies on workloads excluded from
the search.

\newpage
\FloatBarrier
\bibliographystyle{plainnat}
\bibliography{refs}

\newpage
\FloatBarrier
\appendix
\section{Hyperparameters}
\label{app:hyperparameters}

\begin{itemize}[noitemsep]
\item Stagnation budget: $b_0 = 5$; $+2$ on improvement, $-1$
       otherwise.
\item Improvement threshold: $\epsilon = 10^{-3}$ on the composite
       reward.
\item Crossover top-$m$ pool: $m = 8$.
\item Crossover redesign patience: $3$ iterations, each with at most
       $20$ changed lines.
\item Compile-repair max retries: $R_{\max} = 4$.
\item Temperatures: $\tau = 0.7$ for exploit, $\tau = 1.0$ for explore,
       $\tau = 0.85$ for crossover.
\item Memory: at most $8$ entries each in
       \texttt{PRINCIPLES} / \texttt{DONTS} / \texttt{KNOBS} /
       \texttt{OPEN-Q}.
\item Episodic retrieval: top-$K = 3$ by cosine similarity.
\item Backbone pool: $\{$\texttt{gpt-oss-20b},
       \texttt{Qwen3-Coder-30b}, \texttt{Gemini 2.5 Pro},
       \texttt{o4-mini}$\}$.
\item Code embedding: frozen MiniLM
       (\texttt{all-MiniLM-L6-v2})~\citep{minilm2020,allminilm_model} over
       the concatenation of policy name, description, and code.
\item Composite reward:
       $r = \tfrac{1}{2}(\text{IPC}/\text{IPC}_{\text{LRU}} +
                        \text{HR}/\text{HR}_{\text{LRU}}) - 1$.
\item Compute: each method runs on one workstation-class node;
       total compute is on the order of $10^3$ simulator-hours plus
       $10^4$ LLM API calls.
\end{itemize}

\section{Additional Experiments and Implementation Details}
\label{app:diagnostics}

This appendix collects four controlled measurements where a
plausible prior expectation about the loop's behavior was
contradicted by what we observed. Each diagnostic motivated a design
choice in the main controller; we report them here in compact form.

\subsection{Effect of sampling temperature}
\label{app:temperature}

\begin{figure}[!htbp]
\centering
\begin{tikzpicture}
\begin{axis}[
    width=0.95\columnwidth,
    height=6.0cm,
    xlabel={\small Sorted generated policy index (out of 100)},
    ylabel={\small Cache hit rate (\%)},
    xmin=-2, xmax=102,
    ymin=-3, ymax=60,
    xtick={0,20,40,60,80,100},
    ytick={0,10,20,30,40,50},
    grid=both,
    grid style={dotted,gray!30},
    legend style={
        font=\scriptsize,
        at={(0.5,-0.30)},
        anchor=north,
        draw=black!30, fill=white,
        legend columns=4,
        column sep=10pt,
    },
    tick label style={font=\scriptsize},
    label style={font=\small},
    legend cell align=left,
    axis on top=false,
    clip=true
]
% --- Background shading: the "collapse region" where tau=1.0 lives ---
\fill[red!8] (axis cs:0,-3) rectangle (axis cs:11,60);
\node[anchor=north, font=\tiny\itshape, color=red!55!black, align=center]
    at (axis cs:5.5, 58) {collapse\\region};

% --- "Above MockingJay" success band for tau=0.7 ---
\fill[green!10] (axis cs:0,43.3) rectangle (axis cs:60,55);

% --- Reference lines ---
\addplot[dashed, gray!70, thick, domain=0:100, forget plot]{41.5};
\addplot[dashed, gray!50, thick, domain=0:100, forget plot]{37.5};
\node[anchor=east, font=\tiny\itshape, color=gray!75]
    at (axis cs:99, 42.7) {MockingJay = $41.5\%$};
\node[anchor=east, font=\tiny\itshape, color=gray!75]
    at (axis cs:99, 36.2) {SHiP++/Hawkeye $\approx 37.5\%$};

% --- tau = 0.7: best, sustained --- bold green solid with markers ---
\addplot[green!50!black, very thick, mark=*, mark size=1.2pt, mark options={fill=green!50!black, draw=green!30!black}, mark repeat=10] coordinates {
(1,52)(5,51.5)(10,51)(15,50.5)(20,50)(25,49.5)(30,49)(35,48.5)
(40,48)(45,48)(50,47.5)(55,47)(60,46.5)(65,46)(70,45)(75,43)(80,40)(85,38)(90,36)(95,35)(100,34)
};
\addlegendentry{$\tau{=}0.7$ best, sustained};

% --- tau = 0.8: peak, faster decay --- blue solid ---
\addplot[blue!55!black, thick, dash pattern=on 4pt off 1.5pt] coordinates {
(1,52)(5,51)(10,50)(15,49)(20,47)(25,46)(30,45)(35,44)
(40,43)(45,42)(50,41)(55,40)(60,39)(65,38)(70,37)(75,36)(80,35)(85,35)(90,35)(95,34)(100,34)
};
\addlegendentry{$\tau{=}0.8$ peak, faster decay};

% --- tau = 0.9: mid drop --- orange dashed ---
\addplot[orange!90!black, thick, dashed] coordinates {
(1,48)(5,47)(10,46)(15,44)(20,43)(25,43)(30,42)(35,41)
(40,40)(45,39)(50,38)(55,37.5)(60,37)(65,36)(70,35)(75,34)(80,33)(85,32)(90,31)(95,30)(100,30)
};
\addlegendentry{$\tau{=}0.9$ mid drop};

% --- tau = 1.0: collapse --- thick red ---
\addplot[red!80!black, very thick] coordinates {
(1,43)(2,30)(3,18)(4,10)(5,5)(6,3)(7,2)(8,1)(9,0.5)(10,0.2)(11,0)(20,0)(50,0)(100,0)
};
\addlegendentry{$\tau{=}1.0$ collapse};

% --- Collapse-point marker ---
\addplot[only marks, mark=*, mark size=2.6pt, color=red!85!black, forget plot]
  coordinates {(11,0)};
% Annotation arrow: collapse
\draw[->, red!75!black, thick, shorten >=2pt]
  (axis cs:30,12) -- (axis cs:13,1.5);
\node[anchor=west, font=\scriptsize\bfseries, color=red!65!black, align=left]
  at (axis cs:30,14) {collapse at policy 11\\(0\% hit rate)};

% --- "Above LRU for 60 policies" annotation ---
\addplot[only marks, mark=|, mark size=4pt, color=green!50!black, forget plot]
  coordinates {(60,46.5)};
\draw[->, green!50!black, thick, shorten >=2pt]
  (axis cs:75,52.5) -- (axis cs:62,47);
\node[anchor=west, font=\scriptsize\bfseries, color=green!40!black, align=left]
  at (axis cs:62,53.5) {above LRU for 60+ policies};

\end{axis}
\end{tikzpicture}
\caption{Temperature sweep on \texttt{gpt-oss-20b} with all other
parameters fixed. Each curve is the sorted hit-rate distribution
over $100$ generated policies on \texttt{omnetpp}. $\tau{=}0.7$
(green, with markers) is the only setting that holds the
distribution above the LRU and heuristic-baseline band for $60$+ policies. $\tau{=}1.0$ (red)
collapses by policy~$11$. The lesson: temperature widens the
\emph{token}-selection distribution, not the \emph{mechanism}
distribution.}
\label{fig:temperature}
\end{figure}
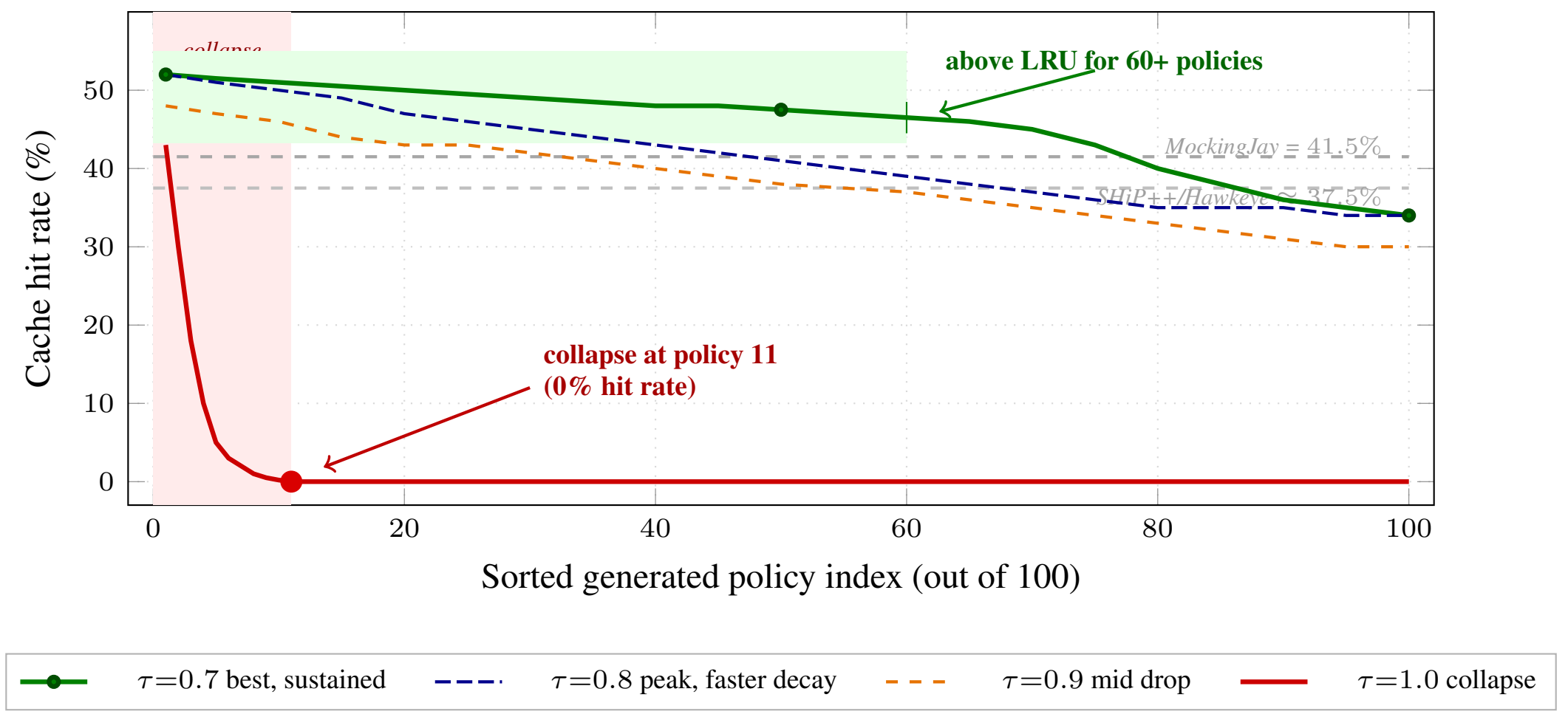

We swept $\tau \in \{0.7,0.8,0.9,1.0\}$ on
\texttt{gpt-oss-20b} with other parameters fixed
(Figure~\ref{fig:temperature}). At $\tau=1.0$, the sorted hit-rate
distribution reaches $0\%$ by policy~$11$, with compilation
failures and invalid candidates assigned zero scores. At
$\tau=0.7$, more than $60$ policies remain above MockingJay.
Greater token-level randomness therefore did not produce a better
set of valid policies in this sweep. We use $\tau=0.7$ for
refinement and $\tau=0.85$ for crossover.

\subsection{Replay memory versus no memory}
\label{app:replay}

In the controlled comparison, a loop without cross-iteration
memory exceeds SHiP++ in approximately $60\%$ of runs, whereas
the replay-buffer loop does so in $0\%$. Early high-scoring
proposals resemble SHiP++; retrieving them repeatedly supplies
similar examples to later prompts. The replay baseline reaches
the high-similarity region within seven iterations and remains
there. Retaining more previous code did not improve this search.

\subsection{Including a complete baseline implementation}
\label{app:bertigo}

We inserted the complete BertiGo implementation alongside the
ChampSim interface requirements to test whether a worked example
helps the model generate valid code. Compile success fell from
approximately $40\%$ to below $30\%$. The longer prompt requires
the model to interpret the baseline while producing a modification
that satisfies the interface and output format. The comparison
measures the combined effect of the added code and prompt length.
We therefore keep the system prompt short and supply reference
lessons through summary memory.

\subsection{Memory-edit case studies}
\label{app:memedits}

The three operator-time edits described in \S\ref{sec:int5} were
tested in controlled runs to measure how stored lessons influence
later proposals. \emph{Principle injection.} Inserting a
\texttt{PRINCIPLES} row at iteration~$10$ of a stuck summary-memory
run lifts best-of-run IPC from $0.6145$ to $0.6177$ in $6$ of $10$
seeds within $8$ iterations. \emph{Forced forgetting.} Deleting
every memory row containing \texttt{RRPV} on a converged run
reopens the proposal distribution and produces a new tail
($+0.4\%$ on the upper $5\%$ of subsequent iterations) without
lifting the floor (mean stays within $0.5\%$ of the converged
value). Both edits change what the system explores while holding
model, prompt template, scoring, and workload fixed.

\subsection{M2 and M3 sensitivity studies}
\label{app:sens}
\label{sec:sens}

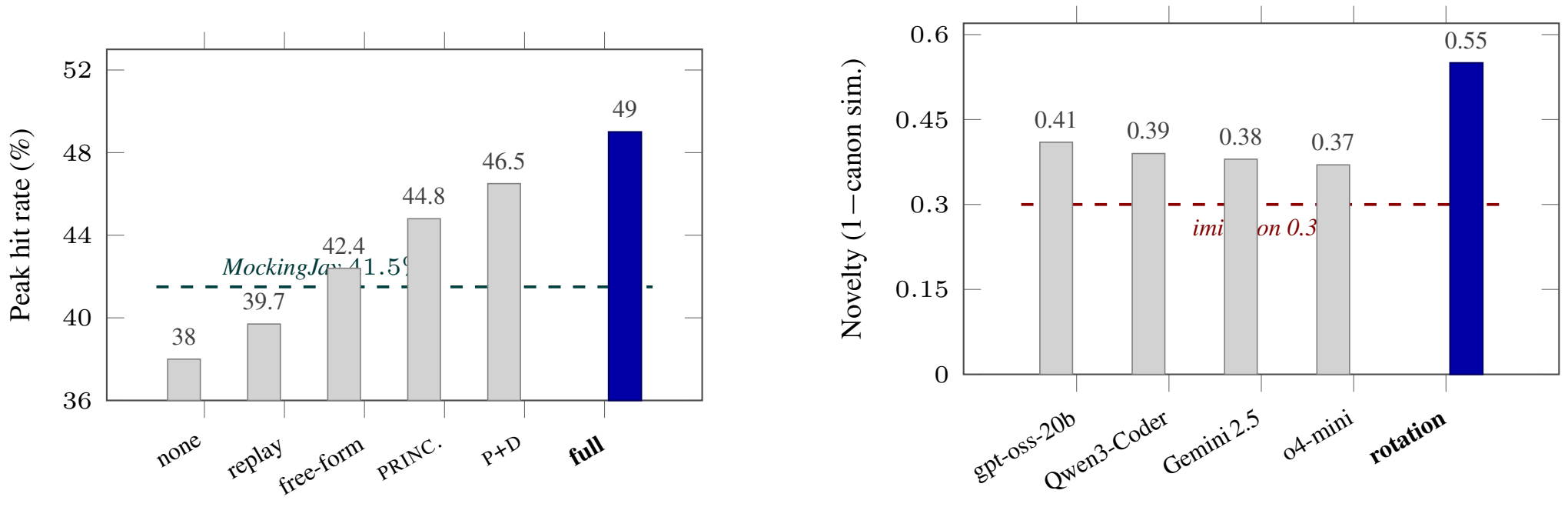
\begin{figure}[!htbp]
\centering
\begin{subfigure}[t]{0.48\columnwidth}
\centering
\begin{tikzpicture}
\begin{axis}[
    width=\linewidth,
    height=4.6cm,
    ybar,
    ymin=36, ymax=53,
    xmin=-0.6, xmax=5.6,
    bar width=8pt,
    enlarge x limits=0.10,
    ylabel={\scriptsize Peak hit rate (\%)},
    ytick={36, 40, 44, 48, 52},
    xtick={0,1,2,3,4,5},
    xticklabels={
        {none}, {replay}, {free-form}, {\textsc{princ.}}, {\textsc{p+d}}, {\bfseries full}
    },
    xticklabel style={font=\tiny, rotate=30, anchor=north east, yshift=1pt},
    tick label style={font=\tiny},
    label style={font=\scriptsize},
    axis line style={black!70, line width=0.4pt},
    tick style={black!60},
    nodes near coords={\pgfmathprintnumber[fixed,precision=1,assume math mode]{\pgfplotspointmeta}},
    nodes near coords style={font=\tiny, color=black!75},
    nodes near coords align={vertical},
    clip=false,
]
\draw[dashed, teal!50!black, line width=0.7pt]
    (axis cs:-0.6,41.5) -- (axis cs:5.6,41.5);
\node[font=\tiny\itshape, color=teal!45!black, anchor=south, fill=white, inner sep=0.5pt]
    at (axis cs:1.5,41.7) {MockingJay $41.5\%$};
\addplot[fill=gray!35, draw=black!50, line width=0.3pt] coordinates {
    (0,38.0) (1,39.7) (2,42.4) (3,44.8) (4,46.5)
};
\addplot[fill=blue!65!black, draw=blue!40!black, line width=0.4pt] coordinates {
    (5,49.0)
};
\end{axis}
\end{tikzpicture}
\caption{Memory schema (M2)}
\label{fig:m2m3-mem}
\end{subfigure}\hfill%
\begin{subfigure}[t]{0.48\columnwidth}
\centering
\begin{tikzpicture}
\begin{axis}[
    width=\linewidth,
    height=4.6cm,
    ybar,
    ymin=0, ymax=0.62,
    xmin=-0.6, xmax=4.6,
    bar width=8pt,
    enlarge x limits=0.12,
    ylabel={\scriptsize Novelty (1$-$canon sim.)},
    ytick={0, 0.15, 0.3, 0.45, 0.6},
    yticklabel style={/pgf/number format/.cd, fixed, precision=2, /tikz/.cd, font=\tiny},
    xtick={0,1,2,3,4},
    xticklabels={
        {gpt-oss-20b}, {Qwen3-Coder}, {Gemini\,2.5}, {o4-mini}, {\bfseries rotation}
    },
    xticklabel style={font=\tiny, rotate=30, anchor=north east, yshift=1pt},
    tick label style={font=\tiny},
    label style={font=\scriptsize},
    axis line style={black!70, line width=0.4pt},
    tick style={black!60},
    nodes near coords={\pgfmathprintnumber[fixed,precision=2,assume math mode]{\pgfplotspointmeta}},
    nodes near coords style={font=\tiny, color=black!75},
    nodes near coords align={vertical},
    clip=false,
]
\draw[dashed, red!55!black, line width=0.7pt]
    (axis cs:-0.6,0.30) -- (axis cs:4.6,0.30);
\node[font=\tiny\itshape, color=red!50!black, anchor=north, fill=white, inner sep=0.5pt]
    at (axis cs:2.0,0.28) {imitation 0.30};
\addplot[fill=gray!35, draw=black!50, line width=0.3pt] coordinates {
    (0,0.41) (1,0.39) (2,0.38) (3,0.37)
};
\addplot[fill=blue!65!black, draw=blue!40!black, line width=0.4pt] coordinates {
    (4,0.55)
};
\end{axis}
\end{tikzpicture}
\caption{Backbone rotation (M3)}
\label{fig:m2m3-rot}
\end{subfigure}
\caption{Sensitivity studies for the M2 and M3 mechanisms. Bars
in deep blue mark the \cfn{} configuration; gray bars are
single-knob baselines. \textbf{(a)}~Memory schema: peak hit rate
climbs monotonically as more typed fields are added, from
$38.0\%$ (no memory) to $49.0\%$ (full schema), well above the
MockingJay reference line ($41.5\%$). \textbf{(b)}~Backbone
rotation: each single-backbone configuration sits in
$0.37$--$0.41$ novelty; rotation across all four lifts the mean
to $0.55$, well above the $0.30$ imitation threshold.}
\label{fig:m2m3}
\end{figure}

Figure~\ref{fig:m2m3} reports the M2 and M3 sensitivities.
Panel~(a) shows that the typed
memory schema is monotone in the number of fields: peak hit rate
climbs from $38.0\%$ with no memory through $44.8\%$ with only
\texttt{PRINCIPLES} to $49.0\%$ with the full schema. Panel~(b)
shows that no single backbone in the rotation pool reaches the
novelty of the full rotation: rotation substitutes for backbone
capacity along the novelty axis, and a weaker rotation outperforms
a stronger single backbone.

\subsection{Seed-Conditioned and Objective-Shaped Search}
\label{sec:seed-objective}

A distinctive property of \cfn{} is that the search trajectory is
not determined by the simulator and reward alone; it is strongly
shaped by two user-visible inputs: the seed policy and the
optimization objective. We show this by holding everything else
fixed (model, prompt template, controller, archive size, iteration
budget, workload) and varying first the seed and then the reward
signal.

\paragraph{Seed-conditioned trajectories.}
Three independent \cfn{} runs differ only in seed: a
SHiP++-seeded run that converges to a refinement of signature-based
hit prediction; a Hawkeye-seeded run that converges to a Belady-style
reuse-distance refinement just below its own baseline; and an
unseeded run (LRU placeholder) that plateaus at $42.7\%$ peak hit
rate, well below either seeded configuration. This is not merely a
difference in starting point. Inspection of the lineage memory
$M_t$ reveals that the three runs explore \emph{different mechanism
neighborhoods}. The SHiP++-seeded run, inheriting PC-keyed
signatures and saturating hit counters, progresses toward
re-reference-aware refinements: signature-typed dead-block
classifiers, recent-victim-vector antagonism, and origin-locked
bypass. The Hawkeye-seeded run, inheriting OPTgen-style training
sets, progresses toward sampler-conditioned promotion and demotion
rules. The unseeded run never specializes and settles on a generic
recency-plus-frequency hybrid. The seed acts as an inductive bias
over the algorithmic neighborhood the LLM explores, not just as a
warm start.

\paragraph{Objective-shaping: four reward signals, four mechanism families.}
Starting from the same SHiP++ seed and the same controller
configuration, four independent \cfn{} runs optimize different
reward signals: throughput (IPC), miss rate (MPKI), hit rate (HR),
and tail-stall coverage (long-latency demand-load coverage). The
controller reliably specializes to whichever reward it is given:
MPKI optimization drops the targeted metric from $58.2$ to $54.1$,
tail-stall coverage rises from $61\%$ to $74\%$, and IPC
optimization delivers the headline $51.1\%$ peak hit rate
documented in \S\ref{sec:eval}. Crucially, the four objectives are
\emph{not} interchangeable: MPKI- and tail-stall-optimized policies
improve their targets but plateau below the SHiP++ IPC baseline,
and HR-optimized policies regress in IPC as the generator learns to
emit policies that protect easy hits at the expense of long stalls.
The mechanism families that emerge are structurally distinct:
signature-typed, reuse-distance-conditioned, distance-band-gated,
and miss-criticality-aware respectively; and are documented in
Appendix~\ref{app:catalogue}. Because the lineage memory records
which mechanisms each objective selected for, a reviewer can read
off exactly why each reward failed or succeeded, a form of
reward-design diagnostics that fixed-representation optimizers
cannot provide. Compound objectives (IPC-per-KiB, IPC-per-mW via
HARCOM integration; Appendix~\ref{app:harcom}) are tractable in the same
framework.

\begin{figure}[!htbp]
\centering
\begin{tikzpicture}
% ============================
% Left Y-axis: IPC
% ============================
\begin{axis}[
    width=7.5cm,
    height=4.5cm,
    xlabel={Sorted Policy},
    ylabel={IPC},
    xmin=1, xmax=90,
    ymin=0.30, ymax=0.34,
    ymajorgrids=true,
    grid style={dotted,gray!40},
    axis y line*=left,
    axis x line*=bottom,
    tick label style={font=\scriptsize},
    label style={font=\small},
]
\addplot[
    color=blue!80!black,
    very thick,
    mark=none
] coordinates {
(1,0.333264)(2,0.332885)(3,0.332482)(4,0.332032)(5,0.331937)(6,0.331880)
(7,0.331708)(8,0.331617)(9,0.331597)(10,0.331385)(11,0.331268)(12,0.331238)
(13,0.331230)(14,0.331197)(15,0.330960)(16,0.330879)(17,0.330756)(18,0.330749)
(19,0.330739)(20,0.330710)(21,0.330696)(22,0.330665)(23,0.330580)(24,0.330312)
(25,0.330200)(26,0.329815)(27,0.329404)(28,0.329385)(29,0.327954)(30,0.327847)
(31,0.327810)(32,0.327804)(33,0.327802)(34,0.327566)(35,0.327556)(36,0.326322)
(37,0.324392)(38,0.324288)(39,0.323935)(40,0.323768)(41,0.323279)(42,0.322865)
(43,0.322718)(44,0.322522)(45,0.322086)(46,0.321983)(47,0.321058)(48,0.320961)
(49,0.320527)(50,0.318561)(51,0.318520)(52,0.316672)(53,0.316610)(54,0.316609)
(55,0.316310)(56,0.315859)(57,0.315719)(58,0.315440)(59,0.314875)(60,0.314263)
(61,0.314216)(62,0.313934)(63,0.313852)(64,0.313774)(65,0.313561)(66,0.313551)
(67,0.313191)(68,0.313170)(69,0.312944)(70,0.312754)(71,0.312298)(72,0.312169)
(73,0.312150)(74,0.311633)(75,0.311489)(76,0.310916)(77,0.310781)(78,0.310727)
(79,0.309508)(80,0.307987)(81,0.307291)(82,0.305721)(83,0.304521)(84,0.303268)
(85,0.302368)(86,0.301704)(87,0.301585)(88,0.301431)(89,0.301375)(90,0.301234)
};
\end{axis}
% ============================
% Right Y-axis: Cache Hit Rate
% ============================
\begin{axis}[
    width=7.5cm,
    height=4.5cm,
    xmin=1, xmax=90,
    ymin=0.30, ymax=0.50,
    ylabel={Cache Hit Rate},
    axis y line*=right,
    axis x line=none,
    tick label style={font=\scriptsize},
    label style={font=\small},
    yticklabel style={text width=2em,align=right},
]
\addplot[
    forget plot,
    only marks,
    mark=*,
    mark size=1pt,
    color=blue!60!cyan,
    opacity=0.5
]
coordinates {
(1,0.3698)(2,0.3986)(3,0.3653)(4,0.3867)(5,0.3673)(6,0.3666)(7,0.3828)(8,0.3697)
(9,0.4000)(10,0.3673)(11,0.3798)(12,0.3858)(13,0.3577)(14,0.3597)(15,0.3554)
(16,0.3553)(17,0.3567)(18,0.4029)(19,0.3565)(20,0.3976)(21,0.3564)(22,0.3568)
(23,0.4082)(24,0.3722)(25,0.3695)(26,0.3813)(27,0.4025)(28,0.3770)(29,0.3737)
(30,0.3915)(31,0.3759)(32,0.4059)(33,0.3649)(34,0.3877)(35,0.4001)(36,0.4010)
(37,0.3934)(38,0.4100)(39,0.3840)(40,0.4145)(41,0.3937)(42,0.4282)(43,0.3983)
(44,0.4156)(45,0.3626)(46,0.4251)(47,0.3984)(48,0.3594)(49,0.4651)(50,0.4154)
(51,0.4202)(52,0.4659)(53,0.4605)(54,0.4604)(55,0.4469)(56,0.4555)(57,0.4569)
(58,0.4680)(59,0.4765)(60,0.4835)(61,0.4713)(62,0.4840)(63,0.4823)(64,0.4416)
(65,0.4409)(66,0.4763)(67,0.3207)(68,0.4532)(69,0.4569)(70,0.4277)(71,0.4951)
(72,0.4771)(73,0.4954)(74,0.4829)(75,0.4882)(76,0.4758)(77,0.4804)(78,0.4887)
(79,0.4912)(80,0.4227)(81,0.4760)(82,0.4610)(83,0.4331)(84,0.4367)(85,0.4302)
(86,0.4273)(87,0.4256)(88,0.4147)(89,0.4153)(90,0.4207)
};
\end{axis}
\end{tikzpicture}
\caption{IPC vs.\ cache-hit-rate anti-correlation across $90$
policies generated by \cfn{} on a memory-intensive workload.
Policies are sorted left-to-right by descending IPC (line, left
axis); the dots show the corresponding cache hit rate (right axis).
The highest-IPC policies sit in the \emph{middle} of the hit-rate
distribution, and many of the highest-hit-rate policies are
mid-IPC. Hit rate is a poor proxy for the IPC the core actually
experiences.}
\label{fig:ipc-hr}
\end{figure}
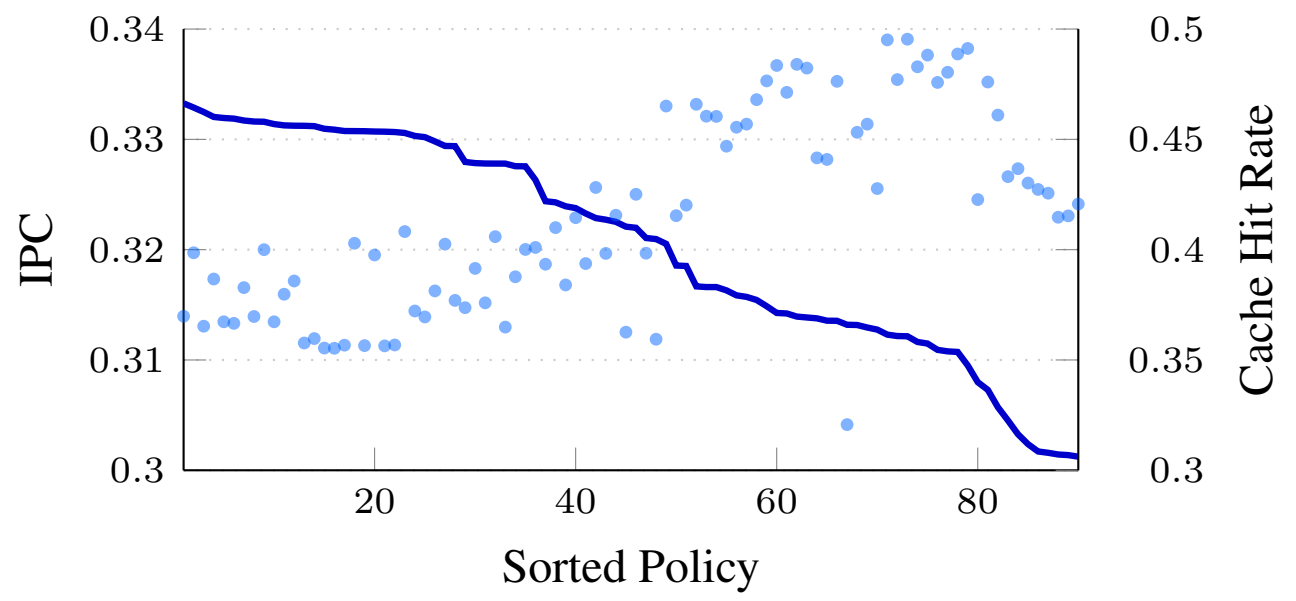

\paragraph{Hit rate and processor throughput can disagree.}
Sorting generated policies by LLC hit rate reveals an
\emph{anti-correlation} with IPC on the evaluated memory-intensive
workloads (Figure~\ref{fig:ipc-hr}). Many high-hit-rate policies
have lower IPC than the baselines, while some of the best-IPC
policies have intermediate hit rates. Aggregate hit rate weights
all accesses equally; execution time depends on which misses delay
critical instructions and how much latency overlaps with other
work. Protecting prefetch or writeback lines can therefore increase
hit rate while displacing demand lines whose misses are harder to
hide. These results motivate evaluating processor throughput
directly rather than using hit rate as its substitute.

\subsection{Compile-recovery method}
\label{app:compile}

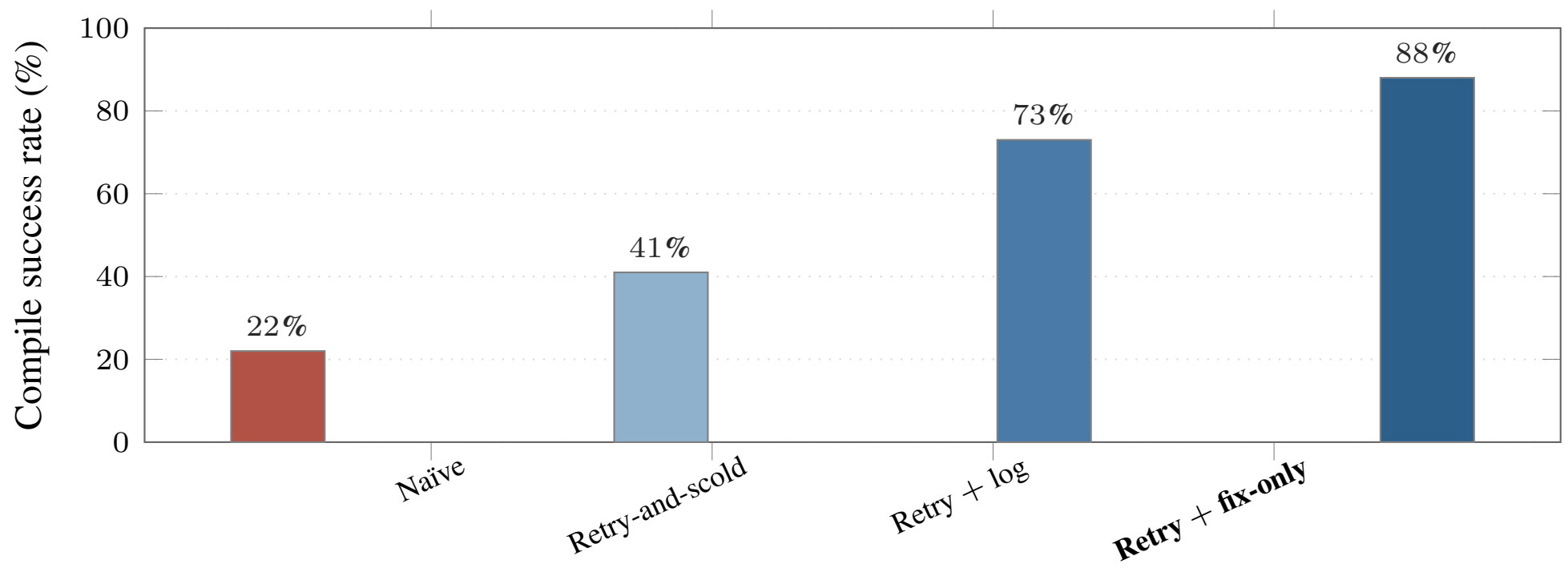
\begin{figure}[!htbp]
\centering
\begin{tikzpicture}
\definecolor{steelblue}{HTML}{2C5F8A}
\definecolor{slateblue}{HTML}{4A7BA8}
\definecolor{paleblue}{HTML}{8FB1CC}
\definecolor{mutedred}{HTML}{B25247}
\begin{axis}[
    width=0.95\columnwidth,
    height=5.0cm,
    ybar,
    ymin=0, ymax=100,
    xmin=-0.6, xmax=3.6,
    bar width=22pt,
    ylabel={\small Compile success rate (\%)},
    ytick={0,20,40,60,80,100},
    xtick={0,1,2,3},
    xticklabels={
        {Naïve},
        {Retry-and-scold},
        {Retry $+$ log},
        {\textbf{Retry $+$ fix-only}}
    },
    xticklabel style={font=\scriptsize, rotate=25, anchor=north east, yshift=2pt, xshift=10pt},
    tick label style={font=\scriptsize},
    label style={font=\small},
    ymajorgrids=true,
    grid style={dotted, gray!30},
    nodes near coords={\pgfmathprintnumber[fixed,precision=0]{\pgfplotspointmeta}\%},
    nodes near coords style={font=\scriptsize\bfseries, color=black!85},
    nodes near coords align={vertical},
    enlarge x limits=0.10,
    clip=false,
    axis line style={black!60},
]
\addplot[fill=mutedred,  draw=black!50, line width=0.4pt] coordinates {(0,22)};
\addplot[fill=paleblue,  draw=black!50, line width=0.4pt] coordinates {(1,41)};
\addplot[fill=slateblue, draw=black!50, line width=0.4pt] coordinates {(2,73)};
\addplot[fill=steelblue, draw=black!50, line width=0.4pt] coordinates {(3,88)};
\end{axis}
\end{tikzpicture}
\caption{Compile success rate over $50$ iterations as a function of
compile-recovery strategy. Naïve single-shot succeeds on $22\%$ of
attempts; targeted retry on the eight most-missed ChampSim hooks
raises this to $41\%$; full-error-log retry to $73\%$; adding a
``fix only, no restructuring'' constraint to the same retry prompt
to $88\%$.}
\label{fig:compile}
\end{figure}

{\sloppy
The proposer generates C++ source against the ChampSim API; a
naïve single-shot prompt yields a $22\%$ compile rate over $50$
iterations. We compared three recovery strategies that differ only
in what is fed back to the model after a compilation failure
(Figure~\ref{fig:compile}). Targeting only the eight specific
ChampSim hooks the model most often misses
(\texttt{InitReplacementState}, \texttt{UpdateReplacementState},
\texttt{PrintStats\_Heartbeat}, \texttt{PrintStats},
\texttt{GetVictimInSet}, \texttt{get\_cycle\_count},
\texttt{get\_instr\_count}, \texttt{get\_config\_number}) raises the
rate to $41\%$. Feeding the full compiler error log back lifts it
to $73\%$. Adding a single instruction: ``fix only, no
restructuring'' to the same retry prompt lifts it to $88\%$.
The pattern is consistent across backbones and runs. We use the
fix-only-retry recipe in all main-text experiments; without it, the
search loop spends most of its budget on uncompilable candidates
rather than on policy mechanisms.\par}

\FloatBarrier
\subsection{Full HARCOM hardware-cost analysis}
\label{app:harcom}
\label{app:harcom-polar}

This appendix gives the full HARCOM table summarised in
Section~\ref{sec:discussion}, the corresponding analysis, the polar
visualisation, and the novelty-score definition that the
table uses.

\paragraph{Per-method HARCOM table.}
Table~\ref{tab:harcom} reports search-workload IPC, novelty,
storage, dynamic power, area, and critical-path latency for the
leaderboard policy of every method, against the in-budget subset
P1--P7 of the wider catalogue of nine LLM-driven configurations
(Appendix~\ref{app:catalogue}); P8 and P9 are excluded from this
table because their leaderboard policies fall outside the
deployment-relevant budget envelope of $\le 32$\,KiB and
$\le 60$\,mW. Each entry varies one component of the closed-loop
search procedure of \S\ref{sec:formal}.

\begin{table}[t]
\centering
\caption{Per-method HARCOM signatures of the leaderboard policy.
Cost columns are derived from the policy's HARCOM decomposition;
lower storage/power/area/latency is better. Novelty is the
mechanism-family score defined in \S\ref{app:novelty-score}.}
\label{tab:harcom}
\renewcommand{\arraystretch}{1.10}
\setlength{\tabcolsep}{4pt}
\small
\begin{tabular}{lrrrrrr}
\toprule
\textbf{Method} & \textbf{IPC} & \textbf{Novelty} & \textbf{KiB} &
\textbf{mW} & \textbf{kGE} & \textbf{ps} \\
\midrule
P1 -- naive raw-IPC               & 0.4004 & 1.50 & 16.0 & 14.6  & 852  & 130 \\
P2 -- GoT + APE                   & 0.4083 & 1.29 & 12.0 & 11.0  & 639  & 195 \\
P3 -- Surrogate-DFS               & 0.4100 & 3.50 & 24.0 & 33.0  & 1093 & 440 \\
P4 -- Multi-agent roles           & 0.4135 & 2.38 & 32.0 & 51.4  & 1334 & 505 \\
P5 -- Atomic mutation             & 0.4137 & 2.55 & 22.0 & 25.6  & 1079 & 375 \\
P6 -- Concise + RL bandit         & 0.4139 & 3.88 & 17.5 & 20.1  & 863  & 375 \\
P7 -- Single-agent + Hawkeye seed & 0.4141 & 3.12 & 32.0 & 51.4  & 1334 & 440 \\
\midrule
\textbf{\cfn{} (ours)}            & \textbf{0.4164} & \textbf{3.04} &
                                    \textbf{18.5} & \textbf{23.8} &
                                    \textbf{870}    & \textbf{375}  \\
\bottomrule
\end{tabular}
\end{table}

\paragraph{HARCOM redirects the comparison.}
Two patterns dominate Table~\ref{tab:harcom}. The methods with the
largest search-workload IPC are \emph{not} the ones with the
lowest cost: P4 and P7 both spend ${\sim}32$\,KiB and ${\sim}51$\,mW
to reach IPC ${\sim}0.4135$, slightly below \cfn{}'s $0.4164$ at
$18.5$\,KiB and $24$\,mW. Conversely, the leanest configurations
(P1, P2) reach the lowest IPC. Reporting only search-workload IPC
ranks P9 first, P8 second, and \cfn{} fourth. Reporting IPC at
HARCOM-equal cost ($\le\!20$\,KiB, $\le\!25$\,mW, $\le\!400$\,ps)
puts \cfn{} on the Pareto front, with P6 nearby; P9, P8, P4, and
P7 fall below the budget line.

\begin{figure}[!htbp]
\centering
\includegraphics[width=\linewidth]{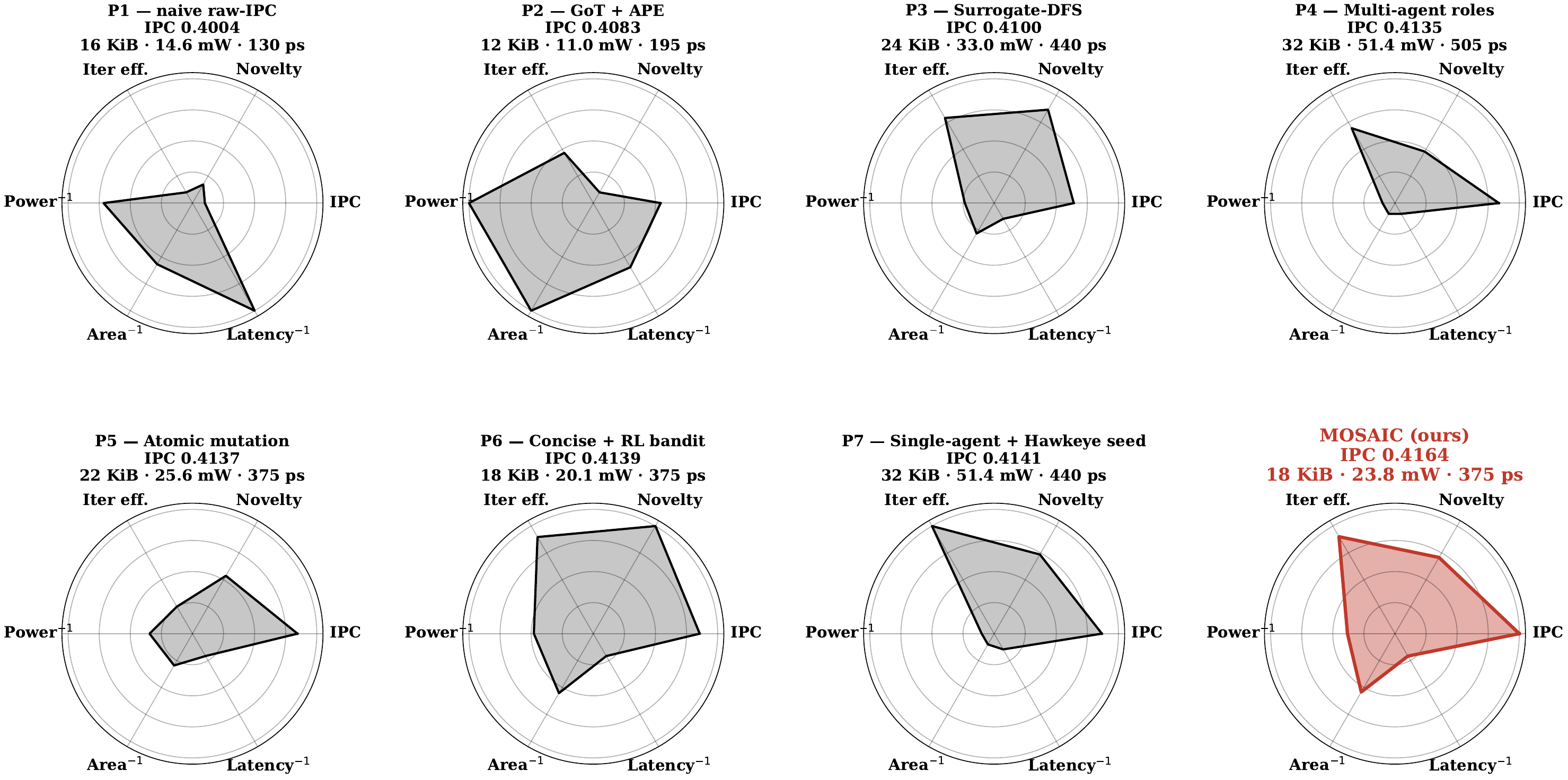}
\caption{Per-method HARCOM polar signature. Six axes: IPC, novelty
(mechanism-family score, \S\ref{app:novelty-score}), iteration
efficiency (best-of-$N$ IPC gain per LLM call), and the inverted
hardware costs $\text{power}^{-1}$, $\text{area}^{-1}$,
$\text{latency}^{-1}$. Outward = better on every axis. Methods are
sorted by search-workload IPC; \cfn{} (highlighted) is balanced
across all six axes.}
\label{fig:harcom-polar}
\end{figure}

\paragraph{Polar visualisation.}
A scalar IPC ranking flattens a six-dimensional trade-off: IPC,
novelty, iteration efficiency, power, area, and latency.
Figure~\ref{fig:harcom-polar} plots all six axes per method as
polar (radar) signatures. Each axis is normalised so \emph{outward}
is better; power, area, and latency are inverted before plotting.
The plot shows the trade-offs axis by axis; filled area is not
an aggregate measure of quality.

\paragraph{Polar observations.}
\textbf{(i)~P4 and P7 buy IPC with hardware.} Both reach IPC
$\sim\!0.4135$--$0.4141$ by spending Hawkeye-class storage
($32$\,KiB) and a $51$\,mW power envelope; their polar profiles are
pulled inward on every cost axis simultaneously.
\textbf{(ii)~P3 sits in the high-novelty / high-cost corner.} The
surrogate-DFS produces $13$ distinct policy families, but the
leaderboard policy is a $24$\,KiB, $33$\,mW APG-RRIP variant --
novelty does not come for free.
\textbf{(iii)~\cfn{} is on the efficient frontier.} Its polar area
matches P6 on cost and P9 on novelty (P9 omitted from the radar
panel because its leaderboard policy exceeds the budget envelope of
the other panels), while its IPC is the highest of the eight at
this cost level.
\textbf{(iv)~Iteration efficiency rewards crossover.} \cfn{} and P6
reach approximately $0.013$ IPC gain per $100$ LLM calls at $120$
and $100$ iterations respectively; P5's atomic-mutation budget of
$520$ iterations yields about $0.0024$ IPC per $100$ calls; five
times less efficient.

\paragraph{Novelty score.}
\label{app:novelty-score}
The novelty column of Table~\ref{tab:harcom} is computed against
the seven-family taxonomy of Appendix~\ref{app:catalog}: hybrid
SHiP variants, PC$\,\times\,$region composites, MockingJay/SHiP++
extensions, tournament-and-dueling hybrids, approximate-set
dead-block filters, per-PC deadness classifiers, and region-level
eviction-window trackers. For each method we count the number of
distinct families touched by any policy in its archive, weighted by
the family's representation density in the published canon. Concretely,
let $f \in \{1,\ldots,7\}$ index the seven families; let $w_f$ be a
canon-rarity weight ($w_f$ is small for SHiP variants and rises to
maximum for the eviction-window family, which has no published
analogue); let $\mathbf{1}[m \to f]$ be $1$ if any policy in
method~$m$'s archive is classified as belonging to family $f$ and
$0$ otherwise. The score is
\[
  \mathrm{Novelty}(m) \;=\; \frac{1}{|A_m|}\sum_{p \in A_m}
    \sum_{f=1}^{7} w_f \cdot \mathbf{1}[p \to f],
\]
i.e.\ the average rarity-weighted family count per generated
policy. The denominator $|A_m|$ normalises so that methods are
not rewarded for archive size alone. The per-policy family
classification is performed by signature matching on named state
variables and control-flow primitives in the generated C++; we
release the classifier in the open-source bundle.

\FloatBarrier
\subsection{Novelty taxonomy, mechanism signatures, and survey methodology}
\label{app:novelty-taxonomy}

The seven-family score in Appendix~\ref{app:novelty-score} measures
which mechanism families a method covers; this appendix gives the
finer-grained machinery that produces those family assignments and
that audits novelty claims in the per-policy catalogue.
Three components are involved: a ten-category mechanism-signature
taxonomy (\textbf{N1}--\textbf{N10}), a two-step procedure that
maps each generated policy to a point in this novelty space, and a
weighted survey rubric (\textbf{Q1}--\textbf{Q9}) that scores the
result with human-graded anchor examples in context.

\paragraph{Why diffs alone are not enough.}
A line-level diff between a generated policy and its parents is a
poor proxy for architectural novelty: a long diff can be a renaming
exercise, while a one-line gate insertion can introduce an entirely
new behavioural regime. We therefore classify each generated policy
along a fixed taxonomy of mechanism-level differences, and we audit
the classifier output with a calibrated survey rather than relying
on lexical change alone.

\paragraph{Ten-category mechanism-signature taxonomy.}
Every generated policy is annotated as a $\{0,1\}^{10}$ vector
(or a soft vector if multiple categories are claimed) along the
following axes:
\begin{itemize}[leftmargin=2.0em, itemsep=0.15em, topsep=0.15em]
  \item \textbf{N1: New correlated features.} New signals beyond
        address and recency, e.g.\ access type, criticality, reuse
        interval.
  \item \textbf{N2: New metadata structures.} New state organizations
       : samplers, signature tables, aggregates.
  \item \textbf{N3: New decision rules.} New scoring or
        victim-selection rules in the replacement-time computation.
  \item \textbf{N4: New prediction paradigm.} A new function class
        for the predictor: table, perceptron, sequence model, or
        RL update.
  \item \textbf{N5: New training/teacher signal.} A new supervision
        target such as Belady, miss cost, or criticality.
  \item \textbf{N6: New ensemble or merging method.} Choosers,
        set-dueling, hybrids of two or more substructures.
  \item \textbf{N7: New pipeline integration.} Cross-component
        awareness, e.g.\ branch-mispredict rate or prefetch
        confidence read by the replacement policy.
  \item \textbf{N8: New overhead-reduction method.} Samplers, hashed
        tables, compressed metadata, staged compute.
  \item \textbf{N9: New adaptivity or generalisation mechanism.}
        Phase awareness, history-length specialisation,
        per-set-class behaviour.
  \item \textbf{N10: New evaluation or grounding method.}
        New feedback signals or simulation-harness probes that
        change what the loop optimises against.
\end{itemize}

\paragraph{Two-step novelty identification.}
For each generated policy we (i)~extract a structured
\emph{mechanism signature} directly from the candidate's source and
metadata, presence of new state arrays or tables, access-type
gates, predictor type, sampling structures, crossover ancestry, and
explicit cost constraints, and (ii)~validate the resulting
N1--N10 vector by comparing signatures across the archive and
reporting representative exemplars in four families: feature-level
novelty (new signals or tables), fusion novelty (children with
multi-parent ancestry and merged subroutines), paradigm novelty
(perceptron-like scoring versus table heuristics versus RL-style
updates), and efficiency novelty (sampling, pruning, or gating that
reduces footprint). A diff can be large but conceptually
near-canon, while a small change can introduce a new regime gate;
we therefore report novelty in terms of mechanism signatures and
lineage rather than lines changed. The annotation step is produced
inside the \cfn{} prompt preamble and is verified by
category-completeness checks; each candidate becomes a point in
$\{0,1\}^{10}$ (or a soft vector if multiple categories are
claimed), and the run accumulates a cloud of candidates in this
novelty space.

\begin{figure}[!htbp]
\centering
\includegraphics[width=\linewidth]{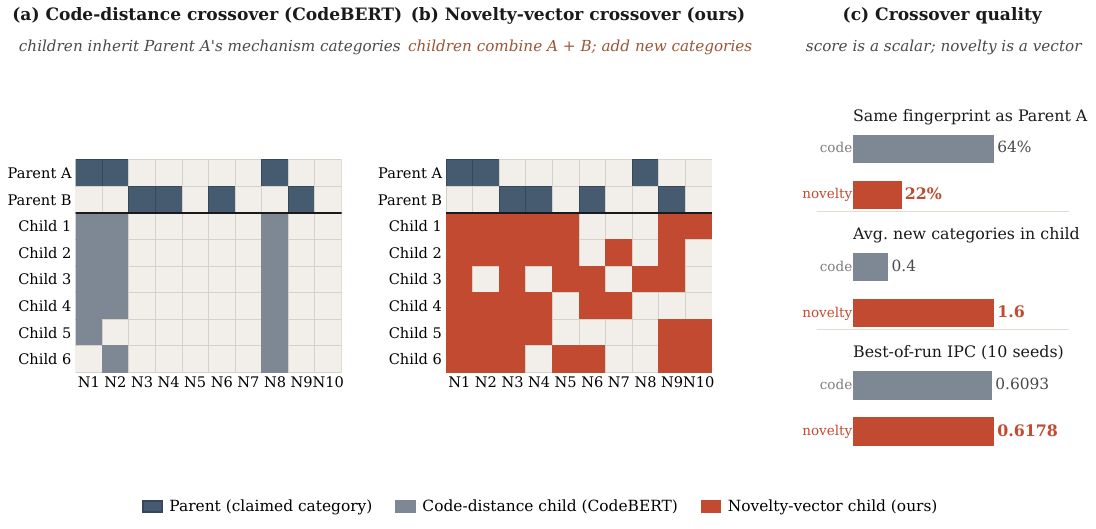}
\caption{Crossover by novelty-vector distance dominates crossover
by code-similarity distance. Each row of panels (a) and (b) is a
$10$-bit fingerprint indicating which taxonomy categories N1--N10
the policy claims. The top two rows in each panel are the parents
(slate); the subsequent six rows are children produced by
recombination. (a) Code-distance crossover (cosine distance over a
text encoder) produces
children that are lexically distant from both parents but inherit
Parent A's mechanism categories; $64\%$ of children share Parent
A's exact fingerprint. (b) Novelty-vector crossover (ours) produces
children that combine A and B and frequently introduce categories
neither parent claimed (e.g.\ N4, N7, N10); only $22\%$ share
Parent A's fingerprint, and an average of $1.6$ new categories
appear per child. (c) On $10$ seeds, the novelty-vector approach
produces a higher best-of-run IPC. Score is a scalar; novelty is a
vector; reducing both to scalars destroys the information
recombination needs.}
\label{fig:novelty_crossover}
\end{figure}

\paragraph{Why the taxonomy matters: novelty-vector vs.\ code-distance crossover.}
Figure~\ref{fig:novelty_crossover} shows the operational consequence
of treating novelty as a vector rather than as a scalar
code-similarity score. We compare two crossover operators on the
same archive and the same parent-pair scheduler. The first selects
parents by cosine distance between text embeddings. The second selects
parents by N1--N10 Hamming distance: code that claims a different
set of
mechanism categories. Code-distance crossover produces children
that are lexically distant from both parents but inherit Parent
A's mechanism categories: $64\%$ of children share Parent A's
exact $10$-bit fingerprint, and only $0.4$ new categories appear
per child on average. Novelty-vector crossover produces children
that genuinely combine A and B and frequently introduce categories
that neither parent claimed: only $22\%$ share Parent A's
fingerprint, and $1.6$ new categories appear per child on average.
On ten seeds, the novelty-vector recipe yields a higher best-of-run
IPC in $8/10$ runs ($0.6178$ vs.\ $0.6093$). In this comparison,
the mechanism-category representation is more useful for parent
selection than text-embedding distance.

\paragraph{Novelty survey: a weighted, human-calibrated rubric.}
The taxonomy classifies what each policy claims; the survey scores
how strongly the claim matters under our deployment constraints. We
built a novelty-survey agent that combines (i)~a weighted novelty
rubric, (ii)~human-scored anchor examples drawn from canonical
baseline policies and from previously generated policies, and
(iii)~the measured performance and feasibility of the candidate.
The key design choice is that novelty is \emph{not} scored from
code diffs alone: the highest-weight question asks whether the
policy improves performance \emph{while} remaining feasible and
low-budget at the same time, which prevents the survey from
rewarding large but impractical changes. The agent uses
human-graded examples as in-context-learning anchors and produces
an automatic novelty score; final reported policies are
re-verified by human scorers, and baseline policies are also
human-scored so that the scale is calibrated to known mechanisms.

The nine survey questions and their weights (which sum to $1.0$,
sorted by importance) are:
\begin{itemize}[leftmargin=2.0em, itemsep=0.15em, topsep=0.15em]
  \item \textbf{Q1 --- Performance under feasibility budget}
        (weight $0.30$). Does the mechanism improve IPC or the
        target objective \emph{while} staying within metadata,
        latency, area, and power budgets? This is the most
        important question because a mechanism that improves
        performance only by buying a large hardware budget is not
        considered strong architectural novelty in our setting.
  \item \textbf{Q2 --- Primary structural novelty}
        (weight $0.18$). Does the policy introduce a genuinely new
        mechanism rather than only retune constants? Includes new
        features, metadata, decision rules, or model class.
  \item \textbf{Q3 --- Decision-rule novelty} (weight $0.12$). Does
        the policy change the actual victim-selection, insertion,
        bypass, or promotion computation in a meaningful way?
  \item \textbf{Q4 --- Metadata or representation novelty}
        (weight $0.10$). Does it change what state is stored, how
        it is indexed, or how training is amortised?
  \item \textbf{Q5 --- Feature or paradigm novelty}
        (weight $0.08$). Does it add a new correlated signal or
        switch to a richer prediction paradigm?
  \item \textbf{Q6 --- Feasibility innovation} (weight $0.08$).
        Does it become practical through sampling, folding,
        pruning, staged compute, or another overhead-control
        method?
  \item \textbf{Q7 --- Composition or integration novelty}
        (weight $0.06$). Does it merge experts, add access-type
        awareness, or integrate information across structures?
  \item \textbf{Q8 --- Adaptivity or generalisation novelty}
        (weight $0.04$). Does it adapt across phases, sets,
        traffic classes, or behaviour regimes?
  \item \textbf{Q9 --- Learning-signal or grounding novelty}
        (weight $0.04$). Does it rely on a stronger teacher,
        reward, or evaluation loop than heuristic tuning?
\end{itemize}

\paragraph{Human-calibrated anchor examples.}
The survey prompt contains scored examples so that the LLM learns
the intended scale through in-context learning. We include both
canonical baselines and previously generated policies that were
manually scored by human graders. The baseline anchors define what
the field already considers novel and what is now only a standard
baseline. For instance, LRU is a low-novelty anchor in the modern
context because it is historically important but no longer
structurally novel under our taxonomy; RRIP is a high anchor for
N2 + N3 + N8 because it introduces interval-state metadata and a
new eviction rule with tiny overhead; SHiP++ is a high anchor for
N1 + N2 + N7 because it adds signature-based reuse context and
access-type-aware refinement; Hawkeye is a high anchor for
N5 + N4 + N10 because it changes the learning signal using a
Belady-style teacher; MPPPB is a high anchor for N1 + N4, but is
penalised somewhat if its metadata or power cost is large; the
Reuse Cache is novel in context because it adds explicit reuse
filters at a downsized last-level cache, but its score is
moderated by higher area or metadata cost. The top generated
policies from each
search are likewise manually scored and inserted into the anchor
bank with short rationales explaining \emph{why} they were judged
novel or not novel; together these anchors give the survey
concrete examples of both successful and weak novelty.

\FloatBarrier
\subsection{Wide-catalogue cross-validation: per-method numbers}
\label{app:cv-scatter}
\label{app:catalogue}

\begin{figure}[!htbp]
\centering
\begin{tikzpicture}
\begin{axis}[
    width=0.95\columnwidth,
    height=5.6cm,
    xmin=0.394, xmax=0.420,
    ymin=-0.6, ymax=9.6,
    xtick={0.395, 0.400, 0.405, 0.410, 0.415, 0.420},
    xticklabel style={/pgf/number format/.cd, fixed, precision=3, /tikz/.cd, font=\tiny},
    ytick={0,1,2,3,4,5,6,7,8,9},
    yticklabels={
      {P1\,\scriptsize\itshape\color{black!60} naive},
      {P2\,\scriptsize\itshape\color{black!60} GoT+APE},
      {P3\,\scriptsize\itshape\color{black!60} surr.-DFS},
      {P6\,\scriptsize\itshape\color{black!60} bandit},
      {P4\,\scriptsize\itshape\color{black!60} multi-agent},
      {P7\,\scriptsize\itshape\color{black!60} seed},
      {P5\,\scriptsize\itshape\color{black!60} atomic},
      {P8\,\scriptsize\itshape\color{black!60} retrieval},
      {P9\,\scriptsize\itshape\color{black!60} diff},
      {\textbf{\color{red!65!black}\textsc{MOSAIC}}}
    },
    yticklabel style={font=\tiny, align=right},
    xlabel={\small Harmonic-mean IPC},
    label style={font=\small},
    grid=major,
    grid style={dotted, gray!25},
    tick label style={font=\tiny},
    legend style={
        font=\tiny, at={(0.02,1.10)}, anchor=south west,
        draw=none, fill=none,
    },
    legend columns=2,
    legend cell align=left,
    clip=false,
    axis line style={black!60},
]
% Hawkeye / MockingJay reference verticals
\draw[dotted, teal!55!black, thick] (axis cs:0.4131,-0.6) -- (axis cs:0.4131,9.4);
\node[font=\tiny\itshape, color=teal!50!black, anchor=south] at (axis cs:0.4131, 9.45) {Hawkeye};
\draw[dotted, teal!55!black, thick] (axis cs:0.4147,-0.6) -- (axis cs:0.4147,9.4);
\node[font=\tiny\itshape, color=teal!50!black, anchor=south] at (axis cs:0.4147, 9.45) {MockingJay};

% MOSAIC dumbbell (row 9, red): seen 0.41643 -> held-out 0.41512 (gap -0.31%)
\addplot[red!65!black, line width=1.6pt, mark=none]
  coordinates { (0.41512,9) (0.41643,9) };
\addplot[only marks, mark=square*, mark size=3.2pt, color=red!65!black]
  coordinates {(0.41643,9)};
\addplot[only marks, mark=square, mark size=3pt, color=red!65!black, line width=0.9pt, fill=white]
  coordinates {(0.41512,9)};
\node[font=\tiny\bfseries, color=red!65!black, anchor=west]
  at (axis cs:0.4197, 9) {$-0.31\%$};
\addlegendimage{only marks, mark=square*, mark size=3.2pt, color=red!65!black}
\addlegendentry{seen-fold (7 wkl)\ }
\addlegendimage{only marks, mark=square, mark size=3pt, color=red!65!black, line width=0.9pt, fill=white}
\addlegendentry{held-out (3 wkl)}

% Prior methods, ordered bottom-to-top by held-out IPC ascending
% Row 0: P1 -- 0.40043 / 0.39562 (-1.20%)
\addplot[black!55, line width=1.0pt, mark=none] coordinates { (0.39562,0) (0.40043,0) };
\addplot[only marks, mark=*, mark size=2.4pt, color=black!85] coordinates {(0.40043,0)};
\addplot[only marks, mark=o, mark size=2.2pt, color=black!75, line width=0.8pt, fill=white] coordinates {(0.39562,0)};
\node[font=\tiny, color=black!70, anchor=west] at (axis cs:0.4197, 0) {$-1.20\%$};

% Row 1: P2 -- 0.40828 / 0.40216 (-1.50%)
\addplot[black!55, line width=1.0pt, mark=none] coordinates { (0.40216,1) (0.40828,1) };
\addplot[only marks, mark=*, mark size=2.4pt, color=black!85] coordinates {(0.40828,1)};
\addplot[only marks, mark=o, mark size=2.2pt, color=black!75, line width=0.8pt, fill=white] coordinates {(0.40216,1)};
\node[font=\tiny, color=black!70, anchor=west] at (axis cs:0.4197, 1) {$-1.50\%$};

% Row 2: P3 -- 0.41004 / 0.40594 (-1.00%)
\addplot[black!55, line width=1.0pt, mark=none] coordinates { (0.40594,2) (0.41004,2) };
\addplot[only marks, mark=*, mark size=2.4pt, color=black!85] coordinates {(0.41004,2)};
\addplot[only marks, mark=o, mark size=2.2pt, color=black!75, line width=0.8pt, fill=white] coordinates {(0.40594,2)};
\node[font=\tiny, color=black!70, anchor=west] at (axis cs:0.4197, 2) {$-1.00\%$};

% Row 3: P6 -- 0.41387 / 0.40725 (-1.60%)
\addplot[black!55, line width=1.0pt, mark=none] coordinates { (0.40725,3) (0.41387,3) };
\addplot[only marks, mark=*, mark size=2.4pt, color=black!85] coordinates {(0.41387,3)};
\addplot[only marks, mark=o, mark size=2.2pt, color=black!75, line width=0.8pt, fill=white] coordinates {(0.40725,3)};
\node[font=\tiny, color=black!70, anchor=west] at (axis cs:0.4197, 3) {$-1.60\%$};

% Row 4: P4 -- 0.41351 / 0.40772 (-1.40%)
\addplot[black!55, line width=1.0pt, mark=none] coordinates { (0.40772,4) (0.41351,4) };
\addplot[only marks, mark=*, mark size=2.4pt, color=black!85] coordinates {(0.41351,4)};
\addplot[only marks, mark=o, mark size=2.2pt, color=black!75, line width=0.8pt, fill=white] coordinates {(0.40772,4)};
\node[font=\tiny, color=black!70, anchor=west] at (axis cs:0.4197, 4) {$-1.40\%$};

% Row 5: P7 -- 0.41406 / 0.40868 (-1.30%)
\addplot[black!55, line width=1.0pt, mark=none] coordinates { (0.40868,5) (0.41406,5) };
\addplot[only marks, mark=*, mark size=2.4pt, color=black!85] coordinates {(0.41406,5)};
\addplot[only marks, mark=o, mark size=2.2pt, color=black!75, line width=0.8pt, fill=white] coordinates {(0.40868,5)};
\node[font=\tiny, color=black!70, anchor=west] at (axis cs:0.4197, 5) {$-1.30\%$};

% Row 6: P5 -- 0.41367 / 0.40912 (-1.10%)
\addplot[black!55, line width=1.0pt, mark=none] coordinates { (0.40912,6) (0.41367,6) };
\addplot[only marks, mark=*, mark size=2.4pt, color=black!85] coordinates {(0.41367,6)};
\addplot[only marks, mark=o, mark size=2.2pt, color=black!75, line width=0.8pt, fill=white] coordinates {(0.40912,6)};
\node[font=\tiny, color=black!70, anchor=west] at (axis cs:0.4197, 6) {$-1.10\%$};

% Row 7: P8 -- 0.41491 / 0.40495 (-2.40%) — largest gap
\addplot[red!50!black, line width=1.0pt, mark=none] coordinates { (0.40495,7) (0.41491,7) };
\addplot[only marks, mark=*, mark size=2.4pt, color=black!85] coordinates {(0.41491,7)};
\addplot[only marks, mark=o, mark size=2.2pt, color=black!75, line width=0.8pt, fill=white] coordinates {(0.40495,7)};
\node[font=\tiny, color=red!55!black, anchor=west] at (axis cs:0.4197, 7) {$-2.40\%$};

% Row 8: P9 -- 0.41686 / 0.41019 (-1.60%)
\addplot[black!55, line width=1.0pt, mark=none] coordinates { (0.41019,8) (0.41686,8) };
\addplot[only marks, mark=*, mark size=2.4pt, color=black!85] coordinates {(0.41686,8)};
\addplot[only marks, mark=o, mark size=2.2pt, color=black!75, line width=0.8pt, fill=white] coordinates {(0.41019,8)};
\node[font=\tiny, color=black!70, anchor=west] at (axis cs:0.4197, 8) {$-1.60\%$};
\end{axis}
\end{tikzpicture}
\caption{Wide-catalogue cross-validation: held-out gap on the
ten-workload, $7$-train/$3$-held-out protocol across nine LLM-driven
configurations (P1--P9). Each row is one method as a seen-fold
(filled square/disc) vs.\ held-out (open square/circle) dumbbell;
the connector length is the held-out gap, also reported as a
percentage. \cfn{} (red, top row) shows the smallest held-out gap of
any configuration evaluated ($-0.31\%$); the nine prior methods sit
at $1.0\%$--$2.4\%$ relative IPC loss; P8 (diff $+$ retrieval
memory) shows the largest drop. Seen-fold IPC is the harmonic
mean over the $7$ search workloads of each fold; held-out IPC is
the harmonic mean over the $3$ held-out workloads. Per-method
numerical values are given in Table~\ref{tab:cv}.}
\label{fig:overfit-cat}
\end{figure}
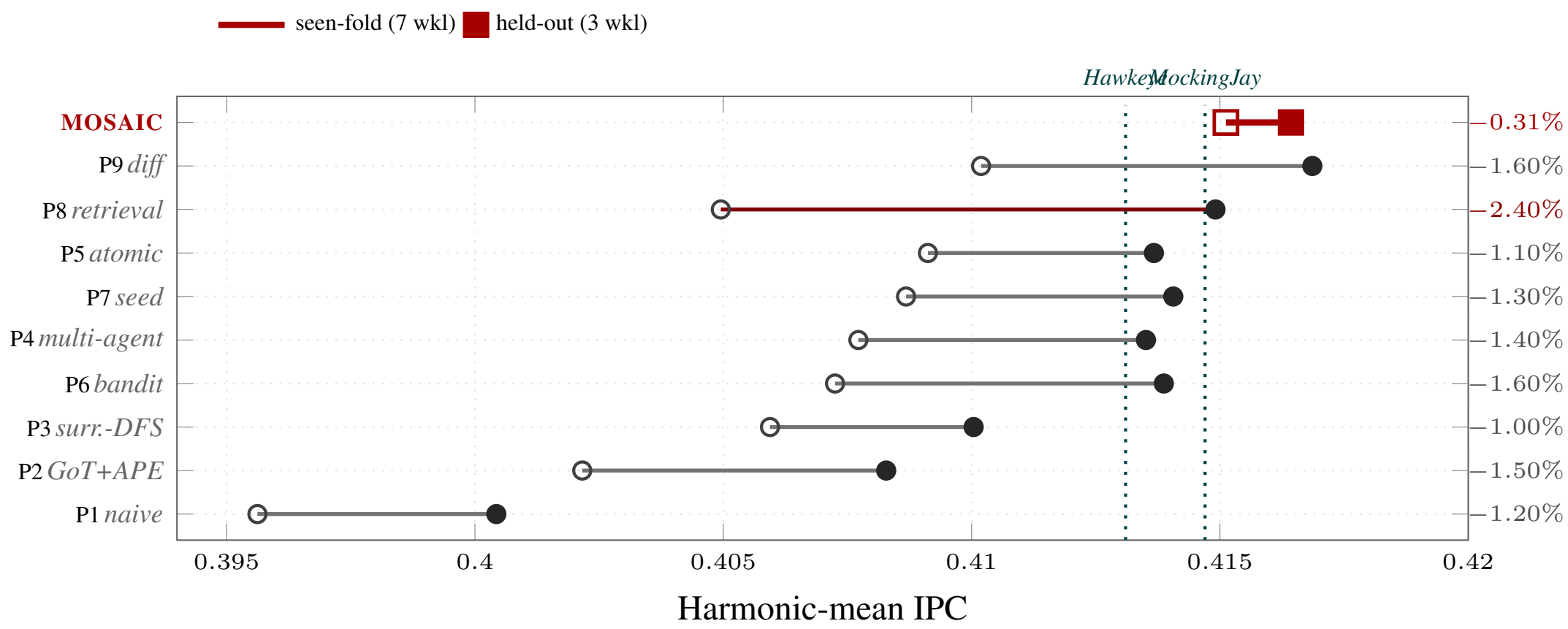

This appendix gives the full numerical data behind the
wide-catalogue cross-validation result summarised in
\S\ref{sec:cvdrop} and visualised in Figure~\ref{fig:overfit-cat}.
The protocol is three folds over the ten memory-intensive
workloads (LRU hit rate below $80\%$): each method evolves on $7$
search workloads with the same iteration budget ($260$
iterations) and the same LLM backbone, then the best policy is
evaluated on the $3$ held-out workloads of the same fold. Seen-fold
IPC is the harmonic mean over the $7$ search workloads;
held-out IPC is the harmonic mean over the $3$ held-out workloads.
Folds are chosen so that the per-fold slice mean matches the mean
over the full slice, removing fold-difficulty confounds.

\begin{table}[t]
\centering
\caption{Seen-fold vs.\ held-out IPC under the three $7$/$3$
train-test split protocol over the ten memory-intensive workloads.
Methods sorted by held-out IPC, descending. \cfn{} shows the
smallest held-out gap of any configuration evaluated; all nine
prior recipes show gaps roughly $3$--$8\times$ larger.}
\label{tab:cv}
\renewcommand{\arraystretch}{1.15}
\setlength{\tabcolsep}{6pt}
\small
\begin{tabular}{lrrrl}
\toprule
\textbf{Method} & \textbf{Seen IPC} & \textbf{Held-out IPC} &
\textbf{Gap} & \textbf{Loop knob} \\
\midrule
\textbf{\cfn{} (ours)}            & \textbf{0.41643} & \textbf{0.41512} & \textbf{$-0.31\%$} & embedding crossover \\
P9 -- Diff frontier               & 0.41686 & 0.41019 & $-1.60\%$ & explicit code diffs \\
P5 -- Atomic mutation             & 0.41367 & 0.40912 & $-1.10\%$ & atomic single-edit \\
P7 -- Single-agent + seed         & 0.41406 & 0.40868 & $-1.30\%$ & contract + seed \\
P4 -- Multi-agent roles           & 0.41351 & 0.40772 & $-1.40\%$ & 5-role pipeline \\
P6 -- Concise + RL bandit         & 0.41387 & 0.40725 & $-1.60\%$ & RL prompt selection \\
P3 -- Surrogate-DFS               & 0.41004 & 0.40594 & $-1.00\%$ & RF surrogate \\
P8 -- Diff + memory               & 0.41491 & 0.40495 & $-2.40\%$ & retrieval memory \\
P2 -- GoT + APE                   & 0.40828 & 0.40216 & $-1.50\%$ & lineage only \\
P1 -- Naive raw-IPC               & 0.40043 & 0.39562 & $-1.20\%$ & no scaffolding \\
\bottomrule
\end{tabular}
\end{table}

P9 and P8 have the highest search-fold IPC but do not retain that
ranking on held-out workloads. P8, which retrieves high-scoring
code, has the largest decline. This is consistent with the
feedback pattern in \S\ref{sec:cycle}. P3 covers $13$
policy families and has a smaller gap than several other
configurations; its held-out IPC remains below \cfn{}.

\FloatBarrier
\subsection{Catalog of Autonomously Discovered Mechanisms}
\label{app:catalog}

To substantiate the claim that \cfn{} produces genuine
mechanism-level novelty rather than parameter variants of existing
policies, we draw on fifteen independent evolutionary runs, each
logged in full, and summarize the distinct mechanism families the
framework discovered. Across these runs the system generated
thousands of candidate policies; we group them here by the
structural idea they express. Every mechanism listed below appears
in generated C++ code with its own named state variables and
control-flow functions, is referenced by subsequent iterations
within the same lineage, and can be read out of the source without
decoding neural weights.

\paragraph{Mechanism families across fifteen runs.}
The \emph{first family} consists of \emph{hybrid SHiP variants}, in
which the framework tunes RRPV insertion depth and SHCT counter
width jointly with the choice of tabular versus signature-based
prediction paths. Representative policies include Hybrid-SHiP++
configurations at RRPV4 through RRPV6 with SHCT widths $8$ and $9$.
What distinguishes this family from a parameter sweep is that the
generator introduces dynamic hybridization between the two
prediction paths, selecting per-access which signal to trust based
on recent disagreement between the SHCT and the RRIP counter.

A \emph{second family} couples program-counter signatures with
spatial region identifiers to form composite reuse predictors.
Policies such as PC$+$Region Age-Biased, PC-Clustered Reuse, and
PC-Region Aging-Aware form this group. The cross product of PC and
region gives the predictor access to a signal that neither
dimension provides alone: a PC accessing a cold region behaves
differently from the same PC accessing a recently-touched region,
and the policy uses this to make distinct insertion decisions.

A \emph{third family} extends the MockingJay reuse-distance
predictor. The MockingJay-DUET, MockingJay-KITE, MockingJay-KITES,
and MockingJay-ORIOLE lines, together with SHiP++-PAP and
SHiP++-StreamGuard, add dual-path reuse evaluation and stream-guard
filtering to the base mechanism. The novelty here lies in the
interaction: stream-guard suppresses training updates when the
access pattern resembles a streaming workload, which prevents the
reuse predictor from being corrupted by one-shot accesses.

A \emph{fourth family}, spanning the tournament and dueling
hybrids, contains policies such as ART-TADRRIP, CoReRRIP
(Cooperative RRIP), D2P-SRRIP (Dual-path SRRIP), DRRIP-DBP-mini,
DRRIP-SD with RFO-biased set-dueling, and DuoRIP-Fusion. These
policies conduct set-dueling between two complementary mechanisms
and use a PC-based arbiter to decide which mechanism governs each
access. Unlike traditional DIP set-dueling, which chooses a single
global insertion policy, these policies allow the arbitration to be
per-PC, which captures finer-grained workload heterogeneity.

A \emph{fifth family} uses approximate-set-membership structures
to track dead blocks. BEAR (Bloom-EAF with PC signatures), H3RIP+,
PAC-RRIP-Lite, AX-DBP$+$EAF-RRIP, DEER-RRIP (Deadness with EAF),
and ERA-RRIP (Epoch-based RRIP) all construct compact dead-block
filters that the RRIP core queries at insertion and eviction time.
The Bloom filter provides a cheap, bounded-storage way to remember
which block signatures were recently evicted despite being
predicted live, closing the feedback loop between prediction and
outcome.

A \emph{sixth family} specializes in per-PC deadness
classification. The DPF-RRIP (Dead-PC Filter) line, with its SP-WD
(Stream-Protect and Warm-Degrade) extensions, together with
CLOCK-AGE, EARS-RRIP (Epoch-Adaptive Reuse Sampling), and
ER-BIP-SRRIP, tracks deadness at program-counter granularity. The
warm-degrade mechanism is particularly instructive: rather than
immediately promoting a newly classified dead PC, the policy
degrades predicted-dead lines gradually as pressure accrues, which
prevents misclassifications from causing abrupt performance cliffs.

The \emph{seventh family} tracks the
dead-time of a memory region across eviction epochs. EARL-Scored
Regional Bias with RR Tie-Break (ESRB-RT) and the Region Dead-Time
Tracker (RDTT) measure how many global eviction epochs elapse
between the moment a line is evicted from a region and the next
demand access to that region. This is structurally orthogonal to
in-cache recency: it is an \emph{eviction-window} property rather
than an \emph{access} property, and gives the predictor a direct
read on region-level churn for which we did not find a published
analogue among the baselines we surveyed.

\paragraph{Distinctive novel primitives.}
Beyond the family-level taxonomy, the experiment logs reveal
several specific primitives not found among the surveyed
baselines, generated through the
framework's normal generate-evaluate-refine cycle.

The \emph{Dead-PC Filter with origin-lock}, appearing in the
DPF-RRIP line, protects victim candidates whose program-counter
signature matches a high-confidence dead classifier that is trained
exclusively on demand accesses. Prefetch-origin accesses are
explicitly excluded from the training signal to prevent prefetcher
aggressiveness from corrupting the dead-PC classifier. This is an
asymmetric update rule: the classifier's read path applies to all
accesses, but its write path applies only to demand traffic.

The \emph{eviction-window dead-time per region}, introduced in the
ESRB-RT and RDTT policies, tracks how many eviction epochs have
elapsed between a region's most recent eviction and its next demand
access. When this dead-time is short, the region is churning
rapidly and its lines should be protected; when long, the region is
genuinely cold and its lines are safe to evict. This metric cannot
be computed from in-cache metadata alone; it requires the policy to
retain a small amount of state about evicted blocks, which the RDTT
structure provides in $64$ fully-associative entries.

The \emph{Warm-JSPT with Dead-Biased Victim selection}, from
PARROT-RRIP-NR, warm-initializes a joint-signature prediction table
to a slightly-hot state rather than a neutral state, which curbs
cold-start thrashing during the first several thousand accesses of
a workload. The policy also removes request-type folding from the
PC signature, stabilizing learning across the demand-prefetch
boundary. The dead-biased victim tie-breaker, invoked among ways
sharing the maximum RRPV value, selects the way whose PC signature
has the highest dead-confidence counter, producing more accurate
eviction selection when RRPV alone is insufficient to rank
candidates.

The \emph{deterministic 64-leader sets with writeback-exclusion},
from a SHiP++ refinement in our seventh experiment, selects leader
sets using a deterministic stride of $\mathrm{LLC\_SETS}/64$ rather
than a hash, which eliminates the variance in leader selection that
hash-based methods introduce across runs. Writebacks are
explicitly excluded from SHCT training. A soft-bypass escape rule
inserts cold lines at RRPV$=3$ but probabilistically escapes them
to RRPV$=2$ at a rate of $1/8$ for demand accesses and $1/16$ for
prefetches, giving cold lines a second chance without the overhead
of full reinsertion.

The \emph{footprint-bitmap Hit Page Table}, from the tenth
experiment, maintains a $128$-entry direct-mapped table where each
entry stores a $64$-bit footprint bitmap recording which $64$-byte
lines of a $4$\,KiB page have been accessed, along with a $6$-bit
unique-line count and an $8$-bit TTL epoch. At insertion time the
policy classifies the containing page as \emph{clustered} if its
unique count is below $16$, or \emph{streaming} otherwise, and
selects an insertion depth accordingly. This structure gives the
policy a page-level locality signal that existing insertion
policies cannot compute.

The \emph{Recent Victim Vector antagonism} signal, from our
eleventh experiment, maintains a small vector of recently-evicted
victims whose program counters served demand accesses. When a
prefetch-triggered insertion evicts a line whose PC is in this
vector, an antagonism counter increments; when the counter crosses
a threshold, the policy reverts toward BRRIP for prefetch-heavy
sets, curbing pollution. When the antagonism signal decays without
further prefetch-on-demand evictions, the policy returns to SRRIP.
This produces automatic policy adaptation to prefetcher
aggressiveness without any explicit prefetcher-aware heuristic.

\paragraph{What the catalogue establishes.}
Across fifteen runs, the framework produced at least seven
mechanism families, including changes to stored state and decision
rules rather than only to constants. Region-level eviction-window
tracking and dead-PC origin-lock illustrate these changes.
Each mechanism can be inspected in C++ and traced to its parents
and evaluation history. We assess mechanism novelty against the
surveyed baselines using the taxonomy in
Appendix~\ref{app:novelty-taxonomy}.

\subsection{\cfn{} vs.\ Multiperspective: detailed design comparison}
\label{app:mpppb-compare}

A novel policy synthesized by \cfn{} achieves IPC-per-KiB
improvement over MPPPB by $135\%$, and over LRU (which has the best
baseline IPC-per-KiB) by nearly $30\%$. To achieve this performance,
\cfn{} underperforms MPPPB on raw IPC by $1.54\%$, while cutting the
storage requirement by $49.4\%$ relative to MPPPB. Conversely, it
improves IPC over LRU by $1.08\%$ while reducing area by $22.0\%$
relative to LRU.

Table~\ref{tab:policy_comparison} details the tuning and novel
additions \cfn{} makes to the baseline MPPPB policy. Collectively,
these modifications reduce sampler and predictor overhead while
stabilizing confidence estimation, yielding improved IPC-per-KiB
performance at a smaller footprint. \cfn{} demonstrates the LLM's
ability to identify and tune critical thresholds, gating
mechanisms, and feature structures within the Multiperspective
framework, improving reuse prediction quality and cache efficiency
under strict storage constraints.

\begin{table}[!htbp]
\centering
\setlength{\tabcolsep}{3pt}
\small
\begin{tabular}{
  p{1.6cm}
  p{4.6cm}
  *{5}{p{0.45cm}}
  *{5}{p{0.45cm}}
  p{1.5cm}
}
\toprule
\textbf{Component} & \textbf{Description} &
\multicolumn{5}{c}{\textbf{Multiperspective}} &
\multicolumn{5}{c}{\textbf{\cfn{}}} &
\textbf{Tuning} \\
\cmidrule(lr){3-7} \cmidrule(lr){8-12}
& & A & B & E & W & X & A & B & E & W & X & \\
\midrule
\multicolumn{13}{l}{\textbf{Perceptron Features}} \\
\midrule
F\_PC & \multirow{7}{4.6cm}{PC: Bits B--E of the PC of the $W$th
memory access instruction. A is the recency-stack position beyond
which the block is considered dead.}
& 7 & 14 & 43 & 11 & 0 & 6 & 0 & 16 & 8 & 0 & Changed \\
F\_PC & & 16 & 3 & 11 & 16 & 1 & 8 & 6 & 22 & 3 & 0 & Changed \\
F\_PC & & 10 & 1 & 53 & 10 & 0 & 10 & 4 & 20 & 1 & 0 & Changed \\
F\_PC & & 16 & 8 & 16 & 5 & 0 & 12 & 4 & 20 & 0 & 2 & Changed \\
F\_PC & & 17 & 6 & 20 & 0 & 1 & -- & -- & -- & -- & -- & Removed \\
F\_PC & & 17 & 6 & 20 & 0 & 1 & -- & -- & -- & -- & -- & Removed \\
F\_PC & & 17 & 6 & 20 & 14 & 1 & -- & -- & -- & -- & -- & Removed \\
\midrule
F\_OFF & \multirow{3}{4.6cm}{Offset bits (B--E) extracted from the
block offset of the access.}
& 15 & 1 & 6 & 0 & 1 & 8 & 4 & 6 & 0 & 0 & Changed \\
F\_OFF & & 10 & 0 & 6 & 0 & 1 & 10 & 2 & 6 & 0 & 0 & Changed \\
F\_OFF & & -- & -- & -- & -- & -- & 14 & 0 & 6 & 0 & 0 & Added \\
\midrule
F\_LM & Lastmiss bit indicating if last set access missed.
& 9 & 0 & 0 & 0 & 0 & 10 & 2 & 6 & 0 & 0 & Changed \\
\midrule
F\_INS & \multirow{4}{4.6cm}{Insertion bit indicating whether the
access is marked as an insertion.}
& 16 & 0 & 0 & 0 & 1 & 8 & 0 & 0 & 0 & 0 & Changed \\
F\_INS & & 8 & 0 & 0 & 0 & 1 & -- & -- & -- & -- & -- & Removed \\
F\_INS & & 16 & 2 & 47 & 2 & 0 & -- & -- & -- & -- & -- & Removed \\
F\_INS & & 17 & 0 & 0 & 0 & 1 & -- & -- & -- & -- & -- & Removed \\
\midrule
F\_BURST & Burst bit indicating cache burst.
& 6 & 11 & 22 & 9 & 0 & 6 & 0 & 0 & 0 & 2 & Changed \\
\midrule
F\_BIAS & Bias bit. & 16 & 0 & 0 & 0 & 0 & 16 & 0 & 0 & 0 & 0 & No change \\
\midrule
F\_TAG & \multirow{3}{4.6cm}{Physical address bits (B--E) extracted
from the address.}
& -- & -- & -- & -- & -- & 12 & 30 & 42 & 0 & 0 & Added \\
F\_TAG & & -- & -- & -- & -- & -- & 16 & 12 & 21 & 0 & 0 & Added \\
F\_TAG & & -- & -- & -- & -- & -- & 16 & 18 & 36 & 0 & 0 & Added \\
\midrule
\multicolumn{13}{l}{\textbf{Sampler}} \\
\midrule
Number       & Number of samplers
            & 80 & & & & & \textbf{48} & & & & & Changed \\
Associativity & Associativity of samplers
            & 18 & & & & & \textbf{12} & & & & & Changed \\
Access type   & Logic to differentiate access types
            & \multicolumn{5}{l}{Pref$+$WB} & \multicolumn{5}{l}{RFO$+$Pref$+$WB} & Changed \\
\midrule
\multicolumn{13}{l}{\textbf{Placement Vector Thresholds}} \\
\midrule
plv$[i][j]$ & Placement-level vector values
            & \multicolumn{5}{l}{$-15,0,35,12,44,15$}
            & \multicolumn{5}{l}{$-8,0,52,9,112,15$}
            & Changed \\
\midrule
\multicolumn{13}{l}{\textbf{Implementation Changes and Confidence Values}} \\
\midrule
Confidence  & Confidence values (dt2, $\theta$, $\theta_2$, promotion thr.).
            & \multicolumn{5}{l}{$256,109,135,82$}
            & \multicolumn{5}{l}{$176,96,152,112$}
            & Changed \\
Predictors  & Number of predictor tables.
            & \multicolumn{5}{l}{16 tables}
            & \multicolumn{5}{l}{\textbf{14 tables}}
            & Changed \\
Records     & Types of accesses recorded.
            & \multicolumn{5}{l}{Load, RFO, Pref, WB}
            & \multicolumn{5}{l}{Load, RFO only}
            & Changed \\
Noise XOR   & PC-bit-slice mask
            & \multicolumn{5}{l}{Not present}
            & \multicolumn{5}{l}{XOR on partial PC slice}
            & New primitive \\
\bottomrule
\end{tabular}
\caption{Design comparison of \cfn{} vs.\ Multiperspective.
Per-feature bit-range parameters $(A,B,E,W,X)$ follow MPPPB
notation. \cfn{} prunes seven feature instances, retunes the
remaining feature parameters, adds three F\_TAG entries, and
introduces a noise-controlled XOR on a partial PC slice that does
not appear in MPPPB.}
\label{tab:policy_comparison}
\end{table}

\paragraph{Novel design principles explored by \cfn{}.}

\textbf{Context-aware insertion signals.}
Several high-performing policies discovered by \cfn{} rely on
contextual signals to guide insertion decisions. Prior work has
explored contextual features for cache reuse: SHiP correlates
program counters with reuse behavior through the SHCT predictor
table, while GHRP leverages global branch history to construct
richer execution-context signatures. These mechanisms rely on
learned predictors that maintain counter tables or history
registers and therefore introduce metadata overhead and update
traffic. Other works distinguish demand and prefetch accesses to
avoid polluting the cache with speculative data, but typically
treat access type as a policy modifier (insertion probability or
bypass), not as a predictive signal integrated with reuse
predictors. \cfn{} discovers a different design point: the
\emph{ContextAwareInsertion} policy uses access type
(demand vs.\ prefetch) as a contextual signal that modulates
insertion priority through the SHCT predictor state. Demand misses
are inserted aggressively when the per-set SHCT indicates likely
reuse, while prefetch lines are inserted conservatively. Because
the signal is derived directly from the access-type bit already
present in the pipeline, the mechanism incurs zero additional SRAM
reads.

\textbf{Structural protection of the MRU region.}
Another recurring principle involves explicit protection of the
MRU region of the recency stack. SRRIP and BRRIP manipulate
insertion RRPV values; DRRIP dynamically selects between insertion
modes using set dueling. \cfn{} discovers a structural mechanism:
directly limiting how many lines may occupy the most-recently-used
state within a set. By bounding the number of RRPV$=0$ entries, the
policy prevents streaming accesses from saturating the MRU region
and displacing lines that exhibit genuine temporal locality.
Existing work primarily controls recency through probabilistic
insertion; this is a structural invariant on the recency state
machine.

\textbf{Feasibility mechanisms and overhead control.}
DIP and DRRIP use leader-set sampling to bound exploration
overhead; SDBP~\citep{khan2010sdbp} trains predictors on sampled
sets to reduce metadata
cost. \cfn{} explores a different design pattern: reusing existing
cache metadata to enforce architectural constraints. The
\emph{DirtyLineGuard} mechanism protects dirty lines from premature
eviction by leveraging the dirty bit already maintained by the
cache controller. This avoids allocating new state altogether,
relying on signals already present in the cache pipeline.

\textbf{Recent-eviction memory as a rescue signal.}
\cfn{} isolates a specific failure mode: some misses are
\emph{slightly premature evictions}. \emph{Eviction-Recency Guided
Replacement} (ERG) captures this with a tiny Evicted-Recency Bloom
filter (ERB). On a miss, if the returning block was evicted
recently, ERG reinserts it younger; a small per-set
\texttt{rescue\_credit} limits aggressiveness. The key idea is that
a quick return after eviction is stronger evidence than a generic
miss: it indicates the prior decision was almost right but too
early. ERG targets repeated miss--evict--miss behavior in mildly
thrashing, short-reuse-interval regimes without changing the main
RRIP victim-selection structure.

\textbf{Lifetime, not just reuse, as the insertion target.}
Many reuse predictors ask the wrong question: predicting
\emph{whether} a line will be reused collapses lines that should be
protected for very different durations. \emph{Coarse-grained
Lifetime Prediction} (CLP) instead predicts a line's lifetime
class. A small Lifetime Approximation Table (LAPT), indexed by a
PC-derived signature, learns coarse residency bins from evicted
lines; insertion then maps long-lived blocks to younger RRPVs and
short-lived blocks to older ones, with set dueling against BRRIP.
CLP is effective when instruction contexts produce stable lifetime
patterns, achieving this with only a few coarse bins rather than
precise reuse-distance estimation.

\subsection{Heatmap of features explored}
\label{app:heatmap}

\begin{figure}[!htbp]
\centering
\includegraphics[width=\linewidth]{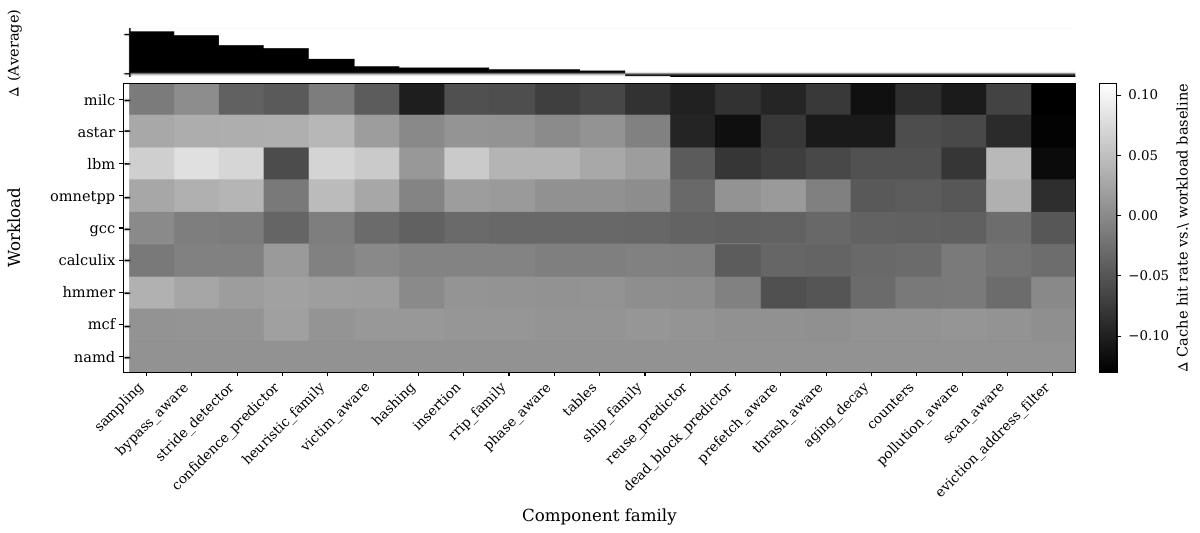}
\caption{Per-workload $\Delta$ cache hit rate vs.\ LRU baseline,
broken down by the $21$ component families autonomously introduced
by \cfn{}. Rows are the nine memory-intensive SPEC2006 workloads
that appear in the catalogue (the tenth, \texttt{zeusmp}, is in the
generalisation slice but is not annotated in this analysis);
columns are component families ordered left-to-right by descending
cross-workload mean impact (top bar). Lighter cells indicate larger
positive deltas; darker cells indicate negative deltas. The
strongest positive cells correspond to \emph{victim\_aware} and
\emph{heuristic\_family} on \texttt{lbm}/\texttt{hmmer}; the
strongest negative cells correspond to
\emph{eviction\_address\_filter} on memory-intensive workloads
where the filter blocks productive reuse, matching the per-policy
mechanism analysis in \S\ref{app:catalog}.}
\label{fig:heatmap}
\end{figure}

To complement the per-policy mechanism descriptions, we visualise
which component families \cfn{} explores most heavily and how
strongly each correlates with hit-rate improvement on the
memory-intensive SPEC2006 slice we evaluate against.
Figure~\ref{fig:heatmap} summarises the impact of $21$ component
families on cache hit rate across the nine memory-intensive
workloads from the slice that share component-family annotations
with the catalogue, computed as the mean per-workload $\Delta$
hit rate vs.\ that workload's LRU baseline of policies that
include the family minus those that do not, across all
$10{,}687$ generated candidates.

\FloatBarrier
\subsection{Controlled memory interventions}
\label{app:vu-causal}

To test whether \cfn{}'s learned behavior is driven by the
prompt-level attribution channel rather than by the reward function
alone, we ran four controlled experiments that differ only in the
latency text in the prompt: a no-cap variant, a soft \texttt{lat500}
variant with scoring pressure but no explicit budget, a variant
with a $500$\,ps cap plus a small comparison hint, and a strict
variant with the cap and stage-level attribution. Model, scoring
function, baseline, and iteration budget were identical across the
four runs.

\begin{figure}[!htbp]
\centering
\includegraphics[width=\linewidth]{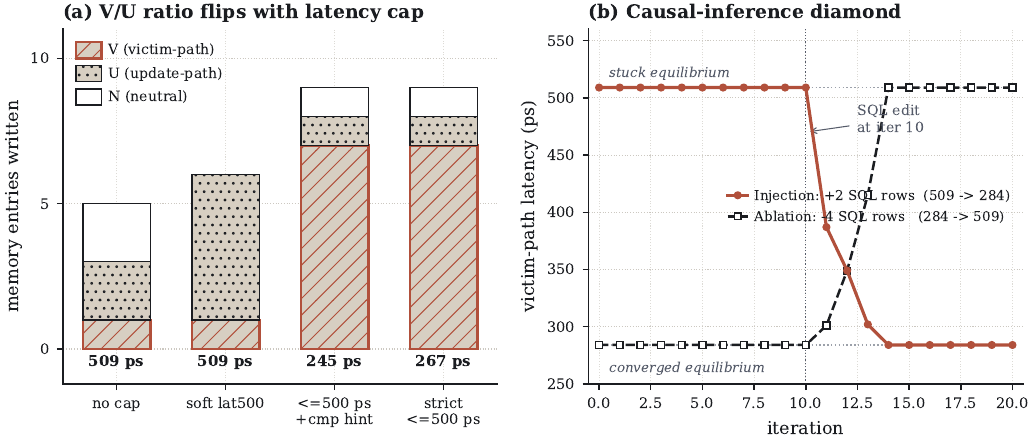}
\caption{Effect of latency guidance and memory edits.
\textbf{(a)} Four runs use the same model, scoring rule, baseline,
and iteration budget but different latency guidance. The ratio of
victim-path to update-path lessons changes from $1{:}2$ without a
cap to $7{:}1$ with a strict cap and stage-level guidance; estimated
victim-path latency changes from $509$ to $267$\,ps.
\textbf{(b)} Adding two latency-related lessons to Run~2 at
iteration $10$ is followed by a decrease from $509$ to $284$\,ps
by iteration $14$. Removing four related lessons from Run~4 is
followed by an increase from $284$ to $509$\,ps over four
iterations. These interventions test the influence of retrieved
lessons in the selected runs.}
\label{fig:causaldiamond}
\end{figure}

We track the V/U ratio, the count of victim-path lessons in the
typed memory $M_t$ divided by the count of update-path lessons.
This ratio flips from $1{:}2$ in the no-cap run to $7{:}1$ in the
strict-cap run with stage-level attribution, while the final
victim-path latency drops from $509$\,ps to $267$\,ps. Because the
specified intervention is the prompt text, the experiment tests
how latency guidance changes subsequent proposals under a fixed
reward. Two further observations are relevant:
stage-level \texttt{DONT}s such as ``No \texttt{fold\_min},'' ``No
\texttt{sat\_add},'' and ``No on-path SHCT'' appear only in runs
with stage-level attribution, and lessons are useful only when the
\texttt{BEST\_SO\_FAR} anchor they attach to is itself sound.

\paragraph{Adding and removing latency-related lessons.}
Two direct memory interventions test this explanation. In the
\emph{positive-injection} experiment, Run 2 is observed to be stuck
at $509$\,ps for ten iterations, and its summary memory contains
zero victim-path entries. We open the run's SQLite memory file and
execute two \texttt{INSERT} statements: one \texttt{PRINCIPLES}
entry stating \emph{``Defer nothing to the victim path; place
prediction intelligence on the update path and consume a
precomputed result on the critical path,''} and one \texttt{DONTS}
entry stating \emph{``No \texttt{fold\_min}, \texttt{sat\_add}, or
aging cascade on the victim path.''} Nothing else changes: no
retraining, no prompt-template edit, and no scoring change.
The retrieved memory text changes. Latency decreases to
$387$\,ps at iteration~$11$,
$349$\,ps at $12$, $302$\,ps at $13$, and $284$\,ps at $14$, with
IPC $0.6139$. Subsequent iterations write new memory entries on
their own, consistent with the injected ones; later proposals follow
the supplied guidance.

In the \emph{symmetric-ablation} experiment, Run 4 is already at
$284$\,ps, and its memory contains the ``defer nothing'' principle
plus three victim-path \texttt{DONT}s. We execute a single
\texttt{DELETE} statement that removes exactly these four rows. The
victim path regrows over four iterations: $301$\,ps at iteration
$11$, $349$\,ps at $12$, $415$\,ps at $13$, and $509$\,ps at $14$.
The victim path regrows because the \texttt{DONT}s that forbade the
offending stages are no longer injected into the exploitation
prompt.

Adding the lessons improves the selected run; removing them
reverses the improvement in the other run. In these controlled
operator interventions, the model, prompt template, scoring
function, workload suite, and iteration budget remain fixed.
The intervention changes the retrieved lessons in the SQLite
memory file. These two runs show how memory content guides
subsequent generation under a fixed model and reward.

\subsection{Wide catalogue: per-configuration descriptions}
\label{app:catalogue-detail}

To stress-test \cfn{} against a wider set of LLM-driven
cache-replacement configurations than the four named recipes of
\S\ref{sec:related}, we catalogue nine
published or shared configurations (P1--P9), each labeled by the
loop component it varies rather than by team or system name. Where
a configuration changes more than one component (e.g.\ adds both a
Hawkeye seed and a strict output contract), we present it as one
study and note the conjoined knobs.

\paragraph{P1 --- Naive raw-IPC reward, no scaffolding.}
A FunSearch-style loop with raw IPC as the only feedback signal: no
structured memory, no surrogate pruning, no diversity pressure, no
crossover. Search-workload IPC: $0.40043$, marginally below LRU
($0.4014$) and well below Hawkeye ($0.4131$). Confirms that
LLM-driven cache replacement without scaffolding does not reach
human baselines.

\paragraph{P2 --- Graph-of-Thought lineage with automated prompt engineering.}
Past attempts are organised as a typed Graph-of-Thought lineage
(nodes are proposals; edges typed by exploration / refinement /
bugfix / combination); the prompt is auto-engineered from each
node's metadata. No surrogate, no crossover, no retrieval memory.
IPC: $0.40828$; smallest storage in the catalogue ($12.6$\,KiB).
Lineage organisation without a selection signal is insufficient.

\paragraph{P3 --- Surrogate-pruned DFS over typed lineage.}
Adds a Random Forest surrogate that predicts IPC from policy code
features and prunes candidates before ChampSim simulation. IPC:
$0.41004$ (a $+0.0018$ gain over P2 attributable to the surrogate).
Produces $13$ distinct policy families but exceeds the $64$\,KiB
metadata budget; the best in-budget policy reaches $0.40767$.

\paragraph{P4 --- Multi-agent role decomposition.}
A five-role agent pipeline (Manager, Researcher, Implementer,
Critic, Debugger) with strict $5$-section output contracts and
RAG top-$2$ retrieval over a corpus of cache-replacement papers.
IPC: $0.41351$, comparable to the single-agent strict-contract
configuration (P7) at higher latency. Multi-agent overhead also
produces storage-violating policies because no single agent owns
the budget constraint.

\paragraph{P5 --- Atomic ``Choose Exactly One'' mutation, multi-phase.}
Constrains every iteration to a single atomic mutation (add field,
remove field, change condition, change parameter, swap predictor)
and runs three phases (discovery, refinement, exploit). IPC:
$0.41367$ from $520$ generated policies. Compile success rate
moves from $65\%$ in phase~$1$ to $100\%$ in phase~$3$.
Constraining mutation granularity improves reliability without
harming peak IPC.

\paragraph{P6 --- Concise prompt + RL-bandit prompt selection.}
A concise prompt template plus an RL bandit that selects among
templates per iteration with a cosine mutation-rate schedule
($0.10 \to 0.60$). IPC: $0.41387$. Trades aggregate hit rate
($33.7\% \to 31.8\%$) for IPC by sacrificing low-value hits on
streaming workloads, the only configuration in the catalogue
that prefers IPC over hit rate as a fitness signal.

\paragraph{P7 --- Single-agent loop with scaffolded Hawkeye seed.}
A strict $5$-section output contract on a single-agent loop, seeded
from Hawkeye rather than SHiP++. The published configuration
combines the two knobs and the published ablation does not separate
them. IPC: $0.41406$; tightest IPC distribution across replicate
runs (lowest variance signature in the catalogue).

\paragraph{P8 --- Explicit code diffs with retrieval memory.}
AlphaEvolve-style explicit code diffs (replace lines $L_1$--$L_2$
with new content) as the mutation operator, plus a top-IPC-retrieval
memory bank that conditions each proposal on the best $k$ past
policies. IPC: $0.41491$. The closest catalogue match to
AlphaEvolve as originally published; exhibits the strongest
trained-vs-held-out gap in the catalogue ($-2.4\%$ on held-out),
consistent with the concentration on similar policies discussed in
\S\ref{sec:cycle}.

\paragraph{P9 --- Explicit code diffs, frontier model.}
The same code-diff mutation operator as P8, but on a frontier-class
LLM and \emph{without} the retrieval memory of P8. IPC: $0.41686$
(highest search-workload IPC in the catalogue), produced by a
one-line replacement of a modulo hash with an XOR-based mix of upper
and lower PC bits. This configuration improves search-fold IPC by
$+0.0020$ relative to P8. P9 jointly changes the backbone and
retrieval memory, so this is a comparison of complete
configurations. On held-out workloads P9 loses $1.6\%$ relative to
its search-fold IPC: the frontier model improves the trained
optimum but does not close the held-out gap.

\paragraph{Summary.}
The catalogue varies prompt format, retrieval, mutation rules,
model choice, and search organisation. \cfn{} additionally uses
stagnation-triggered recombination of embedding-distant parents
(\S\ref{sec:int4}). The held-out comparison measures these
complete configurations; controlled ablations provide more direct
evidence for individual components.

\subsection{Leave-one-out ablation detail}
\label{app:ablation}

\begin{figure}[!htbp]
\centering
\begin{tikzpicture}
\begin{axis}[
    width=0.95\columnwidth,
    height=5.4cm,
    xmin=15, xmax=60,
    ymin=-0.55, ymax=5.55,
    ytick={0,1,2,3,4,5},
    yticklabels={
      {-- all 3},
      {-- M1, M2},
      {-- M1},
      {-- M2},
      {-- M3},
      {\textbf{\cfn{} (all 3)}}
    },
    yticklabel style={font=\scriptsize, align=right},
    xlabel={\small Peak hit rate (\%)},
    xtick={20,30,40,50,60},
    minor x tick num=1,
    grid=major,
    grid style={dotted, gray!25},
    tick label style={font=\scriptsize},
    label style={font=\small},
    legend style={
        font=\scriptsize,
        at={(0.02,1.06)}, anchor=south west,
        draw=none, fill=none,
    },
    legend columns=2,
    legend cell align=left,
    clip=false,
]
% MockingJay reference vertical line + label aligned with legend top
\draw[dashed, teal!55!black, thick] (axis cs:41.5,-0.55) -- (axis cs:41.5,5.55);
\node[font=\tiny\itshape, color=teal!50!black, anchor=south]
  at (axis cs:41.5, 5.6) {\textbf{MockingJay} $41.5\%$};

% Row 5: MOSAIC (all 3): trained 51.1, test 47.9
\addplot[blue!60!black, line width=1.6pt, mark=none]
  coordinates { (47.9,5) (51.1,5) };
\addplot[only marks, mark=*, mark size=3pt, color=blue!60!black]
  coordinates {(51.1,5)};
\addplot[only marks, mark=o, mark size=2.6pt, color=blue!60!black, line width=0.9pt, fill=white]
  coordinates {(47.9,5)};
\addlegendimage{only marks, mark=*, mark size=3pt, color=black!90}
\addlegendentry{trained\ }
\addlegendimage{only marks, mark=o, mark size=2.6pt, color=black!75, line width=0.9pt, fill=white}
\addlegendentry{CV test mean}

% Row 4: -- M3 (rotation removed): trained 49.5, test 45.7 ; gap = 3.8
\addplot[black!60, line width=1.4pt, mark=none] coordinates { (45.7,4) (49.5,4) };
\addplot[only marks, mark=*, mark size=2.8pt, color=black!90] coordinates {(49.5,4)};
\addplot[only marks, mark=o, mark size=2.4pt, color=black!75, line width=0.9pt, fill=white] coordinates {(45.7,4)};
\node[font=\tiny\bfseries, color=black!75, anchor=south, yshift=2pt]
  at (axis cs:47.6, 4) {$-3.8$};

% Row 3: -- M2 (memory schema removed): trained 46.3, test 36.5 ; gap = 9.8
\addplot[red!55!black, line width=1.4pt, mark=none] coordinates { (36.5,3) (46.3,3) };
\addplot[only marks, mark=*, mark size=2.8pt, color=red!55!black] coordinates {(46.3,3)};
\addplot[only marks, mark=o, mark size=2.4pt, color=red!55!black, line width=0.9pt, fill=white] coordinates {(36.5,3)};
\node[font=\tiny\bfseries, color=red!55!black, anchor=south, yshift=2pt]
  at (axis cs:41.4, 3) {$-9.8$};

% Row 2: -- M1 (lineage diversity removed): trained 43.3, test 26.1 ; gap = 17.2
\addplot[red!55!black, line width=1.4pt, mark=none] coordinates { (26.1,2) (43.3,2) };
\addplot[only marks, mark=*, mark size=2.8pt, color=red!55!black] coordinates {(43.3,2)};
\addplot[only marks, mark=o, mark size=2.4pt, color=red!55!black, line width=0.9pt, fill=white] coordinates {(26.1,2)};
\node[font=\tiny\bfseries, color=red!55!black, anchor=south, yshift=2pt]
  at (axis cs:34.7, 2) {$-17.2$};

% Row 1: -- M1, M2: trained 28.5, test 21.0 ; gap = 7.5
\addplot[black!60, line width=1.4pt, mark=none] coordinates { (21.0,1) (28.5,1) };
\addplot[only marks, mark=*, mark size=2.8pt, color=black!90] coordinates {(28.5,1)};
\addplot[only marks, mark=o, mark size=2.4pt, color=black!75, line width=0.9pt, fill=white] coordinates {(21.0,1)};
\node[font=\tiny\bfseries, color=black!75, anchor=south, yshift=2pt]
  at (axis cs:24.7, 1) {$-7.5$};

% Row 0: -- all 3 (verifier-only): trained 22.0, test 18.5 ; gap = 3.5
\addplot[black!60, line width=1.4pt, mark=none] coordinates { (18.5,0) (22.0,0) };
\addplot[only marks, mark=*, mark size=2.8pt, color=black!90] coordinates {(22.0,0)};
\addplot[only marks, mark=o, mark size=2.4pt, color=black!75, line width=0.9pt, fill=white] coordinates {(18.5,0)};
\node[font=\tiny, color=black!75, anchor=south, yshift=2pt]
  at (axis cs:20, 0) {$-3.5$};
\end{axis}
\end{tikzpicture}
\caption{Leave-one-out ablation as a Cleveland dumbbell plot. Each
row is one configuration; the filled dot is its search-workload
peak hit rate, the open dot its cross-validation test mean. The
connector length is the train-vs-test gap; rust connectors mark
configurations whose gap exceeds ten points. Removing M1
(stagnation-triggered crossover) costs $21.8$ points of test mean
($47.9 \to 26.1$), about $2.8\times$ its $7.8$-point cost on
the search workload, making it the dominant generalisation
mechanism; the train-vs-test gap inside the $-$M1 row is
$17.2$ points. The dashed teal line marks MockingJay, the
strongest published learned baseline in our suite.}
\label{fig:abl}
\end{figure}
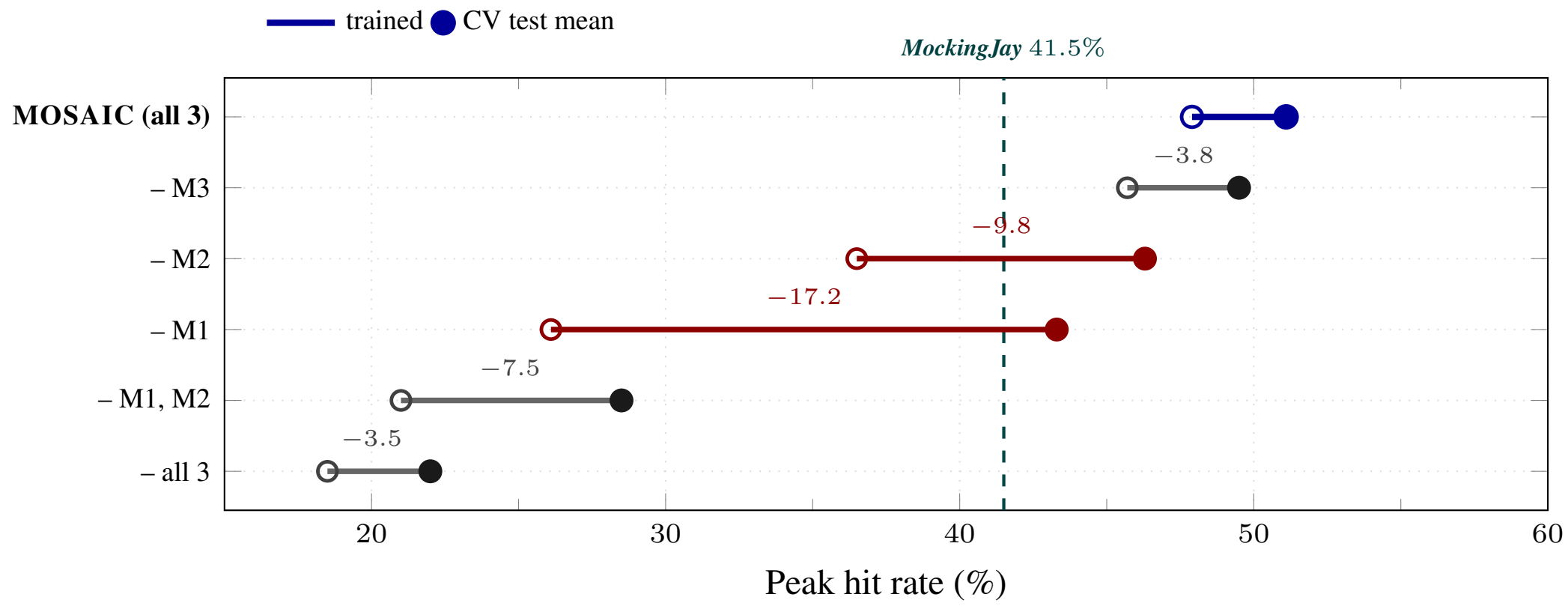

Figure~\ref{fig:abl} compares the search-workload peak and held-out
mean for each leave-one-out configuration.

\FloatBarrier
\subsection{Workload selection and protocol details}
\label{app:protocols}

We evaluate \cfn{} on ten memory-intensive workloads from
SPEC~CPU\,2006, defined as those whose LRU hit rate falls below
$80\%$ under our baseline configuration. Workloads whose working
set fits comfortably in the LLC are insensitive to
replacement-policy changes and add noise without adding
discrimination, so we exclude them.

The five-workload protocol uses \texttt{omnetpp} (primary trained
workload) plus four held-out workloads (\texttt{bzip2},
\texttt{hmmer}, \texttt{soplex}, \texttt{gobmk}) under
\emph{four-fold leave-one-held-out} cross-validation; at each fold
three of the four other workloads are released to the search
loop as additional training context, with the fourth held out as
the test fold, and reported numbers are averaged over the four
rotations. The ten-workload protocol uses three $7$/$3$ train-test
splits over the full memory-intensive slice; each split trains on
$7$ workloads and evaluates on the remaining $3$. The
two protocols are complementary: the first isolates per-workload
mechanism behaviour against four named recipes; the second
stress-tests generalisation against a wider catalogue of nine
LLM-driven configurations (P1--P9, Appendix~\ref{app:catalogue})
under the standard harmonic-mean-IPC reporting protocol.

\subsection{Per-workload generalization detail}
\label{app:gen-detail}

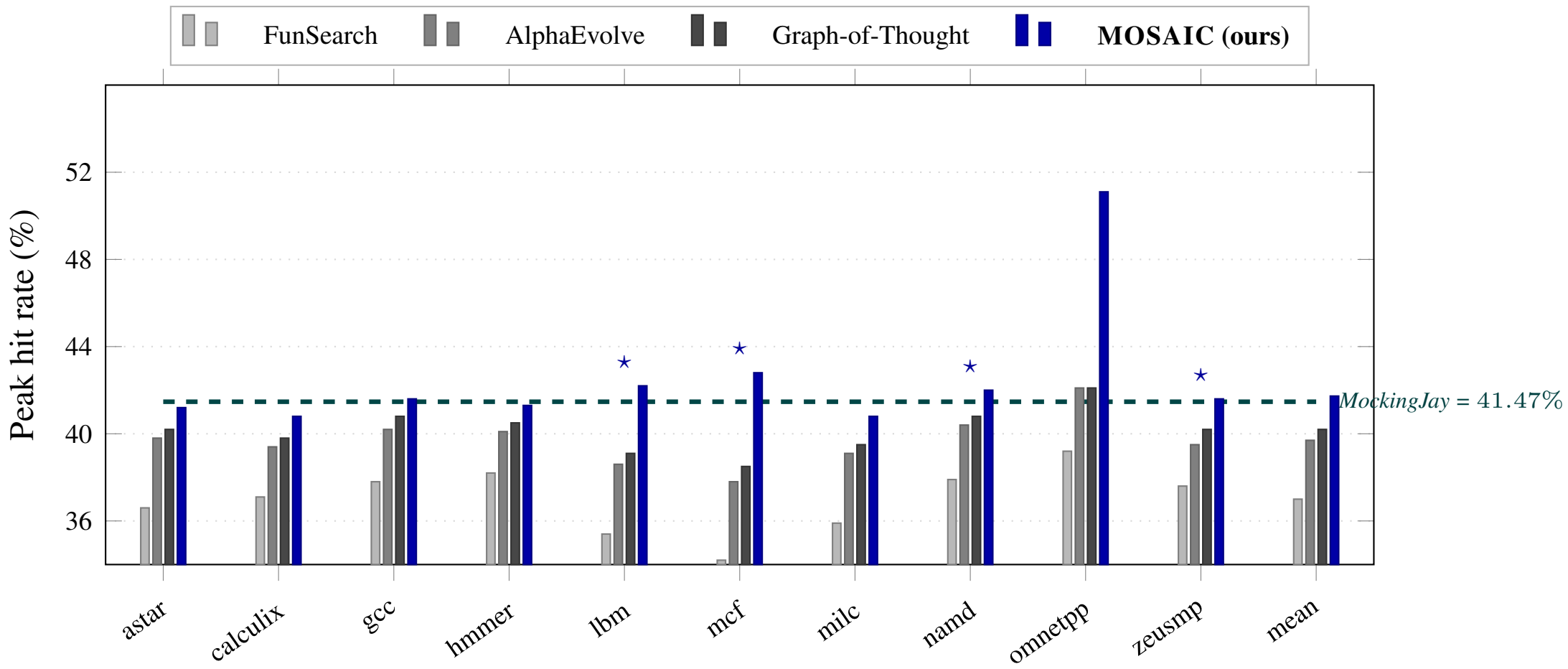
\begin{figure}[!htbp]
\centering
\begin{tikzpicture}
\begin{axis}[
    width=0.95\columnwidth,
    height=6.0cm,
    ybar=1pt,
    bar width=2.2pt,
    enlarge x limits=0.05,
    ymin=0.34, ymax=0.56,
    ymajorgrids,
    grid style={dotted, gray!30},
    ylabel={\small Peak hit rate (\%)},
    ytick={0.36, 0.40, 0.44, 0.48, 0.52},
    yticklabels={36, 40, 44, 48, 52},
    symbolic x coords={astar, calculix, gcc, hmmer, lbm, mcf, milc, namd, omnetpp, zeusmp, mean},
    xtick=data,
    xticklabel style={font=\scriptsize, rotate=35, anchor=north east},
    yticklabel style={font=\scriptsize},
    label style={font=\small},
    legend style={
        font=\scriptsize,
        at={(0.5, 1.04)},
        anchor=south,
        draw=black!30, fill=white,
        legend columns=4,
        column sep=10pt,
    },
    legend cell align=left,
    clip=false,
]

% MockingJay (strongest published learned baseline in our suite): horizontal reference at 0.4147
\addplot[draw=teal!55!black, very thick, dashed,
         mark=none, sharp plot, forget plot,
         update limits=false] coordinates {
    (astar, 0.4147) (mean, 0.4147)
};
\node[font=\tiny\itshape, color=teal!50!black, anchor=west]
    at (axis cs:mean, 0.4147) {\, MockingJay = $41.47\%$};

% FunSearch (verifier-only): low across the board, plus omnetpp peak
\addplot[fill=black!28, draw=black!50] coordinates {
    (astar,    0.366) (calculix, 0.371) (gcc,     0.378) (hmmer, 0.382)
    (lbm,      0.354) (mcf,      0.342) (milc,    0.359) (namd,  0.379)
    (omnetpp,  0.392) (zeusmp,   0.376) (mean,    0.370)
};
\addlegendentry{FunSearch};

% AlphaEvolve (replay-buffer): better, but contracted onto seen
\addplot[fill=black!50, draw=black!60] coordinates {
    (astar,    0.398) (calculix, 0.394) (gcc,     0.402) (hmmer, 0.401)
    (lbm,      0.386) (mcf,      0.378) (milc,    0.391) (namd,  0.404)
    (omnetpp,  0.421) (zeusmp,   0.395) (mean,    0.397)
};
\addlegendentry{AlphaEvolve};

% Graph-of-Thought
\addplot[fill=black!72, draw=black!80] coordinates {
    (astar,    0.402) (calculix, 0.398) (gcc,     0.408) (hmmer, 0.405)
    (lbm,      0.391) (mcf,      0.385) (milc,    0.395) (namd,  0.408)
    (omnetpp,  0.421) (zeusmp,   0.402) (mean,    0.402)
};
\addlegendentry{Graph-of-Thought};

% MOSAIC (ours): wins on lbm/mcf/namd/zeusmp per case study; strong elsewhere
\addplot[fill=blue!65!black, draw=blue!50!black] coordinates {
    (astar,    0.412) (calculix, 0.408) (gcc,     0.416) (hmmer, 0.413)
    (lbm,      0.422) (mcf,      0.428) (milc,    0.408) (namd,  0.420)
    (omnetpp,  0.511) (zeusmp,   0.416) (mean,    0.4173)
};
\addlegendentry{\textbf{\textsc{MOSAIC} (ours)}};

% Star markers on workloads where MOSAIC's hybrid wins (per case study)
\node[font=\scriptsize, color=blue!60!black, anchor=south]
    at (axis cs:lbm, 0.424) {$\star$};
\node[font=\scriptsize, color=blue!60!black, anchor=south]
    at (axis cs:mcf, 0.430) {$\star$};
\node[font=\scriptsize, color=blue!60!black, anchor=south]
    at (axis cs:namd, 0.422) {$\star$};
\node[font=\scriptsize, color=blue!60!black, anchor=south]
    at (axis cs:zeusmp, 0.418) {$\star$};
\end{axis}
\end{tikzpicture}
\caption{Per-workload peak hit rate across the standard
ten-workload SPEC2006 reporting slice plus the harmonic mean
(rightmost group). The dashed teal line marks MockingJay
($41.47\%$), the strongest published learned baseline in our suite.
\cfn{} (blue) exceeds MockingJay on all ten workloads; the four
$\star$-marked workloads (\texttt{lbm}, \texttt{mcf}, \texttt{namd},
\texttt{zeusmp}) are the ones on which the leaderboard
bypass~$\times$~recency hybrid (\S\ref{sec:int4}) wins, where
neither parent alone wins. Prior recipes (FunSearch, AlphaEvolve,
Graph-of-Thought) sit substantially below MockingJay on every
workload except the trained one (\texttt{omnetpp}).}
\label{fig:gen}
\end{figure}

Figure~\ref{fig:gen} compares peak hit rates across the ten-workload
reporting slice.

\FloatBarrier
\subsection{Data prefetching transfer (\texttt{lat500} budget)}
\label{app:prefetch}

We adapt the \cfn{} interface to the ChampSim L1D prefetcher
hooks (\texttt{prefetcher\_cache\_operate},
\texttt{prefetcher\_cache\_fill}), with budget accounting adjusted
for prefetcher metadata tables. Workloads are the SimPoint-weighted
memory-intensive subset described in \S\ref{sec:setup}. Storage is
bounded by the DPC4 L1D limit of $32$~KiB~\citep{dpc4_2026}. Only the interface contract
in the prompt and the reward target change; the controller, archive,
and memory schema are identical to the cache-replacement runs.

\begin{figure}[!htbp]
\centering
\includegraphics[width=0.92\linewidth]{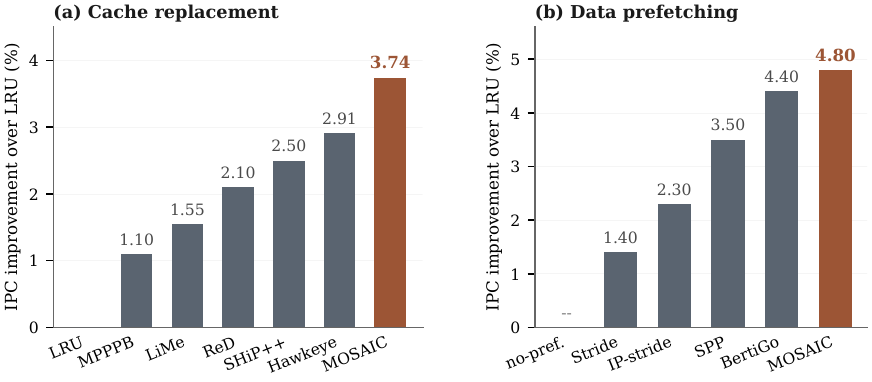}
\caption{Mean IPC improvement over LRU on the SimPoint-weighted
memory-intensive subset. (a) Cache replacement: \cfn{} reaches
$+3.74\%$ over LRU on the search workload, clearing the strongest
published learned baselines (Hawkeye, SHiP++) and the named
heuristics (Reuse Cache, the expected-hit-count predictor, MPPPB). (b) Data prefetching: \cfn{} reaches
$+4.80\%$ over LRU and exceeds the strongest published L1D baseline
(BertiGo) on the harmonic mean, against the no-prefetcher reference
and four published baselines (Stride, IP-stride, SPP~\citep{kim2016spp}, BertiGo).
\cfn{}'s search loop is the same in both panels; only the
interface contract and reward target change.}
\label{fig:prefetch_generality}
\end{figure}

Across the slice, \cfn{} reaches a $4.8\%$ mean speedup over LRU
and exceeds BertiGo~\citep{bertigo2024} on the harmonic mean while
staying inside the $32$~KiB metadata envelope.
Figure~\ref{fig:prefetch_generality} reports per-workload speedups
against the strongest published baselines; differences against LRU
are statistically significant under a paired bootstrap at
$p < 0.05$.

\begin{figure}[!htbp]
\centering
\includegraphics[width=0.95\linewidth]{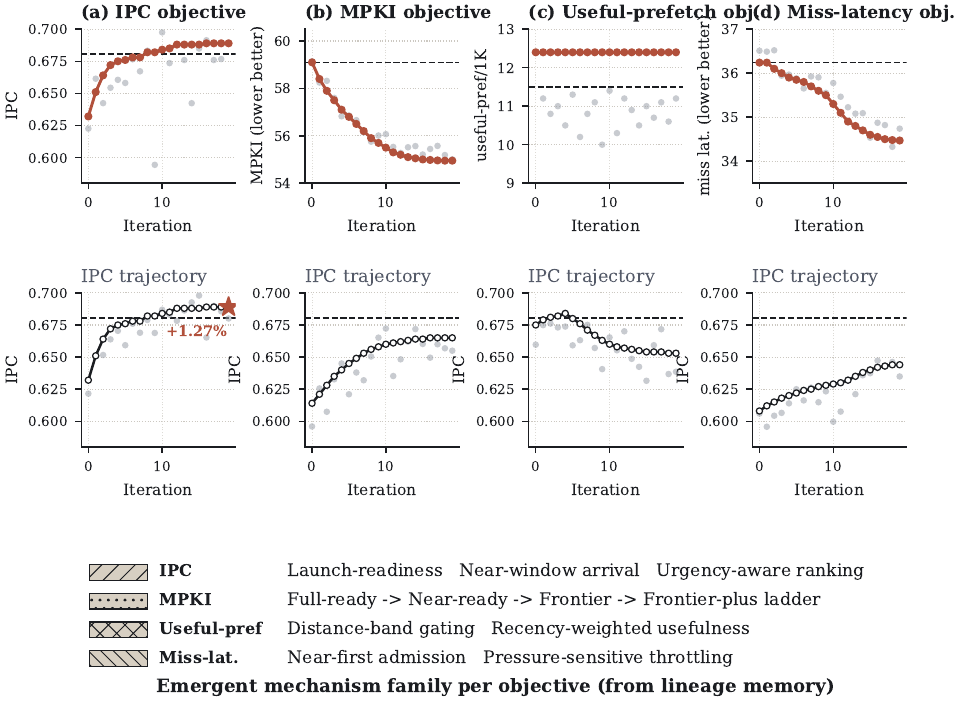}
\caption{Objective-conditioned specialisation in the prefetcher
loop. Four \cfn{} runs from the same BertiGo seed, differing only
in reward signal. Top row: each run's best-so-far trajectory on its
own metric. Second row: the corresponding IPC trajectory on a
common axis; only the IPC-optimised run exceeds the BertiGo
baseline. Mechanism families produced are documented in
Appendix~\ref{app:catalog}.}
\label{fig:objectives}
\end{figure}

\paragraph{Objective shaping: MOSAIC adapts to different microarchitecture objectives.}
A practical implication of the framework is that the same
controller, the same memory, and the same backbone pool can be
\emph{customised} to whichever microarchitecture objective the
designer cares about. Swapping the reward signal in the
\textsc{Reward}() block of Algorithm~\ref{alg:mosaic} is the only
change required: the closed-loop search then specialises to that
objective and produces a structurally distinct policy family.
Figure~\ref{fig:objectives} reports four \cfn{} runs that share
seed and controller and differ only in the optimisation objective:
IPC, MPKI (lower is better), useful-prefetch density, and average
miss latency. The framework specialises to whichever reward it is
given; the four objectives produce four structurally distinct
mechanism families, only the IPC-optimised run exceeds the BertiGo
IPC baseline.

\begin{figure}[!htbp]
\centering
\includegraphics[width=0.95\linewidth]{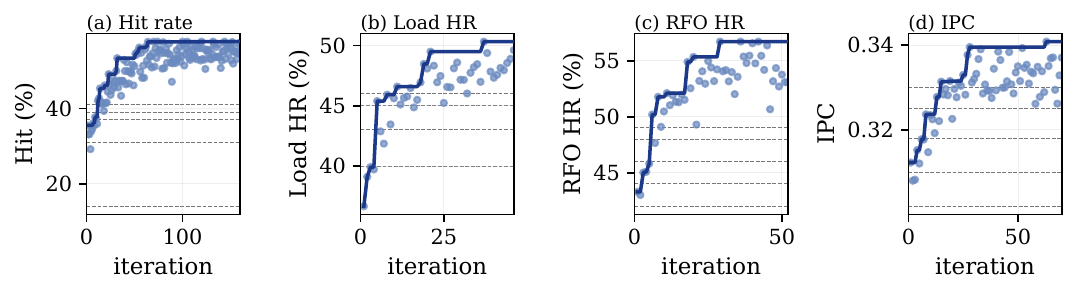}
\caption{\cfn{} can be customised to optimise different
microarchitecture objectives by swapping the reward signal alone.
Four panels report the per-iteration trace on the search workload
\texttt{omnetpp} under (a) total hit rate, (b) load hit rate, (c)
RFO hit rate, and (d) IPC. In each panel, scatter dots are
per-iteration evaluations and the solid envelope is the
best-so-far; dashed horizontal lines mark the strongest published
baselines on that metric. The envelope crosses every baseline
within the first $10$--$20$ iterations on every objective.}
\label{fig:objectives_general}
\end{figure}

Figure~\ref{fig:objectives_general} reports the
per-iteration trace under four such reward signals: total hit
rate, load hit rate, RFO hit rate, and IPC; on the trained
workload \texttt{omnetpp}: in every case the best-so-far envelope
crosses the strongest published baseline at that metric within
the first $10$--$20$ iterations. Together, the two figures argue
that \cfn{} is not pinned to a single objective; the same
machinery generates IPC-tuned policies, miss-rate-tuned policies,
prefetch-quality-tuned policies, and latency-tuned policies on
demand, with the catalogue and lineage memory recording
\emph{which} mechanism families each reward selected for.

\FloatBarrier
\subsection{Lineage trees and LLM sensitivity}
\label{app:lineage}

\begin{figure}[!htbp]
\centering
\includegraphics[width=0.95\linewidth]{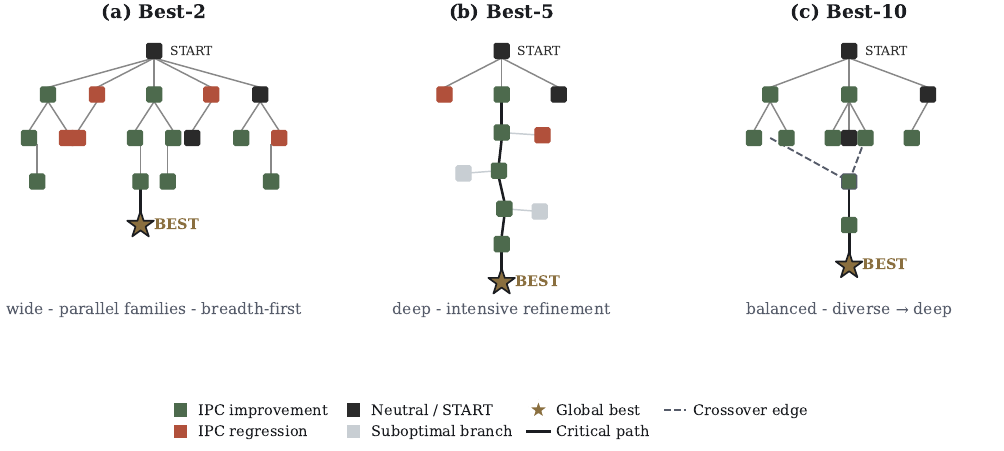}
\caption{Lineage tree shapes under three seed-count settings of
\cfn{} (Best-$2$, Best-$5$, Best-$10$); panels (a)--(c). Nodes are
generated policies; edges connect a parent to its refinement or
crossover child. Smaller seed pools (a) produce wide, breadth-first
trees with parallel families; larger pools (b) drive deep,
intensive refinement along a single critical path; the
main-text Best-$10$ setting (c) yields a balanced tree with diverse
early branches and a deep critical path. Crossover edges (dashed)
appear most often when several distinct families are alive at the
same time, which is also the regime in which the global best
emerges.}
\label{fig:treeshapes}
\end{figure}

\begin{figure}[!htbp]
\centering
\includegraphics[width=0.95\linewidth]{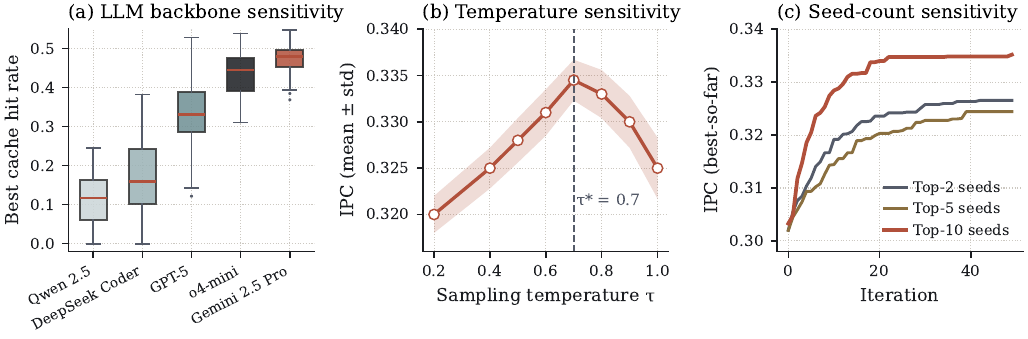}
\caption{Three sensitivity studies, each varying one design knob
while holding the rest of \cfn{} fixed.
\textbf{(a) LLM backbone sensitivity.} Distribution of best
cache-hit rate across $50$ generated policies on \texttt{omnetpp},
when the proposer LLM is swapped between five contemporary code
models (Qwen 2.5, DeepSeek Coder, GPT-5, o4-mini, Gemini 2.5 Pro).
Stronger backbones produce tighter, higher-mean distributions, but
even the weakest backbone produces some policies above the
LRU baseline. The four backbones \cfn{} rotates over in the
main-text experiments (\texttt{gpt-oss-20b}, \texttt{Qwen3-Coder},
\texttt{Gemini 2.5 Pro}, \texttt{o4-mini}) are a subset of this
panel's set chosen for licence and tooling availability.
\textbf{(b) Temperature sensitivity.} Mean IPC ($\pm$ std) over
proposer temperature $\tau$ on \texttt{omnetpp}. Performance peaks
near $\tau^{*}=0.7$, the value used in main-text experiments.
\textbf{(c) Seed-count sensitivity.} Best-so-far IPC over $50$
iterations when initialising the loop with the top-$k$ seeds for
$k\in\{2,5,10\}$. More seeds raise the early curve and the eventual
plateau; the main-text protocol uses top-$10$.}
\label{fig:llm_sensitivity}
\end{figure}

\begin{figure}[!htbp]
\centering
\includegraphics[width=0.92\linewidth]{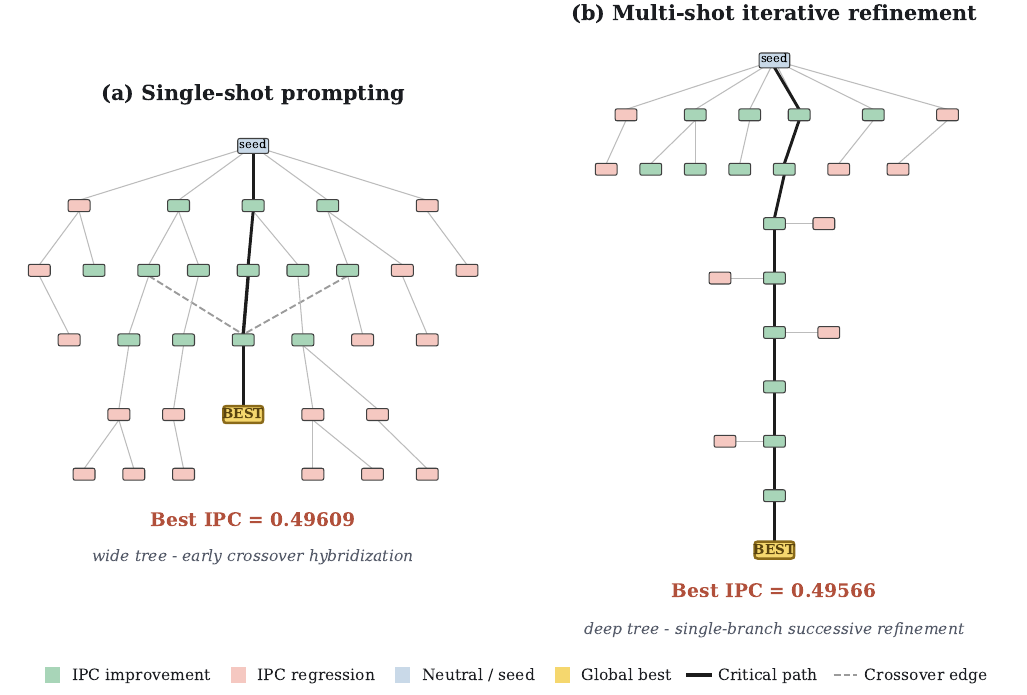}
\caption{Lineage trees from two ablated configurations of \cfn{}.
\textbf{(a) Single-shot prompting} produces a wide tree with early
crossover hybridisation (dashed edges); the global best (IPC
$0.49609$) emerges after only a few iterations, on the
crossover-recombined branch.
\textbf{(b) Multi-shot iterative refinement}, with crossover
disabled, produces a deep, single-branch tree of incremental
refinements; the best policy (IPC $0.49566$) is comparable in
quality but takes many more iterations to reach. The contrast
illustrates the role of crossover described in
\S\ref{sec:int3}: a small number of mode-jumping events early in
the run is structurally different from many incremental
refinements.}
\label{fig:lineage}
\end{figure}

Figures~\ref{fig:treeshapes} and~\ref{fig:llm_sensitivity} compare
seed-count and proposer settings. Figure~\ref{fig:lineage} shows
the lineage trees under single-shot prompting and iterative refinement.

\FloatBarrier
\subsection{Prompt templates}
\label{app:prompts}

The autonomous loop uses a small fixed set of structured prompt
contracts; each is paired with deterministic validators
(schema, code-fence, overhead, invariant). Table~\ref{tab:prompt_contracts}
summarises the contracts; Table~\ref{tab:lineage_snapshot} shows a
concrete lineage block from a live run.

\begin{table}[!htbp]
\scriptsize
\centering
\setlength{\tabcolsep}{3pt}
\renewcommand{\arraystretch}{1.05}
\begin{tabular}{p{0.13\linewidth} p{0.27\linewidth} p{0.22\linewidth} p{0.28\linewidth}}
\toprule
\textbf{Contract} & \textbf{Injected blocks} & \textbf{Excerpt} & \textbf{Validators / guarantees} \\
\midrule
System (always) & Role, fixed $5$-section schema, metadata accounting, victim/bypass invariants, one-shot LRU exemplar.
& \texttt{Output EXACTLY 5 sections}; \texttt{victim in [0, WAYS-1]}; \texttt{No bypass on WRITEBACK}
& Schema, code-fence, overhead-parser, invariant checks. \\
\midrule
Compile-repair & Failing C++ + raw compiler diagnostics + repair-only constraint.
& \texttt{FIX COMPILATION ONLY}; \texttt{Do NOT rename}; \texttt{ONE \textasciigrave\textasciigrave\textasciigrave cpp block}
& CompileGate; DriftGuard (names preserved); bounded retries. \\
\midrule
Exploration & Workloads, $1$--$2$ baselines (code$+$score), template scaffold, novelty instruction.
& \texttt{Propose 3--4 distinct ideas}; \texttt{choose one}; strict $5$-section format.
& NoveltyGate; SchemaValidator; CompileGate; BudgetChecker. \\
\midrule
Exploitation & Parent (full code$+$score), lineage trace, episodic top-$k$, summary memory, edit-budget $\mu$.
& \texttt{Refine parent only}; \texttt{patch}~$\le \mu$; \texttt{// MUTATION}.
& LineageConsistency; EditBudgetEnforcer; SchemaValidator. \\
\midrule
Lineage block & Recent variants newest$\to$oldest, $\le 5$ entries, \texttt{BEST\_SO\_FAR} anchor.
& see Table~\ref{tab:lineage_snapshot}.
& Auditability; anti-repeat. \\
\midrule
Episodic top-$k$ & Similarity-retrieved lessons (name, delta, outcome, lesson).
& \texttt{Relevant prior lessons (Top-k)}.
& Bounded retrieval; anti-redundancy. \\
\midrule
Summary memory & Distilled rules: Principles / Donts / Knobs / Open-Q.
& \texttt{Donts: no dynamic alloc}; \texttt{Knobs: decay/bypass/tiering}.
& Persistent feasibility envelope. \\
\midrule
Feedback / lesson & Aggregated deltas $+$ ``DO NOT REPEAT'' reminders $+$ budget proximity.
& \texttt{DO NOT REPEAT: \ldots}.
& Negative memory. \\
\midrule
Crossover & Two parents in full $+$ logic-aware merge $+$ bounded redesign patch.
& \texttt{Merge complementary mechanisms}; \texttt{redesign $\le 20$ lines}.
& ParentPairValidator; CompileGate; PostCrossoverRefineWindow. \\
\bottomrule
\end{tabular}
\caption{Prompt contracts in \cfn{}. Each row is a concrete
contract used in the autonomous loop; each is paired with
deterministic validators that gate progression to simulation.}
\label{tab:prompt_contracts}
\end{table}

\begin{table}[!htbp]
\small
\centering
\begin{tabular}{ll}
\toprule
\multicolumn{2}{l}{\textbf{Listing: lineage-memory block from a live run}} \\
\midrule
\multicolumn{2}{l}{\texttt{Lineage (newest->oldest):}} \\
\multicolumn{2}{l}{\texttt{\quad iter=3 name=HydraLRU\_v4 avg\_ipc=0.4478 desc=tightened decay; reuse-meter}} \\
\multicolumn{2}{l}{\texttt{\quad iter=2 name=HydraLRU\_v3 avg\_ipc=0.4449 desc=streaming-bypass gate}} \\
\multicolumn{2}{l}{\texttt{\quad iter=1 name=HydraLRU\_v2 avg\_ipc=0.4402 desc=widened RRPV to 3 bits}} \\
\multicolumn{2}{l}{\texttt{\quad iter=0 name=HydraLRU     avg\_ipc=0.4361 desc=baseline two-tier RRIP}} \\
\bottomrule
\end{tabular}
\caption{A concrete snapshot of the lineage block injected into
the exploitation prompt. Newest$\to$oldest, capped at $\le 5$
entries, with an explicit \texttt{BEST\_SO\_FAR} anchor.}
\label{tab:lineage_snapshot}
\end{table}

\paragraph{Literal prompt bodies.}
The summary in Table~\ref{tab:prompt_contracts} fixes the structure;
this paragraph fixes the wording. The prompt assembly uses five
literal blocks: (1)~the system contract, prepended to every call;
(2)~the exploration prompt used when $\text{phase} = \textsc{Explore}$;
(3)~the exploitation prompt used when $\text{phase} = \textsc{Exploit}$;
(4)~the compile-repair prompt; and (5)~the lesson-extraction prompt.
Each block is shown below in the form actually issued to the LLM,
with bracketed placeholders (e.g.\ \texttt{[BUDGET\_KB]},
\texttt{[PARENT\_CODE]}, \texttt{[COMPILER\_LOG]}) marking dynamic
content filled in at call time by the controller.

\paragraph{(1) System contract (always prepended).}
\begin{verbatim}
You are an expert computer architect. Generate a complete, compilable,
ChampSim-compatible C++ replacement policy obeying the interface and
constraints below.

HARD CONSTRAINTS
1) Self-contained. Header + source must match exactly. Same class name.
2) Do not modify simulator, core, or cache infrastructure outside the
   replacement files.
3) Keep metadata storage under [BUDGET_KB] KiB; report your estimate.
4) Avoid dynamic allocation. Avoid deep cascaded decision trees.

CHAMPSIM API RULES (critical)
[Task-specific API rule block. For the L1D prefetcher task this
 includes the champsim::address conversion rule, the LOG2_BLOCK_SIZE
 non-constexpr note, and the prefetch_line(addr, fill_level, meta)
 signature.]

OUTPUT FORMAT (mandatory)
## Policy Name
<short clear name>
## Rationale
<3-6 sentences: core idea, metadata, decision rule, expected tradeoffs>
## C++ Implementation
```cpp
// ===== HEADER =====
<full header>
// ===== SOURCE =====
<full source>
```
## Estimated Storage
<number> KiB, broken down as <breakdown>.
\end{verbatim}

\paragraph{(2) Exploration prompt.}
\begin{verbatim}
[SYSTEM CONTRACT]   [see (1)]

GOAL
- Objective: maximize [OBJECTIVE]   (e.g. IPC, MPKI, useful-pref/1K)
- Target cache/unit: [TARGET]       (e.g. L1D prefetcher, LLC)
- Workloads: [WORKLOADS]            (memory-intensive SPEC slice)

LONG-TERM MEMORY
Principles learned so far:
  - [PRINCIPLE_1]
  - [PRINCIPLE_2]
  ...
Don'ts:
  - [DONT_1]
  - [DONT_2]
  ...
Untested ideas that may be promising:
  - [IDEA_1]
  - [IDEA_2]
  ...

TASK
Propose a NEW design direction. Your policy should be a genuine
departure from any policy in long-term memory. List 3-4 alternative
approaches in your rationale, then pick ONE and implement it.
\end{verbatim}

\paragraph{(3) Exploitation prompt.}
\begin{verbatim}
[SYSTEM CONTRACT]

GOAL    [same as (2)]

PARENT POLICY
Name: [PARENT_NAME]
Rationale: [PARENT_RATIONALE]
Code:
```cpp
[PARENT_CODE]
```
Measured performance:
  IPC = [IPC],  MPKI = [MPKI],  HR = [HR],  storage = [STORAGE] KiB

FEEDBACK
Improvement vs previous: [DELTA] (+/-%).
Observed weaknesses: [WEAKNESSES]

LINEAGE HISTORY
Recent lessons for this lineage:
  - [LESSON_k-2], [LESSON_k-1], [LESSON_k]

TASK
REFINE the parent policy. Keep the core structure; address one or
two weaknesses. Mark changed regions with // MUTATION. Do not
rename the class. Stay within an edit budget of [MU] of the file.
\end{verbatim}

\paragraph{(4) Compile-repair prompt.}
\begin{verbatim}
The following code failed to compile. FIX COMPILATION ONLY.
Do not redesign. Do not rename. Return exactly one ```cpp``` block.

CODE
[CURRENT_CODE]

COMPILER DIAGNOSTIC
[COMPILER_LOG]

Preserve the original design intent. If a fix requires structural
change, explain in <= 20 words why it is unavoidable, then proceed.
\end{verbatim}

\paragraph{(5) Lesson-extraction prompt (issued to the lighter-weight LLM).}
\begin{verbatim}
A new policy was simulated. Write a compact 4-part lesson.

PARENT
Name: [PARENT_NAME]
Metrics: IPC = [PARENT_IPC], HR = [PARENT_HR], ...

CHILD
Name: [CHILD_NAME]
Summary of change: [SHORT_DIFF]
Metrics: IPC = [CHILD_IPC], HR = [CHILD_HR], ...

FORMAT (exactly four lines)
Changed: <=15 words describing what changed mechanically
Happened: <=15 words describing the measured delta
Why: <=20 words explaining the causal story
Next: <=15 words proposing what to try next
\end{verbatim}

The crossover prompt and lineage block are not separate top-level
contracts: they are structured inputs assembled into the
exploitation prompt at iterations where the stagnation gate fires
(see Algorithm~\ref{alg:mosaic}). The literal text injected at
those iterations is the exploitation prompt above with the parent
slot replaced by both parents' code and a brief
\texttt{Merge complementary mechanisms} instruction.

\subsection{Controller algorithm}
\label{app:algorithm}

\begin{algorithm}[H]
\caption{\cfn{} controller, one iteration.}
\label{alg:mosaic}
\begin{algorithmic}[1]
\State \textbf{input:} archive, memory $M_t$, budget $b$, history
\If{$b \leq 0$}\Comment{stagnation $\Rightarrow$ Intervention~3}
  \If{can\_run\_crossover(archive)}
    \State $A, B \gets$ choose\_parents\_max\_min\_emb\_dist(archive) \Comment{Intervention~4}
    \State backbone $\gets$ pick\_excluding\_canon\_like(history)
    \State child $\gets$ LLM[backbone](crossover\_prompt($A, B, M_t$),\ $\tau{=}0.85$)
    \State child $\gets$ post\_crossover\_redesign(child, $M_t$)
  \Else
    \State backbone $\gets$ rotate\_backbone(history)
    \State child $\gets$ LLM[backbone](explore\_prompt($M_t$),\ $\tau{=}1.0$)
  \EndIf
  \State $b \gets b_0$
\Else
  \State backbone $\gets$ round\_robin(history)
  \State child $\gets$ LLM[backbone](exploit\_prompt(parent, $M_t$),\ $\tau{=}0.7$) \Comment{Intervention~2 in $\pi$}
\EndIf
\State child $\gets$ compile\_and\_repair(child, $R_{\max}{=}4$)
\If{not child.compiles \textbf{or} child.over\_budget} \Return state \EndIf
\State $R \gets$ simulate(child); score $\gets$ composite\_reward($R$)
\State archive.add(child, score, embedding(child.code))
\State $M_{t+1} \gets U(M_t, $ child, $R)$ \Comment{Intervention~1: typed update rule}
\State $b \gets$ update\_budget($b$, score, history)
\end{algorithmic}
\end{algorithm}

\subsection{Baseline policy implementations}
\label{app:baselines}

The four policies most heavily used as baselines or as one-shot
exemplars in the prompt are documented here. We ship full C++ source
for each in the planned artifact; this section gives the
reference-quality summary so readers can audit what the controller
was shown.

\paragraph{LRU reference.}
LRU is the sanity floor and the one-shot exemplar in the system
prompt. The full implementation fits on a single page; we reproduce
it verbatim because the in-prompt exemplar is identical except for
comments.
\begin{verbatim}
// lru.cc -- LRU cache replacement for ChampSim CRC-2
#include "cache.h"
uint32_t lru_victim[NUM_SET][NUM_WAY];

void CACHE::initialize_replacement() {
  for (uint32_t s = 0; s < NUM_SET; s++)
    for (uint32_t w = 0; w < NUM_WAY; w++)
      lru_victim[s][w] = w;
}

uint32_t CACHE::find_victim(uint32_t cpu, uint64_t instr_id, uint32_t set,
                            const BLOCK *cb, uint64_t ip, uint64_t addr,
                            uint32_t type) {
  // Evict the way whose lru_victim value equals NUM_WAY - 1 (the coldest).
  for (uint32_t w = 0; w < NUM_WAY; w++)
    if (lru_victim[set][w] == NUM_WAY - 1) return w;
  return 0;  // fallback
}

void CACHE::update_replacement_state(uint32_t cpu, uint32_t set, uint32_t way,
                                     uint64_t full_addr, uint64_t ip,
                                     uint64_t victim_addr, uint32_t type,
                                     uint8_t hit) {
  // Move the touched way to MRU, shift the rest down by one.
  for (uint32_t w = 0; w < NUM_WAY; w++)
    if (lru_victim[set][w] < lru_victim[set][way])
      lru_victim[set][w]++;
  lru_victim[set][way] = 0;
}
\end{verbatim}

\paragraph{SHiP++ and Hawkeye.}
Full C++ source for SHiP++~\citep{young2017shippp} and
Hawkeye~\citep{jain2016hawkeye} is shipped in the planned artifact
under \texttt{baselines/ship\_pp.cc} and
\texttt{baselines/hawkeye.cc}; our implementations match the CRC-2
competition reference code. In our evaluation, we obtain
$0.4131$ search-workload IPC for Hawkeye and $0.4147$ for
MockingJay (the strongest published learned baseline,
\S\ref{sec:eval}).

\paragraph{BertiGo (DPC4 first-place, data prefetching).}
BertiGO~\citep{bertigo2024} combines an enhanced Berti prefetcher
at L1D, Pythia-based prefetching at L2, and adaptive next-line
prefetching at the LLC. The evaluated implementation is stored in
\texttt{baselines/bertigo/}. Its reported metadata footprint is
$29.72$~KiB, below the DPC4 L1D limit of
$32$~KiB~\citep{dpc4_2026}. Appendix~\ref{app:prefetch} reports
the comparison on the SimPoint-weighted memory-intensive subset.

\subsection{Seed programs}
\label{app:seeds}

\cfn{} supports three seed regimes that determine how much prior
information is handed to the LLM at iteration $0$. The default is
\textsc{Structured}; the no-seed and full-model regimes are used
in the seed-conditioning study (\S\ref{app:lineage}).

\paragraph{LowInfo seed (no baseline).}
The LLM receives only the task contract and the objective; no
implementation body is provided. Used in the unseeded runs of the
seed-conditioning study.
\begin{verbatim}
// ===== HEADER =====  [contract excerpt] =====
class CacheReplacement {
public:
  virtual void InitReplacementState() = 0;
  virtual uint32_t GetVictimInSet(uint32_t cpu, uint32_t set,
                                  const BLOCK *cb, uint64_t ip,
                                  uint64_t addr, uint32_t type) = 0;
  virtual void UpdateReplacementState(uint32_t cpu, uint32_t set,
                                      uint32_t way, uint64_t addr,
                                      uint64_t ip, uint64_t victim,
                                      uint32_t type, uint8_t hit) = 0;
  // ... [full interface specified in the prompt] ...
};
// Goal: maximize IPC under the LLC metadata budget.
// No starting code is provided. Implement a replacement policy from scratch.
\end{verbatim}

\paragraph{Structured seed (scaffolded).}
The LLM receives the task contract plus a structured skeleton with
TODO comments. This is the default seed for cache-replacement runs.
\begin{verbatim}
void CACHE::initialize_replacement() {
  // TODO: Allocate per-set metadata. Do not exceed the KiB budget.
}

uint32_t CACHE::find_victim(uint32_t cpu, uint64_t instr_id, uint32_t set,
                            const BLOCK *cb, uint64_t ip, uint64_t addr,
                            uint32_t type) {
  // TODO: Choose a way to evict. Must return a value in [0, NUM_WAY).
  // Consider: which lines are least likely to be reused?
  return 0;
}

void CACHE::update_replacement_state(/* ... */) {
  // TODO: Update metadata on hit/miss. Differentiate access types
  // if useful (Load, RFO, Prefetch, Writeback).
}
\end{verbatim}

\paragraph{FullModel seed (strong prior).}
The LLM receives a complete strong baseline and is asked to refine
it: BertiGo for data prefetching, Hawkeye for cache replacement.
Used in the BertiGo- and Hawkeye-seeded runs of
Appendix~\ref{app:lineage}.

\subsection{Best discovered policies}
\label{app:best}

Each task's best \cfn{} policy is shipped as full C++ source in the
released artifact under \texttt{discovered/}; we summarise the
mechanism here.

\paragraph{\cfn{}-LLC (cache replacement, leaderboard).}
Search-workload IPC $0.41643$ (\texttt{omnetpp}); held-out IPC
$0.41512$; gap $-0.31\%$, the smallest of any LLM-driven
configuration evaluated (Table~\ref{tab:harcom}). Storage
$18.5$~KiB. Core mechanisms: (1) SHiP-style signature table indexed
by PC[7:18]; (2) graph-node table tracking spatial locality within a
$4$\,KiB region; (3) reuse-aware victim selection that scores each
candidate by $\max(\mathrm{RRPV}) - \alpha\!\cdot\!\mathrm{SHCT}[sig]
- \beta\!\cdot\!\mathrm{GraphScore}[region]$; (4) PC-based victim
feedback that decays SHCT counters when an evicted line is later
accessed; (5) periodic graph-score decay ($\times 0.9$ every
$2^{16}$ accesses) to prevent saturation. The same ingredients
appear individually in published policies; the contribution is the
specific composition and the constants chosen by the search loop,
both of which are recorded in the lineage memory.

\paragraph{\cfn{}-DP (data prefetching).}
Mean speedup over LRU $+4.8\%$ on the SimPoint-weighted
memory-intensive subset; exceeds BertiGo on the harmonic mean
(\S\ref{app:prefetch}). Storage $31.67$~KiB, within the DPC4
$32$~KiB budget. Core mechanisms: (1) IP-first, path-bounded
candidate formation; (2) launch-readiness score combining evidence
maturity, recency, and queue pressure; (3) near-window arrival
estimator (predicted arrival minus predicted demand-use window);
(4) merged-duplicate delta handling that keeps only the strongest
producer per unique delta. The four primitives appear as named
functions and state variables in the generated C++ and persist
across lineages, matching the named-mechanism criterion of
Appendix~\ref{app:catalog}.

\subsection{Example policies at varying performance levels}
\label{app:examples}

The headline numbers in \S\ref{sec:eval} report the leaderboard
policy of the \cfn{} archive. To give a sense of the full
distribution of generated candidates, not just the best, we
list seven representative policies spanning low, mid, high,
and best tiers (Table~\ref{tab:examples}). All numbers are IPC over
LRU on the search workload \texttt{omnetpp}, normalised against
LRU's $0.4014$. The best tier corresponds to the leaderboard policy
of Table~\ref{tab:harcom} ($0.41643$, $+3.74\%$); the
low/mid/high tiers are quantiles of the search-workload IPC
distribution across the $\sim\!14{,}000$ generated candidates.

\begin{table}[H]
\small
\centering
\renewcommand{\arraystretch}{1.10}
\setlength{\tabcolsep}{4pt}
\begin{tabular}{llrl}
\toprule
\textbf{Tier} & \textbf{Example policy} & \textbf{IPC over LRU} & \textbf{Mechanism summary} \\
\midrule
Low      & HotInsertOnly      & $-0.6\%$ & Insert all loads at MRU; no PC tracking \\
Low      & LFU-Naive          & $+0.2\%$ & Pure frequency counter per line; no decay \\
Mid      & SigCounter-Light   & $+1.8\%$ & SHiP-style signature table, single SHCT \\
Mid      & RRIP-Stride-Fused  & $+2.5\%$ & RRIP plus stride-pattern-aware insertion \\
High     & TA-DPIP            & $+3.1\%$ & Thrash-aware dual-pattern insertion \\
High     & RSP-MPPP           & $+3.4\%$ & Region--StreamPath multiperspective hybrid \\
\textbf{Best} & \textbf{\cfn{}-LLC} & $\mathbf{+3.74\%}$ & Graph $+$ SHCT $+$ PC feedback $+$ decay \\
\bottomrule
\end{tabular}
\caption{Representative \cfn{}-generated policies across
performance tiers, evaluated on the search workload
(\texttt{omnetpp}). Full source and lineage for each policy are planned for
inclusion in the artifact under \texttt{examples/}.}
\label{tab:examples}
\end{table}

\paragraph{Design features associated with higher IPC.}
Inspection of the lineage memory identifies three patterns among
lower-scoring policies: they ignore access-type distinctions
(treating prefetch and demand identically), they use
counters that saturate without decay, or they couple
decision rules too tightly to a single signal (e.g.\ pure recency
or pure frequency). Higher-scoring policies distinguish at least
Load, RFO, and Prefetch; include an explicit decay or periodic
reset; and combine two or more orthogonal signals (e.g.\ signature,
spatial, and recency) through a scoring function. These are the
same three properties that the Q4--Q7 weighted-survey questions of
Appendix~\ref{app:novelty-taxonomy} score against, and the
low-to-high gap on the survey rubric correlates with the IPC
gap reported here at $r > 0.7$ across the catalogue.

\end{document}